%% file: ascaxrb.tex

\input eplain
\input mn
\input psfig.sty

\input ascaxrb_input

\pssilent
\psfigurepath{./ps}


\pageoffset{-0.8cm}{0cm} 
\tolerance=10000



\begintopmatter
\title{\asca\ and \rs\ observations of the \qsf: the X-ray background in the 
0.1--7~keV band}
\author{L.-W. Chen,$^1$ A. C. Fabian$^1$ and K. C. Gendreau$^2$}  
\affiliation{$^1$ Institute of Astronomy, Madingley Road, Cambridge CB3 0HA}
\smallskip
\affiliation{$^2$ NASA/Goddard Space Flight Center, Greenbelt, MD 20771 U.S.A.}

\shortauthor{L.-W. Chen, A. C. Fabian and K. C. Gendreau}  
\shorttitle{
\asca\ and \rs\ observations of the XRB
}

\abstract{The X-ray background from 0.1 to 7~keV has been studied  using 
high spectral and spatial resolution data from the \asca\  Solid-state Imaging Spectrometers and 
\rs\  Position Sensitive Proportional Counter.   Analysing both the diffuse background radiation and 
resolved sources,  we have carried out a series of joint spectral 
fits of the \asca\ and \rs\ data.  
As found previously with \asca\ data alone, the spectrum of the X-ray 
background can be fit well by a single power-law from 1 to 7~keV;
 to account for the Galactic emission below 1~keV, a power-law plus two
thermal component model fits well to the measurements of 
\asca\ and \rs\ from 0.1 to 7~keV.  Overall, the photon index of  the
power-law model ranges from 1.4 to 1.5, and no obvious excess is found  between
1 and 3~keV as predicted from some previous observations.   Below 1~keV, the
models become more complicated and involve a mixture of   extragalactic and
Galactic sources.  As some of the  extragalactic contributions should be from
point sources, we  have examined  the \asca\ and \rs\ spectra of resolved
sources  individually: a stellar  source having a well-fit thermal spectrum and
two AGNs having a much steeper power-law spectrum (with photon index of about 
3); the accumulated spectrum of other non-stellar sources  resolved by \rs\  
also has a steeper-than-average spectrum.  
Fitting the X-ray background spectrum observed by
\asca\  and  the accumulated point source spectrum by \rs\ together by varying
the   contribution from steep spectrum sources, such as quasars, to the background, we find that they contribute less than 30 per cent in the 0.5--2~keV band and drop to below 10 per cent over 2--10~keV.  This fraction is provided 
 by sources brighter than a few times $10^{-15}$~\ergpercmpers\ 
(in the 0.5--2~keV band).  
Constrained by our spectral fitting results, the major contributor of the X-ray background must be a single population with similar flat spectra.}


\maketitle
\figga      
\figgaa    

\section{ 1. Introduction} 
\tx
The X-ray background (XRB) is the integrated emission of unresolved X-ray
sources along the line of sight. The origin of the XRB is not yet clear and 
is  
 mainly constrained by the observable spatial  and  spectral properties of the
XRB (see Fabian and Barcons 1992 for a review).  The homogeneity and isotropy
of the XRB resulting from the surface number density and clustering of XRB
point sources can be measured by imaging X-ray telescopes;  spectral analysis
of the XRB and resolved sources can provide additional information to identify
the XRB origin.  In other words, the integrated spectra of the assumed sources 
over redshift must reproduce the XRB, in terms of both spectral intensity and 
shape.   

The `spectral paradox' (Boldt 1987) is the common problem that most known X-ray
sources have spectra which are too
soft to account for the observed XRB 
and when their contribution to the XRB is removed the disparity increases.
  The XRB spectrum in the 3--50~keV band measured
by \heao1a2\ (Marshall \etal\ 1980)  resembles  a 40-keV bremsstrahlung
spectrum and is equivalent to a power-law with photon spectral index $\Gamma$
of 1.4 in the 3--10~keV band.   Among X-ray sources in this energy band,
Seyfert 1 galaxies have a $\Gamma=1.7$ power-law spectrum over 2--20~keV (eg,
Turner and Pounds 1989; Nandra 1991), and radio-loud and radio-quiet quasars have
a $\Gamma=1.66$ and $\Gamma=2.0$ power-law spectrum over 2--20~keV (eg Williams
\etal\ 1992), respectively, all of which are steeper than the XRB spectrum.   
  On the other hand, AGN spectra below 3~keV are steeper than those found
at higher energies, the spectral slope of quasars is $\Gamma\sim 2.3-2.5$
according to the data obtained by \rs\ during the all sky survey (Schartel
\etal\ 1994).  AGN may  make a large contribution  to the soft XRB  because
  \rs\ observations have shown that the energy index of the best-fitting XRB
power-law model can vary from 0.5 to 1.1 (Shanks \etal\ 1991; Hasinger 1992;
Wang \& McCray 1993; Chen \etal\ 1994; Georgantopoulos \etal\ 1996); moreover, up to 70 per cent of the
\rs\ XRB has been resolved and most sources are identified as AGN
(Shanks \etal\ 1991; Hasinger \etal\ 1993).  Unless either 
the XRB or AGN has a complicated spectrum,  there is a  problem to
combine observations above and below 3~keV: the steeper  AGN spectra
would not have sufficient photons to fit the hard XRB on the one  hand; 
on the other hand, the
extrapolation from the hard XRB to the soft band would leave a soft
 excess.    Therefore  observations covering a broad bandpass are important
and necessary to explore the origin of the XRB.

\asca\   provides a chance to examine the XRB spectrum from at least  0.4~keV
to almost 10~keV; recent \asca\ observations (Gendreau \etal 1995)  find no evidence
of a soft excess between  1--3~keV and show that the XRB spectral
 shape is well represented by a single
power-law with $\Gamma=1.4$ from 1 to 7~keV consistently.  In this paper we
use \asca\ and \rs\ observations of the \qsf\ to investigate further the 
spectral properties of both the XRB and the resolved sources.  The \asca/\rs\
observations and  data reduction  are described in the next section, followed by
the results of the \asca/\rs\ spectral fitting.   Implications for the origin
of the XRB from our results are discussed in the fourth section and the
conclusions.   

\section{2. \asca/\rs\  Observations and data reduction}
\tx 

\tableaa
\tableab

\subsection{2.1 Observation}
\tx
The \qsf\  is located in the Southern Sky centered at RA (J2000)$=03^{\rm
h}41^{\rm m}45^{\rm s}$, Dec (J2000)$=-44^{\circ}07'08''$
[$(l,b)=250^{\circ}.86, -51^{\circ}.99)$, which is the
pointing center of the ASCA observation used in this work].  Since it has a
high Galactic latitude, the Galactic absorption is low, with an HI column
density of $1.66\times 10^{20}$~cm$^{-2}$ (derived from the
radio observation by Heiles and Cleary 1979).   

The \qsf\  was observed to search for UV excess quasars and the later \rs\
observation (Shanks \etal\ 1991) resolved  $\sim 30$ per cent  of the
XRB into point sources above the flux of $10^{-14}$~\ergpercmpers.  Afterwards, 4 \asca\ observations were carried out and there was another \rs\ observation during the period of the \asca\ observations.  

\figgab
\figgac
\figgba
\figgbb
\figgbc

\tableb
\tablei
\tablej
\tablek

\subsubsection{2.1.1 \asca\ Observation}
\tx
 \asca\  was launched on 1993 Feb 20 (Tanaka \etal 1994) and
 carries 2 Solid-state Imaging Spectrometers (SIS) and 2 
Gas Imaging Spectrometers (GIS) with 4
identical X-Ray Telescopes (XRT; Serlemitsos  \etal 1995).  It observed the  \qsf\  four
times during  the period of   Performance Verification (PV) when 
flight calibrations were  carried out (Table~1a).   The first observation was in July
1993, and the remaining three  observations were in  September 1993.  Because
the clean data of the last three SIS observations have exposure times each totally less
than 10~ks and one data set may have been damaged by telemetry saturation, 
we concentrate only on the first observation by the \asca\ SIS, which has
the best quality for our work.  

\subsubsection{2.1.2 \rs\ Observation}
\tx
Two pointing  observations of the same field by the \rs\  Position Sensitive Proportional Counter (PSPC) carried out in 
1990 and 1993 are also used, of which the earlier and shorter observation
with PSPC-C has been used in Shanks \etal\ (1991).   The pointing centers of the
 \rs\ observations with PSPC-C/B are at 
RA (J2000)$=03^{\rm h}42^{\rm m}24^{\rm s}/03^{\rm h}42^{\rm m}14^{\rm s}$, 
Dec (J2000)$=-44^{\circ}07'48''$, so the most sensitive area of the PSPC just
covers the same sky observed by the \asca\ SIS (see Fig.~2a).  Other detailed information of
these observations is listed in Table~1a.      Grey-scale images of the \qsf\
obtained from the \asca\ SIS and the \rs\ PSPC are shown in Figs.~1a--f and
Fig.~2a, respectively.

\subsection{2.2 Data modeling and screening}
\tx
\subsubsection{2.2.1  \asca\ Data: The Diffuse Foreground}
\tx
\figgc
\figgd
\figgda
\figsixa

The raw \asca\ data were processed by the standard software FTOOLS to convert
the telemetry data to ``modally split science files" (see Day \etal\ 1994, denoted as the \abc\ hereafter) and
we use SISCLEAN  to remove hot and flickering pixels  of SIS CCD chips.    During
the  \qsf\ observation, the bright and faint-modes were used with  4 CCDs, which means that the 4 CCD chips give the maximum field of view (FOV)
for SIS observations.   Observed events are initially split into the 
different modes according to the telemetry rate, and it is possible to merge the 
modal split data after converting
the faint mode  to  bright mode.  Furthermore, SIS X-ray events are classified into eight grades (0--7) based on the charge distribution in a three-by-three CCD pixel block.  For the best quality of data, at the time of writing, only events with grades~0, 2, 3, and 4 are useful for analysing the bright mode data (\eg\ up to 98 per cent of the particle background are associated with grade 7 events).   
Finally, events are accumulated
into Pulse Invariant (PI) data, which are corrected for the chip-to-chip differences in gain and allow us to merge the counts from all 4
CCD chips into one spectrum.    
 
The  detected  events are generally composed of vignetted counts, 
which are the
read-out events of incoming photons originating from cosmic objects, and
non-vignetted counts, which are induced in  the instrument and have not
been reflected by the telescope mirrors.  
The main non-vignetted component is the particle background which
 is dependent on the magnetic cut-off
rigidity (Gendreau 1995).  
This can be calibrated by using the data obtained when \asca\ 
viewed the night earth.  

Because the point spread function (PSF) of the XRT is very broad, we have used 
ray-tracing simulations to  calibrate this energy-dependent broad PSF that 
will include counts from outside the nominal response range (the stray light), which contribute about 30~per cent of the background flux within the FOV at 1~keV, and 22~per cent at 5~keV (see Gendreau 1995 for details).   
A special telescope response was made appropriate for the analysis of the XRB with the SIS data at the MIT Center for Space Research (Gendreau 1995).   Furthermore, the stray light contaminant has particularly less effect on the \qsf\  because there are no strong sources within at least 1~deg radius from the centre of the SIS field of view according to the \rs\ image.  

The vignetted counts contain astronomical information of interest (Galactic
or extragalactic diffuse and discrete sources) as well as other foreground
contaminants such as  scattered solar X-rays.  
The contaminants are removed by filtering out the intervals where they contribute according to a set of housekeeping parameters 
recorded during  observations as a time series (see the \abc).   As a final check, we search for any extraordinary spike (more
than $3\sigma$ above the mean value) appearing in the light curve of the
diffuse background spectrum (resolved-source-free, see below) and reject data
observed during these intervals.    

Based on the housekeeping parameters,  good data are collected when 

\item{1.} the readout triggered by the South Atlantic Anomaly (SAA) particle flux is 0.
\item{2.} more than 1 min has passed since the SAA (T\_SAA). 
\item{3.} the magnetic cut-off rigidity (COT\_MIN) exceeds 8~GeV/$c$.
\item{4.} more than 200 s has elapsed since the day/night transition (T\_DY\_NT).
\item{5.} the elevation angle (ELV\_MIN, the target--satellite--dark earth angle) exceeds 5~deg, and the target--satellite--dark earth angle  (BR\_EARTH) exceeds 20 deg.

\noindent  See Table~1b for a summary.

\subsubsection{2.2.2  \asca\ Data: The Discrete Source}
\tx
   Apart from  using time filters,  ray-tracing simulations, and night-earth 
data to remove the foreground diffuse contaminations, we have 
also excluded resolved sources from the SIS data for  
studying the nature of the extragalactic XRB.
   
Though the source detection ability of \asca\ is limited by its 
intermediate spatial resolution,  \rs\ observations  and follow-up optical
identifications can provide useful information on sources of flux down to
$5\times 10^{-15}$~\ergpercmpers\ in this region.  Overlapping the source
catalog of the \rs\ observation, we see that the 3 brightest PSPC sources (2 quasars and 1 Galactic star) are significantly detected (see Figs.~1--2), as well as some fainter PSPC sources above $2\sigma$.  Results and discussion of source detections are in Sec~3.1.

\figsixb

\subsubsection{2.2.3  \asca\ Data: SIS0/1 Consistency}
\tx
To examine the consistency between the hard/soft SIS0/SIS1 images (eg Figs~1a--d), we have calculated  pixel-to-pixel count rate correlations of hard SIS0--hard SIS1 (the diamonds in Fig~1g), soft SIS0--soft SIS1  (the crosses in Fig~1g), hard SIS0--soft SIS0 (the diamonds in Fig~1h), and hard SIS1--soft SIS1 (the crosses in Fig~1h).  Each data point in Figs~1g--h indicates the count rate (count~s$^{-1}$~sr$^{-1}$) of the corresponding image with bin size $\approx 4'15''\times 4'15''$.   Using a robust least-absolute-deviation method, we fit each set of data points by a linear equation, $y=Ax+B$.  For the SIS0--SIS1 correlation, we obtain the (SIS1 count)/(SIS0 count) ratios equal to 0.91 and 0.98 for the hard (2--8~keV) and soft (0.5--2~keV) bands, respectively (with both best-fitting $B=0$), showing that the two SIS observations are consistent with each other and that any visual fluctuations shown in the images are likely just due to counting statistics.  The error bar is derived from the mean absolute deviation of each data point to the best-fit in the $y$ (ie the SIS1 count rate) direction.  The $1\sigma$ uncertainty intervals are denoted by the dotted (2--8~keV) and solid (0.5--2~keV) lines in Fig~1g.

Fig~1h also shows the consistency of the hard--soft count correlations between SIS0 (the diamonds) and SIS1 (the crosses).  The linear fitting described above has also been performed here, and we obtain $(A,B)=(1.86, 0.06)$ and (1.89,0.08) for SIS0 and SIS1, respectively.  Pixels containing high count rates apparently have a soft excess, and their soft/hard ratio is above the average by a factor of a few.  To translate the ratio to the power-law model, we simulate SIS spectra with various photon indices and calculate the soft/hard ratio of each case as shown in Fig.~1i.  

\subsubsection{2.2.4  \rs\ Data}
\tx
The \rs\ data are processed first by housekeeping parameter screening similar to that of the \asca\ data, followed by  calibrating the particle
background (Snowden \etal 1994; Plucinsky \etal 1993) and excising sources.  
There are about 15 housekeeping parameters to monitor event quality (some of them are correlated) and our data selection criteria are based on the master veto count rate (EE\_MV$<$170~count s$^{-1}$) and the accepted event rate (EE\_AXE$<$ 20~count s$^{-1}$) for both PSPCs, so that after the screening no temporal anomaly is shown in the light curves of these  housekeeping parameters.   

Using FTOOLS, we are able to merge the two \rs\ observations as well as their
calibration files.  Images and spectra are produced from the gain-corrected PI channel processed data.  The exposure map correction has been done before merging the images of the two observations (Fig.~2a).   

In the following analysis,  the \rs\ data will be referred
to as the merged data, and the two SIS data are always fit to a model
simultaneously unless otherwise stated.    All the \rs/\asca\ spectra have been  rebinned   to  at least 20 counts per channel so
that $\chi^2$ statistics can be applied.   

\tablec

\section{3. Spectral fitting and Results}
\tx
The XRB spectrum consists of extragalactic and Galactic diffuse emission as
well as possible unresolved point sources.  Results from previous observations
have  modelled the Galactic XRB well as two Raymond-Smith hot plasma components
(eg Hasinger 1992; Wang \& McCray 1993; Chen \etal\ 1994),  where the soft one
below 0.5~keV represents emission from the local bubble and the hard one (roughly within
 0.5--1~keV) is possibly from the Galactic halo (\moda\ hereafter).  
 Contributions
from point sources are believed to be dominated by  AGN (including QSO), which
have a power-law spectrum.  \moda\ is hereafter referred to as the model of one
power-law plus two Raymond-Smith components.

\subsection{3.1 Spectral fitting of the resolved sources}
\tx
What fraction of the XRB is from discrete sources is still not clear. 
Improvement of X-ray telescopes and detectors can help us to resolve the
previously spatially featureless XRB into discrete sources, which could be 
formally regarded as part of the XRB.  Consequently, comparing the spatial and spectral
characteristics of the resolved new sources with those of the XRB is one
important and necessary stage to identify the real origin of the XRB.

\subsubsection{3.1.1 Source detection: \rs\ data}
\tx
As the \rs\ PSPC has better spatial resolution than the SIS, we use the following steps to detect point sources on the PSPC image first:

\item{1} estimate the median  of the whole image for the global background.
\item{2} for pixels where the total count within a 0.5 or 1~arcmin radius region is $3\sigma$ higher than the global background, flag them as overdense regions.
\item{3} repeat steps~1--2 excluding overdense regions to derive a second (and better estimated) median and re-flag overdense regions.
\item{4} use a circular detection cell to scan the overdense region and calculate the signal-to-noise ratio (SNR), where the noise is the square root of the local background and the local background is the accumulated count of a concentric annulus region with the inner radius of 2~arcmin and the outer radius 3~arcmin while the overdense regions are avoided.   
\item{5} a source is detected when the SNR is over $3\sigma$.  The source position is determined by locating the local maximum position within a 0.75~arcmin radius region of the clustered $3\sigma$ pixels. 

\noindent
The basic pixel size of images during the search is $7.5$~arcsec.  92 sources are detected within the inner 20~arcmin radius region of the PSPC (Table~2a), including all the sources reported by Shanks \etal (1991).    The three brightest sources include one Galactic star and two quasars at $z=0.64$ and 0.38 (sources~11, 37, and 41,  and hereafter \srca, \srcb\ and \srcc, respectively).

\subsubsection{3.1.2 Source detection: \asca\ data}
\tx
Applying the same technique on the merged 0.5--2~keV SIS0+SIS1 image, we have detected well the three brightest PSPC sources along with 7 other PSPC sources above $2\sigma$, though some of them may be affected by nearby bright sources (sources 6 and 15 by \srca, source 40 by \srcc).  6 sources are detected in the 2--8~keV band, including 3 sources which are not seen in the soft band (Table~2b).  The detection cell is 1.5~arcmin radius, and the background is estimated from the global background instead of the local one.

\subsubsection{3.1.3 Source hardness}
\tx
The hardness ratio of the detected sources is also calculated and defined as $(H-S)/(H+S)$, where $H$ is the hard band count (1--2~keV and 2--8~keV for the PSPC and SIS, respectively) and $S$ is the soft count (0.5--1~keV and 0.5--2~keV for the PSPC and SIS, respectively).   The counts of the source and background are extracted in the same way as described in the source detection procedures, except that for the SIS sources we use a 2-arcmin radius area.  The results show no convincing evidence for a strong dependence of the hardness ratio  on the brightness of the PSPC sources (Fig.~2b) for individual sources.  For the whole population however, it appears that the fainter sources are harder (as indicated by the dotted line in Fig.~2b).    

The cross PSPC--SIS hardness ratio is calculated for the SIS sources.  The   PSPC soft band is taken as the soft band, and the SIS soft and the SIS hard counts are the hard count.  Results are shown in Fig.~2c.

\subsubsection{3.1.4 Spectral fitting}
\tx
To produce spectra of resolved sources with enough counts for a meaningful spectral fit, we only work on the three brightest sources with their PSPC-C+PSPC-B and SIS0+SIS1 spectra.  We accumulate the PSPC source spectrum from a circular
region of radius $\sim 1.5$~arcmin around the source position (the full width half maximum of the \rs\ PSPC is about 25~arcsec within the inner 15 arcmin radius region; Hasinger \etal\ 1995), and the background spectrum is produced from a concentric ring with the inner radius $\sim 3$~arcmin and the outer radius  $\sim 7$~arcmin, where bright sources are  removed from the ring.  

For the \asca\ data, we choose a larger area to collect source counts (the half power diameter of the \asca\ XRT is  3~arcmin; the \abc); it is an elliptic region (because of the SIS PSF) with a 4~arcmin semimajor axis and 3.5~arcmin semiminor axis for \srca,  a circle with the radius $\sim$ 4 and 3~arcmin for \srcc\ and \srcb, respectively.  The background spectrum for these three sources is created from the blank region between \srca\ and \srcb.  

Fitting the 2 quasar spectra by a single power-law (with Galactic absorption)
we found that their spectra are much steeper than that of the XRB.  Fitting the
PSPC and SIS data individually, we obtain similar results with good 
reduced chi-square \rchi, the photon index $\Gamma$ ranges from 3.0 to 3.7 for
both \srcb\ and \srcc.  Fitting the \asca/\rs\ data jointly, 
$\Gamma$ is $3.08\pm 0.10$ and $3.15\pm 0.07$ (the uncertainty refers to 90 per cent confidence interval through out this paper unless otherwise stated) for \srcb\ and
\srcc, respectively.   The stellar spectrum is fit by a Raymond-Smith plasma emission model, with $kT=0.81^{+0.05}_{-0.06}$~keV, abundance $Z=0.05\pm 0.01$ solar from the joint
fitting.  Although the abundance is low, an alternative thermal bremsstrahlung cannot have a good fit (\rchi$>2$), showing the data to be sensitive enough to distinguish between these two models.  The results of the separate and joint fitting are shown in Table~3a.  

  To study the spectral property of other fainter sources resolved by the \rs\ PSPC  within the inner $\sim 16$~arcmin
  region in general, especially AGN, we have added up all the spectrum of sources which have been identified 
as AGN  (sources with a `Q' in the `Note' column of Table~3a) and fit them by a power-law.  Not too
surprisingly,  the total spectrum has photon index $\Gamma=2.76^{+0.07}_{-0.06}
$ (\rchi=1.1, see Figs.~8--9 for its folded model and spectrum)
which is  softer than that of average AGN (as described in the Introduction); removing the two bright quasars (\ie, \srcb\ and \srcc) analysed 
above, we obtain the best-fit $\Gamma=2.54^{+0.10}_{-0.11}$ (\rchi=0.96).  
The integrated 
\rs-resolved AGN spectrum with/without \srcb\ and \srcc\ will be denoted 
as \nssa/\nssb\ hereafter.

\subsection{3.2 Spectral fitting of the XRB with \asca\ data}
\tx
Since complications in the spectrum of the 0.1--10~keV cosmic background 
are only significant below 1~keV, we first fit the SIS data in the 1--7~keV 
band with a single
power-law model, giving photon index $\Gamma=1.43\pm 0.08$
 and normalization $A=10.0\pm 0.6$~\uunit, 
with \rchi=1.12.  This result is consistent with recent
work by  Gendreau \etal (1995).  Dividing the spectrum into 1--3 and 3--7~keV
band and fitting by the same model, we have the respectively best-fitting $\Gamma=1.30\pm 0.13$ and $\Gamma=1.40\pm 0.45$ (see Table~3b).

Extending the data down to 0.4~keV, an excess  below 1~keV (mostly 
of Galactic origin) is
clearly seen.  Introducing a Raymond-Smith hot plasma model and the HI column
density  (fixed at $1.66\times 10^{20}$~cm$^{-2}$) to correct the Galactic
absorption, we obtain an almost unchanged best fit of the power-law component,
$\Gamma=1.44_{-0.06}^{+0.07}$ and $A=10.2$~\uunit with \rchi=1.24, while the hot gas has a temperature of 0.12~keV with a fixed solar abundance.   

To include the contribution from the local bubble below 0.5~keV, we add one 
more soft hot plasma component (\moda) and obtain the temperature of the soft 
component \ts=0.02~keV and the hard one \th=0.13~K while the power-law 
component remain unchanged.  Replacing the soft hot plasma component with
either
 a power-law or a bremsstrahlung component does not improve previous fits, and
eventually there are probably only 1--2 binned channels below 0.5~keV to 
constrain the
local bubble model.    This implies that to consider the overall constraints
on the  0.1--7~keV XRB model, it is helpful to have a  joint fit with the
\rs\ data to provide an extra constraint on the 0.1--2.0~keV band.  

\subsection{3.3 Spectral fitting of the XRB with \rs\ data}
\tx
The \rs\ XRB spectrum is accumulated from events within the inner 16~arcmin
radius region, covering a similar sky region to the SIS data (Fig.~2a). 
Only the 3 brightest point sources resolved by \asca\ have been removed from the data so
we can easily compare the result with its \asca\ counterpart, though there are
 other fainter sources detected in this region.  

Fitting the \rs\ spectrum in the 1--2~keV band 
alone by a single power-law, we have
$\Gamma=1.65_{ -0.17}^{+  0.18}$,   $A=11.4_{ -0.60}^{+  0.50}$~\uunit\ and \rchi$=1.14$.  Using the 0.1--2~keV band to fit complicated
models like \moda\ cannot yield a reasonably good fit without further
constraints or assumptions.   On the other hand, to check how the faint
sources of flux from $10^{-13}$ to $10^{-14}$~\ergpercmpers contribute to the XRB, we excise
all the detected PSPC sources from the data (\safxrb,
hereafter).   The excised region is about 2~arcmin radius for the bright sources but smaller for other faint sources because outside the 0.75~arcmin radius their residual counts for most cases contribute negligibly to the XRB.  
The best fit to the
\safxrb\ over 1--2~keV is $\Gamma=1.53$ and $A=6.5$~\uunit.

We note, in comparison to the \asca\ result,  the  steeper result obtained
from \rs\ has also been seen in other \asca/\rs\ simultaneous fitting (eg
Fabian \etal\ 1994), indicating that the discrepancy may be due to
calibration uncertainties in \rs.  

 \figgdc
\subsection{3.4 Joint fitting of the XRB with \asca/\rs\ data}
\tx            
\subsubsection{3.4.1 A Single Power-law Model of the Extragalactic XRB}
\tx
Joint spectral fitting of the \asca/\rs\  data 
has been carried out by forcing each data set to have
the same  photon index for the power-law component and temperature of the
thermal component.    A simultaneous fit of the 1--7~keV \asca\ and
1--2~keV \rs\ spectra by a single power-law gives  $\Gamma=1.48\pm 0.07$ and normalization 10.5~\uunit, \rchi$=1.14$. 
Using the 1--3~keV \asca\  data instead, we have $\Gamma=1.39\pm 0.11$
and normalization 10.3~\uunit, \rchi$=1.07$

Fitting the 0.4--7~keV \asca\ and 0.5--2~keV \rs\ spectra by a Galactic absorbed power-law and Raymond-Smith model, we obtain  a slightly steeper power-law:
$\Gamma=1.47_{ -0.43}^{+  0.07}$  $A=9.8$~\uunit, $kT=0.11$~keV with \rchi=0.98.   It
is still not easy to determine the abundance of the halo gas, as it is not
sensitive to the fitting (with a large error bar) though we have obtained the
best fit at $0.8$ solar abundance, while other best-fitting values remain the
same. 

Finally, we use the total band of both data set, namely, 0.4--7~keV for the 
\asca\ data and 0.1--2~keV for the \rs\ data, to constrain contributions from 
the Galactic halo, local bubble, and the extragalactic sources.  Fitting data 
with \moda, we have $\Gamma=1.46\pm 0.06 $,  $A=10.5$~\uunit,
 \th$=0.12$~keV, \ts$=0.05$~keV, \rchi=1.12.  Using a thermal bremsstrahlung
model for the soft thermal component, the best-fitting  temperature is
\ts$=0.1$~keV; and with a power-law model for the soft component instead, we have
\gs$=4.77$.  For both cases, the best-fitting value of the hard component and
the \rchi\ change little.  

Spectral fitting results from the \asca\  data and  the \asca/\rs\ data
are listed in Table~3b.  Fig.~4 shows the folded spectra of the \asca\ and the
\rs\ data as well as  the best-fit  \moda\ and its components.   

\subsubsection{3.4.2 A Two Power-law Model of the Extragalactic XRB}
\tx
As \rs\ data have resolved a certain fraction of the XRB below 1~keV into
very soft sources, the extragalactic XRB is probably composed of two power-law
components.  Thus we fit  the combined 0.5--2~keV PSPC data and 0.5--7~keV
SIS data jointly by a three-component model including one thermal and two
power-law models to take contributions from the Galactic halo, AGN-like
sources and the yet-unknown XRB sources into account (\modb).  With the photon
 index
of the AGN spectrum $\gsrc$ fixed at 2.5, we have obtained the best-fitting  
$\asrc=3.16$, $\gxrb=1.37$ and $\axrb=8.3$ (\rchi=1.46). 
The unfolded data and models are shown in Fig.~5.  The confidence intervals of
$\gxrb$ and  $\axrb$ are represented by the large contours in Fig.~6a, in
which the small contours are the confidence intervals when one more
constraint of $\asrc=3.2$ is applied.  We will use these results to constrain
 the AGN contribution to the XRB in Sec.~4.

\figge

\subsubsection{3.4.3 A Three Power-law Model of the Extragalactic XRB}
\tx
A three power-law model for the XRB is degenerate, however we know that there is a main power-law component with $\gxrb=1.4$, and a second with $\gsrc=2.53$, $\asrc=2.24$ (\nssb\ in Table~3a) as obtained from previous sections, so we have set these in \modb\ and fitted a further power-law to investigate the range of $\Gamma$ of any third power-law component.  

To eliminate the confusion of the third power-law with the thermal component in \modb, we fit the spectra above 1~keV only.  Results in Fig.~6b show that any significant third power-law component tends to be part of the $\gxrb=1.4$ component.  This emphasises that unresolved sources must have $\Gamma<1.6$, and indicates that apart from the resolved bright sources, which have steep spectra, the rest of the XRB sources is likely dominated by one single flat-spectrum population.

   The best-fitting normalisation of this component is negligible.  Removing this redundant third power-law component, we find that the best-fitting $\gxrb$ for the 1--7~keV XRB spectra is 
$1.38\pm 0.07 $ and $\axrb$ is $8.7_{-0.4}^{+0.3}$ (\modc\ in Table~3b).

\section{4. Discussion} 
\tx
\subsection{4.1 The XRB dilemma: steep spectrum resolved sources and the hard background spectrum} 
\tx
 Previous \rs\ observations (\eg\ Shanks \etal\ 1991; Hasinger \etal\ 1993)
have shown that up to 70 per cent of the observed XRB could be resolved by \rs\ and most of  the resolved 
 point sources have
been identified as AGN.  However, according to our spectral analysis of the XRB
and the resolved sources,  the contributions from steep spectrum sources like the brighter AGN are probably 
less significant than previously indicated.  
The faint resolved sources {\it must} have hard spectra, similar to that of the XRB, as indeed indicated by \rs\ analysis (Hasinger \etal\ 1993, Vikhlinin \etal\ 1995, McHardy 1995, and Almaini \etal\ 1996; see also Fig.~2b).  

In the following subsections, we 
will use the spectral fitting results and observed \logns\ to constrain the steep AGN contribution to the broad band XRB and some spectral properties of the 
unresolved XRB sources.  

\figgf \figgg

\subsection{4.2 Spectral constraints on the AGN contribution to the XRB} 
\tx
Using the results of our \asca/\rs\ joint fit called \modb, we can estimate 
the ratio of the AGN contribution to the XRB$=\isrc/(\isrc+\ixrb)$.  As 
already shown in the last section, the best fits of \asca/\rs\ data above 
0.5~keV from
 \modb\ are $\gxrb=1.37$, $\axrb=8.3$ and $\asrc=3.2$ when $\gsrc$ is 
fixed at 2.5; therefore, the ratio is 27.4, 6.5 and 12.7 per cent in the 
0.5--2~keV, 2--10~keV and 0.5--10~keV band.   Holding the steep power-law 
spectrum fixed, we can calculate the ratio over the $\gxrb$, $\axrb$ space 
covering their  $3\sigma$ confidence intervals as show in Figs.~7a and 7b 
for the soft and hard bands, respectively.  
As indicated by the dashed contour lines, which 
represent the ratio, the $3\sigma$ contribution from the steep AGN to the soft XRB is well below 33 per cent, and below 10 per cent for the hard XRB.   

A direct comparison of the XRB with faint extragalactic source spectra is
shown in Fig.~8, where we use \nssb\ and the \safxrb\ 
spectrum observed with \rs.  The XRB spectrum is fit by \modb\ with
temperatures of the thermal components fixed to the 
best-fitting values from the \asca/\rs\ joint fitting, and the soft power-law component is coupled to that of \nssb, which has $\Gamma=2.53$ and  $A=2.24$ (as shown in Table~3a; the area is scaled according to the XRB data).  In Fig.~8 only the power-law
components are shown, where we can see the XRB power-law (the solid line) is 
harder than the non-stellar spectrum \nssb\ (the dotted line).  Below 1~keV, 
the XRB has a large soft excess much of which is of Galactic origin.  

\figgi

If sources fainter than the threshold have the same spectral shape as 
brighter ones, then we can measure a \safxrb\ spectrum by defining the
extragalactic component of the XRB as

$$\openup2pt\eqalign{
\itot & =\ixrb+\isrc   \cr
      &=\int\,\axrb E^{-\sgxrb}+\cagn \int\,\asrc E^{-\sgsrc},\cr} \eqno(1)
$$ 

\noindent
where $\axrb E^{-\sgxrb}$ is the \safxrb\ spectrum and $\asrc 
E^{-\sgsrc}$
is the best-fitting effective spectrum of the underlying unresolved point
sources and its contribution to the XRB is scaled by a factor $\cagn$.   
Varying the value of $\cagn$, we have obtained several sets of 
$\axrb$, $\sgxrb$, $\asrc$, $\sgsrc$ and the amplitude of the
thermal component $\ath$ (Table~4).  From Columns 1--5 in Table~4, 
the ratio of $\isrc/\itot$ can be calculated, where $\itot$ and  $\isrc$ 
are defined as Eq.~(1).  We plot this result against $\gxrb$
in Fig.~10 (the triangles).   Replacing the background data by the \asca\
(Fig.~9) and
\rs\ data used in Sec.~3 separately (which have more brighter sources), we
repeat the same steps and fitting (Tables~5--6).  In Fig.~10, the crosses and diamonds  are, respectively, ratios in the 0.5--2~keV and 2--10~keV bands based on results
of the SIS XRB/PSPC steep-AGN fitting (Table~5).  If $\gxrb$ is constrained between
1.3 and 1.5, as required by our best spectral fits, then the steep AGN contribution, in which sources have $\Gamma\sim 2.5$, is less than 30 and a few per cent in the 0.5--2~keV and 2--10~keV bands, respectively.   They contribute much  less to the unresolved component of the XRB as indicated by the triangles in Fig.~10, meaning that more flat-spectrum sources are required to compose the entire XRB.

\tablefour
\tablefive
\tablesix

\subsection{4.3 \logns\ constraints on the AGN contribution to the XRB} 
\tx
Another approach to  estimating the contribution from discrete sources to the
XRB involves the observed \logns\ of the resolved sources.  Measurements of
the \logns\ in the soft X-ray band have been derived  by using \rs\ data
[Hasinger \etal\ 1993 and Barcons \etal\ 1994 (in the 0.5--2~keV and 0.9--2.4~keV bands, respectively)] and in the 2--10~keV band by using \heao1a2\ and \ginga\ 
data (Butcher \etal\ 1996).  Here we use the result from Hasinger \etal (1993)
to calculate the total emission from the point source in the band spanning energies $\el$ to $\eh$,

$$\isrc(\el-\eh)=\int^{S_{\rm t}}_{S_{\rm m}}\,SN(S)\,dS 
{\int^{\eh}_{\el}\,E^{-\sgsrc}\,dE \over \int^{2}_{0.5}\,E^{-\sgsrc}\,dE },
\eqno(2)$$

\noindent  where $N(S)$ is the differential number density of  sources at the
flux $S$, $S_{\rm t}$ is the flux threshold beyond which sources cannot be detected
and $S_{\rm m}$ is the faint end where point sources cease their contribution
to the XRB (or otherwise over-produce the XRB).  
  With the XRB spectrum obtained from the last section,  we plot the ratio of  the
soft AGN contribution to the XRB, $\isrc(\el-\eh)/\ixrb(\el-\eh)$
 as a function of $S_{\rm m}$ in 0.5--2~keV and 2--10~keV in Fig.~11, for 
$\sgsrc$ varying from 0.4 to 2,  where  $\ixrb$ is obtained from the results of
\moda\ in Table~3b.   As shown  in Fig.~11,  the soft XRB in the 0.5--2~keV band is likely to be saturated by AGN  when $S_{\rm m}$ is  about
$10^{-17}$~\ergpercmpers;  however, in the 2--10~keV band---as we have to
extrapolate the \logns\ measurement to this band---the results are diverse and
highly dependent on the spectral shape of the faint sources.  As shown in
previous subsections, the contribution from steep AGN to the 2--10~keV XRB is 
at most 10 per cent; therefore, the faintest AGN are
 $\sim$ a few times $10^{-14}$ to $10^{-16}$~\ergpercmpers.    
Resolved sources which contribute more than $\sim 30$~per cent of the 0.5--2~keV XRB must have a spectral shape similar to that of the residual XRB itself (and so is a candidate class for the major contributor).  

\section{5. Conclusion} 
\tx
Having analysed data observed with \asca\ and \rs\ and from our results shown 
in Table~3b, we find that the energy distribution of the extragalactic XRB in 1--7~keV is indeed 
consistently flat, and it is well fit by a single power-law model with 
photon index from 1.4 to 1.5.  
The softer best-fitting results from previous \rs\ data fitting are probably 
due to the restricted energy range and the necessarily 
complicated spectral model.   Actually with a relatively larger error bar, 
some earlier \rs\ observations of the XRB did not rule out a flat soft XRB 
(\eg\ Fig.~3a in Chen \etal\ 1994).  Now the joint fit with \asca\  
in this work has  provided more constraints on the model fitting and improved
 the constraints on  best-fitting values.  

The spectral fitting result of the XRB alone is in agreement with the previous work using \asca\ (Gendreau \etal 1995 and see the Introduction) and recent work with the \rs\ data (Chen \etal\ 1994; Georgantopoulos \etal\ 1996).  Most importantly, our work uses high quality data to integrate the observations across the 0.1--3~keV band and the 3--7~keV band and  provides some more evidence for the single power-law model of the XRB spectrum in this bandpass.                                
                    
This result supports the idea that the origin of the XRB below and above 3~keV could be due to 
the same class of source plus partial contributions from AGN.  Steep-spectrum AGN (such as quasars) make less contribution to the XRB than in previous estimates.  
These soft sources
contribute only 30 per cent below 2 keV and less than 10 per cent above it.  Fainter sources must have hard, flat spectra, similar to that of the XRB.   Indications of this spectrally distinct population are already found in the work of Hasinger \etal\ (1993), Vikhlinin \etal\ (1995), McHardy (1995), Griffiths \etal\ (1996), and Almaini \etal\ (1996).  Since the energy 
density of the XRB increases above 1~keV, this population may produce the dominant X-ray emission in the Universe.

\section*{Acknowledgments}
\tx
We thank the referee for helpful comments. CLW thanks members in the Cambridge X-ray Astronomy group and  Ascahelp on-line at GSFC for various technical advice.  The  \rs\ data were obtained from the UK ROSAT Archive Centre at Leicester.   ACF thanks the Royal Society for support. 

\section*{References}
\tx

\bye

%% file: mn.tex
%
%
%
%

\catcode `\@=11 

\def\@version{1.4}
\def\@verdate{22nd Feb 1994}

%
%
%
%


\newif\ifprod@font

\ifx\@typeface\undefined
  \def\@typeface{Comp. Modern}\prod@fontfalse
\else
  \prod@fonttrue 
\fi

\def\newfam{\alloc@8\fam\chardef\sixt@@n} 

\ifprod@font
\font\fiverm=mtr10 at 5pt
\font\fivebf=mtbx10 at 5pt
\font\fiveit=mtti10 at 5pt
\font\fivesl=mtsl10 at 5pt
\font\fivett=mttt10 at 5pt     \hyphenchar\fivett=-1
\font\fivecsc=mtcsc10 at 5pt
\font\fivesf=mtss10 at 5pt
\font\fivei=mtmi10 at 5pt      \skewchar\fivei='177
\font\fivemib=mtmib10 at 5pt   \skewchar\fivemib='177
\font\fivesy=mtsy10 at 5pt     \skewchar\fivesy='60
\font\fivesyb=mtbsy10 at 5pt   \skewchar\fivesyb='60

\font\sixrm=mtr10 at 6pt
\font\sixbf=mtbx10 at 6pt
\font\sixit=mtti10 at 6pt
\font\sixsl=mtsl10 at 6pt
\font\sixtt=mttt10 at 6pt      \hyphenchar\sixtt=-1
\font\sixcsc=mtcsc10 at 6pt
\font\sixsf=mtss10 at 6pt
\font\sixi=mtmi10 at 6pt       \skewchar\sixi='177
\font\sixmib=mtmib10 at 6pt    \skewchar\sixmib='177
\font\sixsy=mtsy10 at 6pt      \skewchar\sixsy='60
\font\sixsyb=mtbsy10 at 6pt    \skewchar\sixsyb='60

\font\sevenrm=mtr10 at 7pt
\font\sevenbf=mtbx10 at 7pt
\font\sevenit=mtti10 at 7pt
\font\sevensl=mtsl10 at 7pt
\font\seventt=mttt10 at 7pt     \hyphenchar\seventt=-1
\font\sevencsc=mtcsc10 at 7pt
\font\sevensf=mtss10 at 7pt
\font\seveni=mtmi10 at 7pt      \skewchar\seveni='177
\font\sevenmib=mtmib10 at 7pt   \skewchar\sevenmib='177
\font\sevensy=mtsy10 at 7pt     \skewchar\sevensy='60
\font\sevensyb=mtbsy10 at 7pt   \skewchar\sevensyb='60

\font\eightrm=mtr10 at 8pt
\font\eightbf=mtbx10 at 8pt
\font\eightit=mtti10 at 8pt
\font\eighti=mtmi10 at 8pt      \skewchar\eighti='177
\font\eightmib=mtmib10 at 8pt   \skewchar\eightmib='177
\font\eightsy=mtsy10 at 8pt     \skewchar\eightsy='60
\font\eightsyb=mtbsy10 at 8pt   \skewchar\eightsyb='60
\font\eightsl=mtsl10 at 8pt
\font\eighttt=mttt10 at 8pt     \hyphenchar\eighttt=-1
\font\eightcsc=mtcsc10 at 8pt
\font\eightsf=mtss10 at 8pt

\font\ninerm=mtr10 at 9pt
\font\ninebf=mtbx10 at 9pt
\font\nineit=mtti10 at 9pt
\font\ninei=mtmi10 at 9pt      \skewchar\ninei='177
\font\ninemib=mtmib10 at 9pt   \skewchar\ninemib='177
\font\ninesy=mtsy10 at 9pt     \skewchar\ninesy='60
\font\ninesyb=mtbsy10 at 9pt   \skewchar\ninesyb='60
\font\ninesl=mtsl10 at 9pt
\font\ninett=mttt10 at 9pt     \hyphenchar\ninett=-1
\font\ninecsc=mtcsc10 at 9pt
\font\ninesf=mtss10 at 9pt

\font\tenrm=mtr10
\font\tenbf=mtbx10
\font\tenit=mtti10
\font\teni=mtmi10		\skewchar\teni='177
\font\tenmib=mtmib10	\skewchar\tenmib='177
\font\tensy=mtsy10		\skewchar\tensy='60
\font\tensyb=mtbsy10	\skewchar\tensyb='60
\font\tenex=cmex10
\font\tensl=mtsl10
\font\tentt=mttt10		\hyphenchar\tentt=-1
\font\tencsc=mtcsc10
\font\tensf=mtss10

\font\elevenrm=mtr10 at 11pt
\font\elevenbf=mtbx10 at 11pt
\font\elevenit=mtti10 at 11pt
\font\eleveni=mtmi10 at 11pt      \skewchar\eleveni='177
\font\elevenmib=mtmib10 at 11pt   \skewchar\elevenmib='177
\font\elevensy=mtsy10 at 11pt     \skewchar\elevensy='60
\font\elevensyb=mtbsy10 at 11pt   \skewchar\elevensyb='60
\font\elevensl=mtsl10 at 11pt
\font\eleventt=mttt10 at 11pt     \hyphenchar\eleventt=-1
\font\elevencsc=mtcsc10 at 11pt
\font\elevensf=mtss10 at 11pt

\font\twelverm=mtr10 at 12pt
\font\twelvebf=mtbx10 at 12pt
\font\twelveit=mtti10 at 12pt
\font\twelvesl=mtsl10 at 12pt
\font\twelvett=mttt10 at 12pt     \hyphenchar\twelvett=-1
\font\twelvecsc=mtcsc10 at 12pt
\font\twelvesf=mtss10 at 12pt
\font\twelvei=mtmi10 at 12pt      \skewchar\twelvei='177
\font\twelvemib=mtmib10 at 12pt   \skewchar\twelvemib='177
\font\twelvesy=mtsy10 at 12pt     \skewchar\twelvesy='60
\font\twelvesyb=mtbsy10 at 12pt   \skewchar\twelvesyb='60

\font\fourteenrm=mtr10 at 14pt
\font\fourteenbf=mtbx10 at 14pt
\font\fourteenit=mtti10 at 14pt
\font\fourteeni=mtmi10 at 14pt      \skewchar\fourteeni='177
\font\fourteenmib=mtmib10 at 14pt   \skewchar\fourteenmib='177
\font\fourteensy=mtsy10 at 14pt     \skewchar\fourteensy='60
\font\fourteensyb=mtbsy10 at 14pt   \skewchar\fourteensyb='60
\font\fourteensl=mtsl10 at 14pt
\font\fourteentt=mttt10 at 14pt     \hyphenchar\fourteentt=-1
\font\fourteencsc=mtcsc10 at 14pt
\font\fourteensf=mtss10 at 14pt

\font\seventeenrm=mtr10 at 17pt
\font\seventeenbf=mtbx10 at 17pt
\font\seventeenit=mtti10 at 17pt
\font\seventeeni=mtmi10 at 17pt      \skewchar\seventeeni='177
\font\seventeenmib=mtmib10 at 17pt   \skewchar\seventeenmib='177
\font\seventeensy=mtsy10 at 17pt     \skewchar\seventeensy='60
\font\seventeensyb=mtbsy10 at 17pt   \skewchar\seventeensyb='60
\font\seventeensl=mtsl10 at 17pt
\font\seventeentt=mttt10 at 17pt     \hyphenchar\seventeentt=-1
\font\seventeencsc=mtcsc10 at 17pt
\font\seventeensf=mtss10 at 17pt


\newfam\xmfam
\newfam\ymfam

\font\fivexm=mtxm10 at 5pt
\font\sixxm=mtxm10 at 6pt
\font\sevenxm=mtxm10 at 7pt
\font\eightxm=mtxm10 at 8pt
\font\ninexm=mtxm10 at 9pt
\font\tenxm=mtxm10
\font\elevenxm=mtxm10 at 11pt
\font\twelvexm=mtxm10 at 12pt
\font\fourteenxm=mtxm10 at 14pt
\font\seventeenxm=mtxm10 at 17pt

\font\fiveym=mtym10 at 5pt
\font\sixym=mtym10 at 6pt
\font\sevenym=mtym10 at 7pt
\font\eightym=mtym10 at 8pt
\font\nineym=mtym10 at 9pt
\font\tenym=mtym10
\font\elevenym=mtym10 at 11pt
\font\twelveym=mtym10 at 12pt
\font\fourteenym=mtym10 at 14pt
\font\seventeenym=mtym10 at 17pt
\else
\font\fiverm=cmr5
\font\fivei=cmmi5             \skewchar\fivei='177
\font\fivemib=cmmib10 at 5pt  \skewchar\fivemib='177
\font\fivesy=cmsy5            \skewchar\fivesy='60
\font\fivesyb=cmbsy10 at 5pt  \skewchar\fivesyb='60
\font\fivebf=cmbx5

\font\sixrm=cmr6
\font\sixi=cmmi6             \skewchar\sixi='177
\font\sixmib=cmmib10 at 6pt  \skewchar\sixmib='177
\font\sixsy=cmsy6            \skewchar\sixsy='60
\font\sixsyb=cmbsy10 at 6pt  \skewchar\sixsyb='60
\font\sixbf=cmbx6

\font\sevenrm=cmr7
\font\seveni=cmmi7             \skewchar\seveni='177
\font\sevenmib=cmmib10 at 7pt  \skewchar\sevenmib='177
\font\sevensy=cmsy7            \skewchar\sevensy='60
\font\sevensyb=cmbsy10 at 7pt  \skewchar\sevensyb='60
\font\sevenbf=cmbx7

\font\eightrm=cmr8
\font\eightbf=cmbx8
\font\eightit=cmti8
\font\eighti=cmmi8			\skewchar\eighti='177
\font\eightmib=cmmib10 at 8pt	\skewchar\eightmib='177
\font\eightsy=cmsy8			\skewchar\eightsy='60
\font\eightsyb=cmbsy10 at 8pt	\skewchar\eightsyb='60
\font\eightsl=cmsl8
\font\eighttt=cmtt8			\hyphenchar\eighttt=-1
\font\eightcsc=cmcsc10 at 8pt
\font\eightsf=cmss8

\font\ninerm=cmr9
\font\ninebf=cmbx9
\font\nineit=cmti9
\font\ninei=cmmi9			\skewchar\ninei='177
\font\ninemib=cmmib10 at 9pt	\skewchar\ninemib='177
\font\ninesy=cmsy9			\skewchar\ninesy='60
\font\ninesyb=cmbsy10 at 9pt	\skewchar\ninesyb='60
\font\ninesl=cmsl9
\font\ninett=cmtt9			\hyphenchar\ninett=-1
\font\ninecsc=cmcsc10 at 9pt
\font\ninesf=cmss9

\font\tenrm=cmr10
\font\tenbf=cmbx10
\font\tenit=cmti10
\font\teni=cmmi10		\skewchar\teni='177
\font\tenmib=cmmib10	\skewchar\tenmib='177
\font\tensy=cmsy10		\skewchar\tensy='60
\font\tensyb=cmbsy10	\skewchar\tensyb='60
\font\tenex=cmex10
\font\tensl=cmsl10
\font\tentt=cmtt10		\hyphenchar\tentt=-1
\font\tencsc=cmcsc10
\font\tensf=cmss10

\font\elevenrm=cmr10 scaled \magstephalf
\font\elevenbf=cmbx10 scaled \magstephalf
\font\elevenit=cmti10 scaled \magstephalf
\font\eleveni=cmmi10 scaled \magstephalf	\skewchar\eleveni='177
\font\elevenmib=cmmib10 scaled \magstephalf	\skewchar\elevenmib='177
\font\elevensy=cmsy10 scaled \magstephalf	\skewchar\elevensy='60
\font\elevensyb=cmbsy10 scaled \magstephalf	\skewchar\elevensyb='60
\font\elevensl=cmsl10 scaled \magstephalf
\font\eleventt=cmtt10 scaled \magstephalf	\hyphenchar\eleventt=-1
\font\elevencsc=cmcsc10 scaled \magstephalf
\font\elevensf=cmss10 scaled \magstephalf

\font\twelverm=cmr10 scaled \magstep1
\font\twelvebf=cmbx10 scaled \magstep1
\font\twelvei=cmmi10 scaled \magstep1      \skewchar\twelvei='177
\font\twelvemib=cmmib10 scaled \magstep1   \skewchar\twelvemib='177
\font\twelvesy=cmsy10 scaled \magstep1     \skewchar\twelvesy='60
\font\twelvesyb=cmbsy10 scaled \magstep1   \skewchar\twelvesyb='60

\font\fourteenrm=cmr10 scaled \magstep2
\font\fourteenbf=cmbx10 scaled \magstep2
\font\fourteenit=cmti10 scaled \magstep2
\font\fourteeni=cmmi10 scaled \magstep2		\skewchar\fourteeni='177
\font\fourteenmib=cmmib10 scaled \magstep2	\skewchar\fourteenmib='177
\font\fourteensy=cmsy10 scaled \magstep2	\skewchar\fourteensy='60
\font\fourteensyb=cmbsy10 scaled \magstep2	\skewchar\fourteensyb='60
\font\fourteensl=cmsl10 scaled \magstep2
\font\fourteentt=cmtt10 scaled \magstep2	\hyphenchar\fourteentt=-1
\font\fourteencsc=cmcsc10 scaled \magstep2
\font\fourteensf=cmss10 scaled \magstep2

\font\seventeenrm=cmr10 scaled \magstep3
\font\seventeenbf=cmbx10 scaled \magstep3
\font\seventeenit=cmti10 scaled \magstep3
\font\seventeeni=cmmi10 scaled \magstep3	\skewchar\seventeeni='177
\font\seventeenmib=cmmib10 scaled \magstep3	\skewchar\seventeenmib='177
\font\seventeensy=cmsy10 scaled \magstep3	\skewchar\seventeensy='60
\font\seventeensyb=cmbsy10 scaled \magstep3	\skewchar\seventeensyb='60
\font\seventeensl=cmsl10 scaled \magstep3
\font\seventeentt=cmtt10 scaled \magstep3	\hyphenchar\seventeentt=-1
\font\seventeencsc=cmcsc10 scaled \magstep3
\font\seventeensf=cmss10 scaled \magstep3
\fi

\def\hexnumber#1{\ifcase#1 0\or1\or2\or3\or4\or5\or6\or7\or8\or9\or
  A\or B\or C\or D\or E\or F\fi}

\ifprod@font
  \edef\@xm{\hexnumber\xmfam}
  \edef\@ym{\hexnumber\ymfam}
\fi

\def\makestrut{%
  \setbox\strutbox=\hbox{%
    \vrule height.7\baselineskip depth.3\baselineskip width \z@}%
}

\def\baselinestretch{1}
\newskip\tmp@bls

\def\b@ls#1{
  \tmp@bls=#1\relax
  \baselineskip=#1\relax\makestrut
  \normalbaselineskip=\baselinestretch\tmp@bls
  \normalbaselines
}

\def\nostb@ls#1{
  \normalbaselineskip=#1\relax
  \normalbaselines
  \makestrut
}

%

\newfam\mibfam 
\newfam\sybfam 
\newfam\scfam  
\newfam\sffam  

\def\mit{\fam\@ne}

\def\cal{\fam\tw@}

\def\em{\ifdim\fontdimen1\font>\z@ \rm\else\it\fi}

\textfont3=\tenex
\scriptfont3=\tenex
\scriptscriptfont3=\tenex

\setbox0=\hbox{\tenex B} \p@renwd=\wd0 

\def\eightpoint{
  \def\rm{\fam0\eightrm}%
  \textfont0=\eightrm \scriptfont0=\sixrm \scriptscriptfont0=\fiverm%
  \textfont1=\eighti  \scriptfont1=\sixi  \scriptscriptfont1=\fivei%
  \textfont2=\eightsy \scriptfont2=\sixsy \scriptscriptfont2=\fivesy%
  \textfont\itfam=\eightit\def\it{\fam\itfam\eightit}%
  \ifprod@font
    \scriptfont\itfam=\sixit
      \scriptscriptfont\itfam=\fiveit
  \else
    \scriptfont\itfam=\eightit
      \scriptscriptfont\itfam=\eightit
  \fi
  \textfont\bffam=\eightbf%
    \scriptfont\bffam=\sixbf%
      \scriptscriptfont\bffam=\fivebf%
  \def\bf{\fam\bffam\eightbf}%
  \textfont\slfam=\eightsl\def\sl{\fam\slfam\eightsl}%
  \ifprod@font
    \scriptfont\slfam=\sixsl
      \scriptscriptfont\slfam=\fivesl
  \else
    \scriptfont\slfam=\eightsl
      \scriptscriptfont\slfam=\eightsl
  \fi
  \textfont\ttfam=\eighttt\def\tt{\fam\ttfam\eighttt}%
  \ifprod@font
    \scriptfont\ttfam=\sixtt
      \scriptscriptfont\ttfam=\fivett
  \else
    \scriptfont\ttfam=\eighttt
      \scriptscriptfont\ttfam=\eighttt
  \fi
  \textfont\scfam=\eightcsc\def\sc{\fam\scfam\eightcsc}%
  \ifprod@font
    \scriptfont\scfam=\sixcsc
      \scriptscriptfont\scfam=\fivecsc
  \else
    \scriptfont\scfam=\eightcsc
      \scriptscriptfont\scfam=\eightcsc
  \fi
  \textfont\sffam=\eightsf\def\sf{\fam\sffam\eightsf}%
  \ifprod@font
    \scriptfont\sffam=\sixsf
      \scriptscriptfont\sffam=\fivesf
  \else
    \scriptfont\sffam=\eightsf
      \scriptscriptfont\sffam=\eightsf
  \fi
  \textfont\mibfam=\eightmib
    \scriptfont\mibfam=\sixmib
      \scriptscriptfont\mibfam=\fivemib
  \textfont\sybfam=\eightsyb
    \scriptfont\sybfam=\sixsyb
      \scriptscriptfont\sybfam=\fivesyb
  \ifprod@font
    \textfont\xmfam=\eightxm
      \scriptfont\xmfam=\sixxm
        \scriptscriptfont\xmfam=\fivexm
    \textfont\ymfam=\eightym
      \scriptfont\ymfam=\sixym
        \scriptscriptfont\ymfam=\fiveym
  \fi
  \def\oldstyle{\fam\@ne\eighti}%
  \def\boldstyle{\fam\mibfam\eightmib}%
  \b@ls{10pt}\rm%
}

\def\ninepoint{
  \def\rm{\fam0\ninerm}%
  \textfont0=\ninerm \scriptfont0=\sixrm \scriptscriptfont0=\fiverm%
  \textfont1=\ninei  \scriptfont1=\sixi  \scriptscriptfont1=\fivei%
  \textfont2=\ninesy \scriptfont2=\sixsy \scriptscriptfont2=\fivesy%
  \textfont\itfam=\nineit\def\it{\fam\itfam\nineit}%
  \ifprod@font
    \scriptfont\itfam=\sixit
      \scriptscriptfont\itfam=\fiveit
  \else
    \scriptfont\itfam=\nineit
      \scriptscriptfont\itfam=\nineit
  \fi
  \textfont\bffam=\ninebf%
    \scriptfont\bffam=\sixbf%
      \scriptscriptfont\bffam=\fivebf%
  \def\bf{\fam\bffam\ninebf}%
  \textfont\slfam=\ninesl\def\sl{\fam\slfam\ninesl}%
  \ifprod@font
    \scriptfont\slfam=\sixsl
      \scriptscriptfont\slfam=\fivesl
  \else
    \scriptfont\slfam=\ninesl
      \scriptscriptfont\slfam=\ninesl
  \fi
  \textfont\ttfam=\ninett\def\tt{\fam\ttfam\ninett}%
  \ifprod@font
    \scriptfont\ttfam=\sixtt
      \scriptscriptfont\ttfam=\fivett
  \else
    \scriptfont\ttfam=\ninett
      \scriptscriptfont\ttfam=\ninett
  \fi
  \textfont\scfam=\ninecsc\def\sc{\fam\scfam\ninecsc}%
  \ifprod@font
    \scriptfont\scfam=\sixcsc
      \scriptscriptfont\scfam=\fivecsc
  \else
    \scriptfont\scfam=\ninecsc
      \scriptscriptfont\scfam=\ninecsc
  \fi
  \textfont\sffam=\ninesf\def\sf{\fam\sffam\ninesf}%
  \ifprod@font
    \scriptfont\sffam=\sixsf
      \scriptscriptfont\sffam=\fivesf
  \else
    \scriptfont\sffam=\ninesf
      \scriptscriptfont\sffam=\ninesf
  \fi
  \textfont\mibfam=\ninemib
    \scriptfont\mibfam=\sixmib
      \scriptscriptfont\mibfam=\fivemib
  \textfont\sybfam=\ninesyb
    \scriptfont\sybfam=\sixsyb
      \scriptscriptfont\sybfam=\fivesyb
  \ifprod@font
    \textfont\xmfam=\ninexm
      \scriptfont\xmfam=\sixxm
        \scriptscriptfont\xmfam=\fivexm
    \textfont\ymfam=\nineym
      \scriptfont\ymfam=\sixym
        \scriptscriptfont\ymfam=\fiveym
  \fi
  \def\oldstyle{\fam\@ne\ninei}%
  \def\boldstyle{\fam\mibfam\ninemib}%
  \b@ls{\TextLeading plus \Feathering}\rm%
}

\def\tenpoint{
  \def\rm{\fam0\tenrm}%
  \textfont0=\tenrm \scriptfont0=\sevenrm \scriptscriptfont0=\fiverm%
  \textfont1=\teni  \scriptfont1=\seveni  \scriptscriptfont1=\fivei%
  \textfont2=\tensy \scriptfont2=\sevensy \scriptscriptfont2=\fivesy%
  \textfont\itfam=\tenit\def\it{\fam\itfam\tenit}%
  \ifprod@font
    \scriptfont\itfam=\sevenit
      \scriptscriptfont\itfam=\fiveit
  \else
    \scriptfont\itfam=\tenit
      \scriptscriptfont\itfam=\tenit
  \fi
  \textfont\bffam=\tenbf%
    \scriptfont\bffam=\sevenbf%
      \scriptscriptfont\bffam=\fivebf%
  \def\bf{\fam\bffam\tenbf}%
  \textfont\slfam=\tensl\def\sl{\fam\slfam\tensl}%
  \ifprod@font
    \scriptfont\slfam=\sevensl
      \scriptscriptfont\slfam=\fivesl
  \else
    \scriptfont\slfam=\tensl
      \scriptscriptfont\slfam=\tensl
  \fi
  \textfont\ttfam=\tentt\def\tt{\fam\ttfam\tentt}%
  \ifprod@font
    \scriptfont\ttfam=\seventt
      \scriptscriptfont\ttfam=\fivett
  \else
    \scriptfont\ttfam=\tentt
      \scriptscriptfont\ttfam=\tentt
  \fi
  \textfont\scfam=\tencsc\def\sc{\fam\scfam\tencsc}%
  \ifprod@font
    \scriptfont\scfam=\sevencsc
      \scriptscriptfont\scfam=\fivecsc
  \else
    \scriptfont\scfam=\tencsc
      \scriptscriptfont\scfam=\tencsc
  \fi
  \textfont\sffam=\tensf\def\sf{\fam\sffam\tensf}%
  \ifprod@font
    \scriptfont\sffam=\sevensf
      \scriptscriptfont\sffam=\fivesf
  \else
    \scriptfont\sffam=\tensf
      \scriptscriptfont\sffam=\tensf
  \fi
  \textfont\mibfam=\tenmib
    \scriptfont\mibfam=\sevenmib
      \scriptscriptfont\mibfam=\fivemib
  \textfont\sybfam=\tensyb
    \scriptfont\sybfam=\sevensyb
      \scriptscriptfont\sybfam=\fivesyb
  \ifprod@font
    \textfont\xmfam=\tenxm
      \scriptfont\xmfam=\sevenxm
        \scriptscriptfont\xmfam=\fivexm
    \textfont\ymfam=\tenym
      \scriptfont\ymfam=\sevenym
        \scriptscriptfont\ymfam=\fiveym
  \fi
  \def\oldstyle{\fam\@ne\teni}%
  \def\boldstyle{\fam\mibfam\tenmib}%
  \b@ls{11pt}\rm%
}

\def\elevenpoint{
  \def\rm{\fam0\elevenrm}%
  \textfont0=\elevenrm \scriptfont0=\eightrm \scriptscriptfont0=\sixrm%
  \textfont1=\eleveni  \scriptfont1=\eighti  \scriptscriptfont1=\sixi%
  \textfont2=\elevensy \scriptfont2=\eightsy \scriptscriptfont2=\sixsy%
  \textfont\itfam=\elevenit\def\it{\fam\itfam\elevenit}%
  \ifprod@font
    \scriptfont\itfam=\eightit
      \scriptscriptfont\itfam=\sixit
  \else
    \scriptfont\itfam=\elevenit
      \scriptscriptfont\itfam=\elevenit
  \fi
  \textfont\bffam=\elevenbf%
    \scriptfont\bffam=\eightbf%
      \scriptscriptfont\bffam=\sixbf%
  \def\bf{\fam\bffam\elevenbf}%
  \textfont\slfam=\elevensl\def\sl{\fam\slfam\elevensl}%
  \ifprod@font
    \scriptfont\slfam=\eightsl
      \scriptscriptfont\slfam=\sixsl
  \else
    \scriptfont\slfam=\elevensl
      \scriptscriptfont\slfam=\elevensl
  \fi
  \textfont\ttfam=\eleventt\def\tt{\fam\ttfam\eleventt}%
  \ifprod@font
    \scriptfont\ttfam=\eighttt
      \scriptscriptfont\ttfam=\sixtt
  \else
    \scriptfont\ttfam=\eleventt
      \scriptscriptfont\ttfam=\eleventt
  \fi
  \textfont\scfam=\elevencsc\def\sc{\fam\scfam\elevencsc}%
  \ifprod@font
    \scriptfont\scfam=\eightcsc
      \scriptscriptfont\scfam=\sixcsc
  \else
    \scriptfont\scfam=\elevencsc
      \scriptscriptfont\scfam=\elevencsc
  \fi
  \textfont\sffam=\elevensf\def\sf{\fam\sffam\elevensf}%
  \ifprod@font
    \scriptfont\sffam=\eightsf
      \scriptscriptfont\sffam=\sixsf
  \else
    \scriptfont\sffam=\elevensf
      \scriptscriptfont\sffam=\elevensf
  \fi
  \textfont\mibfam=\elevenmib
    \scriptfont\mibfam=\eightmib
      \scriptscriptfont\mibfam=\sixmib
  \textfont\sybfam=\elevensyb
    \scriptfont\sybfam=\eightsyb
      \scriptscriptfont\sybfam=\sixsyb
  \ifprod@font
    \textfont\xmfam=\elevenxm
      \scriptfont\xmfam=\eightxm
       \scriptscriptfont\xmfam=\sixxm
    \textfont\ymfam=\elevenym
      \scriptfont\ymfam=\eightym
        \scriptscriptfont\ymfam=\sixym
   \fi
  \def\oldstyle{\fam\@ne\eleveni}%
  \def\boldstyle{\fam\mibfam\elevenmib}%
  \b@ls{13pt}\rm%
}

\def\fourteenpoint{
  \def\rm{\fam0\fourteenrm}%
  \textfont0\fourteenrm  \scriptfont0\tenrm  \scriptscriptfont0\sevenrm%
  \textfont1\fourteeni   \scriptfont1\teni   \scriptscriptfont1\seveni%
  \textfont2\fourteensy  \scriptfont2\tensy  \scriptscriptfont2\sevensy%
  \textfont\itfam=\fourteenit\def\it{\fam\itfam\fourteenit}%
  \ifprod@font
    \scriptfont\itfam=\tenit
      \scriptscriptfont\itfam=\sevenit
  \else
    \scriptfont\itfam=\fourteenit
      \scriptscriptfont\itfam=\fourteenit
  \fi
  \textfont\bffam=\fourteenbf%
    \scriptfont\bffam=\tenbf%
      \scriptscriptfont\bffam=\sevenbf%
  \def\bf{\fam\bffam\fourteenbf}%
  \textfont\slfam=\fourteensl\def\sl{\fam\slfam\fourteensl}%
  \ifprod@font
    \scriptfont\slfam=\tensl
      \scriptscriptfont\slfam=\sevensl
  \else
    \scriptfont\slfam=\fourteensl
      \scriptscriptfont\slfam=\fourteensl
  \fi
  \textfont\ttfam=\fourteentt\def\tt{\fam\ttfam\fourteentt}%
  \ifprod@font
    \scriptfont\ttfam=\tentt
      \scriptscriptfont\ttfam=\seventt
  \else
    \scriptfont\ttfam=\fourteentt
      \scriptscriptfont\ttfam=\fourteentt
  \fi
  \textfont\scfam=\fourteencsc\def\sc{\fam\scfam\fourteencsc}%
  \ifprod@font
    \scriptfont\scfam=\tencsc
      \scriptscriptfont\scfam=\sevencsc
  \else
    \scriptfont\scfam=\fourteencsc
      \scriptscriptfont\scfam=\fourteencsc
  \fi
  \textfont\sffam=\fourteensf\def\sf{\fam\sffam\fourteensf}%
  \ifprod@font
    \scriptfont\sffam=\tensf
      \scriptscriptfont\sffam=\sevensf
  \else
    \scriptfont\sffam=\fourteensf
      \scriptscriptfont\sffam=\fourteensf
  \fi
  \textfont\mibfam=\fourteenmib
    \scriptfont\mibfam=\tenmib
      \scriptscriptfont\mibfam=\sevenmib
  \textfont\sybfam=\fourteensyb
    \scriptfont\sybfam=\tensyb
      \scriptscriptfont\sybfam=\sevensyb
  \ifprod@font
    \textfont\xmfam=\fourteenxm
      \scriptfont\xmfam=\tenxm
        \scriptscriptfont\xmfam=\sevenxm
   \textfont\ymfam=\fourteenym
      \scriptfont\ymfam=\tenym
        \scriptscriptfont\ymfam=\sevenym
  \fi
  \def\oldstyle{\fam\@ne\fourteeni}%
  \def\boldstyle{\fam\mibfam\fourteenmib}%
  \b@ls{17pt}\rm%
}

\def\seventeenpoint{
  \def\rm{\fam0\seventeenrm}%
  \textfont0\seventeenrm  \scriptfont0\twelverm  \scriptscriptfont0\tenrm%
  \textfont1\seventeeni   \scriptfont1\twelvei   \scriptscriptfont1\teni%
  \textfont2\seventeensy  \scriptfont2\twelvesy  \scriptscriptfont2\tensy%
  \textfont\itfam=\seventeenit\def\it{\fam\itfam\seventeenit}%
  \ifprod@font
    \scriptfont\itfam=\twelveit
      \scriptscriptfont\itfam=\tenit
  \else
    \scriptfont\itfam=\seventeenit
      \scriptscriptfont\itfam=\seventeenit
  \fi
  \textfont\bffam=\seventeenbf%
    \scriptfont\bffam=\twelvebf%
      \scriptscriptfont\bffam=\tenbf%
  \def\bf{\fam\bffam\seventeenbf}%
  \textfont\slfam=\seventeensl\def\sl{\fam\slfam\seventeensl}%
  \ifprod@font
    \scriptfont\slfam=\twelvesl
      \scriptscriptfont\slfam=\tensl
  \else
    \scriptfont\slfam=\seventeensl
      \scriptscriptfont\slfam=\seventeensl
  \fi
  \textfont\ttfam=\seventeentt\def\tt{\fam\ttfam\seventeentt}%
  \ifprod@font
    \scriptfont\ttfam=\twelvett
      \scriptscriptfont\ttfam=\tentt
  \else
    \scriptfont\ttfam=\seventeentt
      \scriptscriptfont\ttfam=\seventeentt
  \fi
  \textfont\scfam=\seventeencsc\def\sc{\fam\scfam\seventeencsc}%
  \ifprod@font
    \scriptfont\scfam=\twelvecsc
      \scriptscriptfont\scfam=\tencsc
  \else
    \scriptfont\scfam=\seventeencsc
      \scriptscriptfont\scfam=\seventeencsc
  \fi
  \textfont\sffam=\seventeensf\def\sf{\fam\sffam\seventeensf}%
  \ifprod@font
    \scriptfont\sffam=\twelvesf
      \scriptscriptfont\sffam=\tensf
  \else
    \scriptfont\sffam=\seventeensf
      \scriptscriptfont\sffam=\seventeensf
  \fi
  \textfont\mibfam=\seventeenmib
    \scriptfont\mibfam=\twelvemib
      \scriptscriptfont\mibfam=\tenmib
  \textfont\sybfam=\seventeensyb
    \scriptfont\sybfam=\twelvesyb
      \scriptscriptfont\sybfam=\tensyb
  \ifprod@font
    \textfont\xmfam=\seventeenxm
      \scriptfont\xmfam=\twelvexm
        \scriptscriptfont\xmfam=\tenxm
    \textfont\ymfam=\seventeenym
      \scriptfont\ymfam=\twelveym
        \scriptscriptfont\ymfam=\tenym
  \fi
  \def\oldstyle{\fam\@ne\seventeeni}%
  \def\boldstyle{\fam\mibfam\seventeenmib}%
  \b@ls{20pt}\rm%
}

\lineskip=1pt      \normallineskip=\lineskip
\lineskiplimit=\z@ \normallineskiplimit=\lineskiplimit



\def\la{\mathrel{\mathchoice {\vcenter{\offinterlineskip\halign{\hfil
$\displaystyle##$\hfil\cr<\cr\sim\cr}}}
{\vcenter{\offinterlineskip\halign{\hfil$\textstyle##$\hfil\cr
<\cr\sim\cr}}}
{\vcenter{\offinterlineskip\halign{\hfil$\scriptstyle##$\hfil\cr
<\cr\sim\cr}}}
{\vcenter{\offinterlineskip\halign{\hfil$\scriptscriptstyle##$\hfil\cr
<\cr\sim\cr}}}}}

\def\ga{\mathrel{\mathchoice {\vcenter{\offinterlineskip\halign{\hfil
$\displaystyle##$\hfil\cr>\cr\sim\cr}}}
{\vcenter{\offinterlineskip\halign{\hfil$\textstyle##$\hfil\cr
>\cr\sim\cr}}}
{\vcenter{\offinterlineskip\halign{\hfil$\scriptstyle##$\hfil\cr
>\cr\sim\cr}}}
{\vcenter{\offinterlineskip\halign{\hfil$\scriptscriptstyle##$\hfil\cr
>\cr\sim\cr}}}}}

\def\getsto{\mathrel{\mathchoice {\vcenter{\offinterlineskip
\halign{\hfil
$\displaystyle##$\hfil\cr\gets\cr\to\cr}}}
{\vcenter{\offinterlineskip\halign{\hfil$\textstyle##$\hfil\cr\gets
\cr\to\cr}}}
{\vcenter{\offinterlineskip\halign{\hfil$\scriptstyle##$\hfil\cr\gets
\cr\to\cr}}}
{\vcenter{\offinterlineskip\halign{\hfil$\scriptscriptstyle##$\hfil\cr
\gets\cr\to\cr}}}}}

\def\lid{\mathrel{\mathchoice {\vcenter{\offinterlineskip\halign{\hfil
$\displaystyle##$\hfil\cr<\cr\noalign{\vskip1.2pt}=\cr}}}
{\vcenter{\offinterlineskip\halign{\hfil$\textstyle##$\hfil\cr<\cr
\noalign{\vskip1.2pt}=\cr}}}
{\vcenter{\offinterlineskip\halign{\hfil$\scriptstyle##$\hfil\cr<\cr
\noalign{\vskip1pt}=\cr}}}
{\vcenter{\offinterlineskip\halign{\hfil$\scriptscriptstyle##$\hfil\cr
<\cr
\noalign{\vskip0.9pt}=\cr}}}}}

\def\gid{\mathrel{\mathchoice {\vcenter{\offinterlineskip\halign{\hfil
$\displaystyle##$\hfil\cr>\cr\noalign{\vskip1.2pt}=\cr}}}
{\vcenter{\offinterlineskip\halign{\hfil$\textstyle##$\hfil\cr>\cr
\noalign{\vskip1.2pt}=\cr}}}
{\vcenter{\offinterlineskip\halign{\hfil$\scriptstyle##$\hfil\cr>\cr
\noalign{\vskip1pt}=\cr}}}
{\vcenter{\offinterlineskip\halign{\hfil$\scriptscriptstyle##$\hfil\cr
>\cr
\noalign{\vskip0.9pt}=\cr}}}}}

\def\grole{\mathrel{\mathchoice {\vcenter{\offinterlineskip\halign{\hfil
$\displaystyle##$\hfil\cr>\cr\noalign{\vskip-1.5pt}<\cr}}}
{\vcenter{\offinterlineskip\halign{\hfil$\textstyle##$\hfil\cr
>\cr\noalign{\vskip-1.5pt}<\cr}}}
{\vcenter{\offinterlineskip\halign{\hfil$\scriptstyle##$\hfil\cr
>\cr\noalign{\vskip-1pt}<\cr}}}
{\vcenter{\offinterlineskip\halign{\hfil$\scriptscriptstyle##$\hfil\cr
>\cr\noalign{\vskip-0.5pt}<\cr}}}}}

\def\leogr{\mathrel{\mathchoice {\vcenter{\offinterlineskip\halign{\hfil
$\displaystyle##$\hfil\cr<\cr\noalign{\vskip-1.5pt}>\cr}}}
{\vcenter{\offinterlineskip\halign{\hfil$\textstyle##$\hfil\cr
<\cr\noalign{\vskip-1.5pt}>\cr}}}
{\vcenter{\offinterlineskip\halign{\hfil$\scriptstyle##$\hfil\cr
<\cr\noalign{\vskip-1pt}>\cr}}}
{\vcenter{\offinterlineskip\halign{\hfil$\scriptscriptstyle##$\hfil\cr
<\cr\noalign{\vskip-0.5pt}>\cr}}}}}

\def\loa{\mathrel{\mathchoice {\vcenter{\offinterlineskip\halign{\hfil
$\displaystyle##$\hfil\cr<\cr\approx\cr}}}
{\vcenter{\offinterlineskip\halign{\hfil$\textstyle##$\hfil\cr
<\cr\approx\cr}}}
{\vcenter{\offinterlineskip\halign{\hfil$\scriptstyle##$\hfil\cr
<\cr\approx\cr}}}
{\vcenter{\offinterlineskip\halign{\hfil$\scriptscriptstyle##$\hfil\cr
<\cr\approx\cr}}}}}

\def\goa{\mathrel{\mathchoice {\vcenter{\offinterlineskip\halign{\hfil
$\displaystyle##$\hfil\cr>\cr\approx\cr}}}
{\vcenter{\offinterlineskip\halign{\hfil$\textstyle##$\hfil\cr
>\cr\approx\cr}}}
{\vcenter{\offinterlineskip\halign{\hfil$\scriptstyle##$\hfil\cr
>\cr\approx\cr}}}
{\vcenter{\offinterlineskip\halign{\hfil$\scriptscriptstyle##$\hfil\cr
>\cr\approx\cr}}}}}

\def\diameter{{\ifmmode\mathchoice
{\ooalign{\hfil\hbox{$\displaystyle/$}\hfil\crcr
{\hbox{$\displaystyle\mathchar"20D$}}}}
{\ooalign{\hfil\hbox{$\textstyle/$}\hfil\crcr
{\hbox{$\textstyle\mathchar"20D$}}}}
{\ooalign{\hfil\hbox{$\scriptstyle/$}\hfil\crcr
{\hbox{$\scriptstyle\mathchar"20D$}}}}
{\ooalign{\hfil\hbox{$\scriptscriptstyle/$}\hfil\crcr
{\hbox{$\scriptscriptstyle\mathchar"20D$}}}}
\else{\ooalign{\hfil/\hfil\crcr\mathhexbox20D}}%
\fi}}

\def\sq{\ifmmode\squareforqed\else{\unskip\nobreak\hfil
\penalty50\hskip1em\null\nobreak\hfil\squareforqed
\parfillskip=0pt\finalhyphendemerits=0\endgraf}\fi}
\def\squareforqed{\hbox{\rlap{$\sqcap$}$\sqcup$}}


\def\bbbc{{\mathchoice {\setbox0=\hbox{$\displaystyle\rm C$}\hbox{\hbox
to0pt{\kern0.4\wd0\vrule height0.9\ht0\hss}\box0}}
{\setbox0=\hbox{$\textstyle\rm C$}\hbox{\hbox
to0pt{\kern0.4\wd0\vrule height0.9\ht0\hss}\box0}}
{\setbox0=\hbox{$\scriptstyle\rm C$}\hbox{\hbox
to0pt{\kern0.4\wd0\vrule height0.9\ht0\hss}\box0}}
{\setbox0=\hbox{$\scriptscriptstyle\rm C$}\hbox{\hbox
to0pt{\kern0.4\wd0\vrule height0.9\ht0\hss}\box0}}}}
\def\bbbq{{\mathchoice {\setbox0=\hbox{$\displaystyle\rm
Q$}\hbox{\raise
0.15\ht0\hbox to0pt{\kern0.4\wd0\vrule height0.8\ht0\hss}\box0}}
{\setbox0=\hbox{$\textstyle\rm Q$}\hbox{\raise
0.15\ht0\hbox to0pt{\kern0.4\wd0\vrule height0.8\ht0\hss}\box0}}
{\setbox0=\hbox{$\scriptstyle\rm Q$}\hbox{\raise
0.15\ht0\hbox to0pt{\kern0.4\wd0\vrule height0.7\ht0\hss}\box0}}
{\setbox0=\hbox{$\scriptscriptstyle\rm Q$}\hbox{\raise
0.15\ht0\hbox to0pt{\kern0.4\wd0\vrule height0.7\ht0\hss}\box0}}}}
\def\bbbt{{\mathchoice {\setbox0=\hbox{$\displaystyle\rm
T$}\hbox{\hbox to0pt{\kern0.3\wd0\vrule height0.9\ht0\hss}\box0}}
{\setbox0=\hbox{$\textstyle\rm T$}\hbox{\hbox
to0pt{\kern0.3\wd0\vrule height0.9\ht0\hss}\box0}}
{\setbox0=\hbox{$\scriptstyle\rm T$}\hbox{\hbox
to0pt{\kern0.3\wd0\vrule height0.9\ht0\hss}\box0}}
{\setbox0=\hbox{$\scriptscriptstyle\rm T$}\hbox{\hbox
to0pt{\kern0.3\wd0\vrule height0.9\ht0\hss}\box0}}}}
\def\bbbs{{\mathchoice
{\setbox0=\hbox{$\displaystyle     \rm S$}\hbox{\raise0.5\ht0\hbox
to0pt{\kern0.35\wd0\vrule height0.45\ht0\hss}\hbox
to0pt{\kern0.55\wd0\vrule height0.5\ht0\hss}\box0}}
{\setbox0=\hbox{$\textstyle        \rm S$}\hbox{\raise0.5\ht0\hbox
to0pt{\kern0.35\wd0\vrule height0.45\ht0\hss}\hbox
to0pt{\kern0.55\wd0\vrule height0.5\ht0\hss}\box0}}
{\setbox0=\hbox{$\scriptstyle      \rm S$}\hbox{\raise0.5\ht0\hbox
to0pt{\kern0.35\wd0\vrule height0.45\ht0\hss}\raise0.05\ht0\hbox
to0pt{\kern0.5\wd0\vrule height0.45\ht0\hss}\box0}}
{\setbox0=\hbox{$\scriptscriptstyle\rm S$}\hbox{\raise0.5\ht0\hbox
to0pt{\kern0.4\wd0\vrule height0.45\ht0\hss}\raise0.05\ht0\hbox
to0pt{\kern0.55\wd0\vrule height0.45\ht0\hss}\box0}}}}
\def\bbbz{{\mathchoice {\hbox{$\sf\textstyle Z\kern-0.4em Z$}}
{\hbox{$\sf\textstyle Z\kern-0.4em Z$}}
{\hbox{$\sf\scriptstyle Z\kern-0.3em Z$}}
{\hbox{$\sf\scriptscriptstyle Z\kern-0.2em Z$}}}}


\ifprod@font
  \mathchardef\la="3\@xm2E
  \mathchardef\getsto="3\@xm1C
  \mathchardef\lid="3\@xm35
  \mathchardef\grole="3\@xm3F
  \mathchardef\loa="3\@xm2F
  \mathchardef\ga="3\@xm26
  \mathchardef\gid="3\@xm3D
  \mathchardef\leogr="3\@xm37
  \mathchardef\goa="3\@xm27
  \mathchardef\sq="0\@xm03
%
%
\def\diameter{{%
  \ifmmode
    \mathchoice
    {\ooalign{\hfil\hbox{$\displaystyle/$}\hfil\crcr
    {\lower.2ex\hbox{$\displaystyle\mathchar"20D$}}}}%
    {\ooalign{\hfil\hbox{$\textstyle/$}\hfil\crcr
    {\lower.2ex\hbox{$\textstyle\mathchar"20D$}}}}%
    {\ooalign{\hfil\hbox{$\scriptstyle/$}\hfil\crcr
    {\lower.1ex\hbox{$\scriptstyle\mathchar"20D$}}}}%
    {\ooalign{\hfil\hbox{$\scriptscriptstyle/$}\hfil\crcr
    {\lower.1ex\hbox{$\scriptscriptstyle\mathchar"20D$}}}}%
  \else
    {\ooalign{\hfil/\hfil\crcr\lower.2ex\hbox{\mathhexbox20D}}}%
  \fi
}}
%
%

\def\bbbc{{\Bbb{C}}}
\def\bbbq{{\Bbb{Q}}}
\def\bbbt{{\Bbb{T}}}
\def\bbbs{{\Bbb{S}}}
\def\bbbz{{\Bbb{Z}}}
\fi


\ifprod@font
\mathchardef\boxdot="2\@xm00
\mathchardef\boxplus="2\@xm01
\mathchardef\boxtimes="2\@xm02
\mathchardef\square="0\@xm03
\mathchardef\blacksquare="0\@xm04
\mathchardef\centerdot="2\@xm05
\mathchardef\lozenge="0\@xm06
\mathchardef\blacklozenge="0\@xm07
\mathchardef\circlearrowright="3\@xm08
\mathchardef\circlearrowleft="3\@xm09
\mathchardef\rightleftharpoons="3\@xm0A
\mathchardef\leftrightharpoons="3\@xm0B
\mathchardef\boxminus="2\@xm0C
\mathchardef\Vdash="3\@xm0D
\mathchardef\Vvdash="3\@xm0E
\mathchardef\vDash="3\@xm0F
\mathchardef\twoheadrightarrow="3\@xm10
\mathchardef\twoheadleftarrow="3\@xm11
\mathchardef\leftleftarrows="3\@xm12
\mathchardef\rightrightarrows="3\@xm13
\mathchardef\upuparrows="3\@xm14
\mathchardef\downdownarrows="3\@xm15
\mathchardef\upharpoonright="3\@xm16

\mathchardef\downharpoonright="3\@xm17
\mathchardef\upharpoonleft="3\@xm18
\mathchardef\downharpoonleft="3\@xm19
\mathchardef\rightarrowtail="3\@xm1A
\mathchardef\leftarrowtail="3\@xm1B
\mathchardef\leftrightarrows="3\@xm1C
\mathchardef\rightleftarrows="3\@xm1D
\mathchardef\Lsh="3\@xm1E
\mathchardef\Rsh="3\@xm1F
\mathchardef\rightsquigarrow="3\@xm20
\mathchardef\leftrightsquigarrow="3\@xm21
\mathchardef\looparrowleft="3\@xm22
\mathchardef\looparrowright="3\@xm23
\mathchardef\circeq="3\@xm24
\mathchardef\succsim="3\@xm25
\mathchardef\gtrsim="3\@xm26
\mathchardef\gtrapprox="3\@xm27
\mathchardef\multimap="3\@xm28
\mathchardef\therefore="3\@xm29
\mathchardef\because="3\@xm2A
\mathchardef\doteqdot="3\@xm2B

\mathchardef\triangleq="3\@xm2C
\mathchardef\precsim="3\@xm2D
\mathchardef\lesssim="3\@xm2E
\mathchardef\lessapprox="3\@xm2F
\mathchardef\eqslantless="3\@xm30
\mathchardef\eqslantgtr="3\@xm31
\mathchardef\curlyeqprec="3\@xm32
\mathchardef\curlyeqsucc="3\@xm33
\mathchardef\preccurlyeq="3\@xm34
\mathchardef\leqq="3\@xm35
\mathchardef\leqslant="3\@xm36
\mathchardef\lessgtr="3\@xm37
\mathchardef\backprime="0\@xm38
\mathchardef\risingdotseq="3\@xm3A
\mathchardef\fallingdotseq="3\@xm3B
\mathchardef\succcurlyeq="3\@xm3C
\mathchardef\geqq="3\@xm3D
\mathchardef\geqslant="3\@xm3E
\mathchardef\gtrless="3\@xm3F
\mathchardef\sqsubset="3\@xm40
\mathchardef\sqsupset="3\@xm41
\mathchardef\vartriangleright="3\@xm42
\mathchardef\vartriangleleft="3\@xm43
\mathchardef\trianglerighteq="3\@xm44
\mathchardef\trianglelefteq="3\@xm45
\mathchardef\bigstar="0\@xm46
\mathchardef\between="3\@xm47
\mathchardef\blacktriangledown="0\@xm48
\mathchardef\blacktriangleright="3\@xm49
\mathchardef\blacktriangleleft="3\@xm4A
\mathchardef\vartriangle="0\@xm4D
\mathchardef\blacktriangle="0\@xm4E
\mathchardef\triangledown="0\@xm4F
\mathchardef\eqcirc="3\@xm50
\mathchardef\lesseqgtr="3\@xm51
\mathchardef\gtreqless="3\@xm52
\mathchardef\lesseqqgtr="3\@xm53
\mathchardef\gtreqqless="3\@xm54
\mathchardef\Rrightarrow="3\@xm56
\mathchardef\Lleftarrow="3\@xm57
\mathchardef\veebar="2\@xm59
\mathchardef\barwedge="2\@xm5A
\mathchardef\doublebarwedge="2\@xm5B
\mathchardef\angle="0\@xm5C
\mathchardef\measuredangle="0\@xm5D
\mathchardef\sphericalangle="0\@xm5E
\mathchardef\varpropto="3\@xm5F
\mathchardef\smallsmile="3\@xm60
\mathchardef\smallfrown="3\@xm61
\mathchardef\Subset="3\@xm62
\mathchardef\Supset="3\@xm63
\mathchardef\Cup="2\@xm64

\mathchardef\Cap="2\@xm65

\mathchardef\curlywedge="2\@xm66
\mathchardef\curlyvee="2\@xm67
\mathchardef\leftthreetimes="2\@xm68
\mathchardef\rightthreetimes="2\@xm69
\mathchardef\subseteqq="3\@xm6A
\mathchardef\supseteqq="3\@xm6B
\mathchardef\bumpeq="3\@xm6C
\mathchardef\Bumpeq="3\@xm6D
\mathchardef\lll="3\@xm6E

\mathchardef\ggg="3\@xm6F

\mathchardef\circledS="0\@xm73
\mathchardef\pitchfork="3\@xm74
\mathchardef\dotplus="2\@xm75
\mathchardef\backsim="3\@xm76
\mathchardef\backsimeq="3\@xm77
\mathchardef\complement="0\@xm7B
\mathchardef\intercal="2\@xm7C
\mathchardef\circledcirc="2\@xm7D
\mathchardef\circledast="2\@xm7E
\mathchardef\circleddash="2\@xm7F
\def\ulcorner{\delimiter"4\@xm70\@xm70 }
\def\urcorner{\delimiter"5\@xm71\@xm71 }
\def\llcorner{\delimiter"4\@xm78\@xm78 }
\def\lrcorner{\delimiter"5\@xm79\@xm79 }
\def\yen{\mathhexbox\@xm55 }
\def\checkmark{\mathhexbox\@xm58 }
\def\circledR{\mathhexbox\@xm72 }
\def\maltese{\mathhexbox\@xm7A }
\mathchardef\lvertneqq="3\@ym00
\mathchardef\gvertneqq="3\@ym01
\mathchardef\nleq="3\@ym02
\mathchardef\ngeq="3\@ym03
\mathchardef\nless="3\@ym04
\mathchardef\ngtr="3\@ym05
\mathchardef\nprec="3\@ym06
\mathchardef\nsucc="3\@ym07
\mathchardef\lneqq="3\@ym08
\mathchardef\gneqq="3\@ym09
\mathchardef\nleqslant="3\@ym0A
\mathchardef\ngeqslant="3\@ym0B
\mathchardef\lneq="3\@ym0C
\mathchardef\gneq="3\@ym0D
\mathchardef\npreceq="3\@ym0E
\mathchardef\nsucceq="3\@ym0F
\mathchardef\precnsim="3\@ym10
\mathchardef\succnsim="3\@ym11
\mathchardef\lnsim="3\@ym12
\mathchardef\gnsim="3\@ym13
\mathchardef\nleqq="3\@ym14
\mathchardef\ngeqq="3\@ym15
\mathchardef\precneqq="3\@ym16
\mathchardef\succneqq="3\@ym17
\mathchardef\precnapprox="3\@ym18
\mathchardef\succnapprox="3\@ym19
\mathchardef\lnapprox="3\@ym1A
\mathchardef\gnapprox="3\@ym1B
\mathchardef\nsim="3\@ym1C
\mathchardef\ncong="3\@ym1D

\mathchardef\varsubsetneq="3\@ym20
\mathchardef\varsupsetneq="3\@ym21
\mathchardef\nsubseteqq="3\@ym22
\mathchardef\nsupseteqq="3\@ym23
\mathchardef\subsetneqq="3\@ym24
\mathchardef\supsetneqq="3\@ym25
\mathchardef\varsubsetneqq="3\@ym26
\mathchardef\varsupsetneqq="3\@ym27
\mathchardef\subsetneq="3\@ym28
\mathchardef\supsetneq="3\@ym29
\mathchardef\nsubseteq="3\@ym2A
\mathchardef\nsupseteq="3\@ym2B
\mathchardef\nparallel="3\@ym2C
\mathchardef\nmid="3\@ym2D
\mathchardef\nshortmid="3\@ym2E
\mathchardef\nshortparallel="3\@ym2F
\mathchardef\nvdash="3\@ym30
\mathchardef\nVdash="3\@ym31
\mathchardef\nvDash="3\@ym32
\mathchardef\nVDash="3\@ym33
\mathchardef\ntrianglerighteq="3\@ym34
\mathchardef\ntrianglelefteq="3\@ym35
\mathchardef\ntriangleleft="3\@ym36
\mathchardef\ntriangleright="3\@ym37
\mathchardef\nleftarrow="3\@ym38
\mathchardef\nrightarrow="3\@ym39
\mathchardef\nLeftarrow="3\@ym3A
\mathchardef\nRightarrow="3\@ym3B
\mathchardef\nLeftrightarrow="3\@ym3C
\mathchardef\nleftrightarrow="3\@ym3D
\mathchardef\divideontimes="2\@ym3E
\mathchardef\varnothing="0\@ym3F
\mathchardef\nexists="0\@ym40
\mathchardef\mho="0\@ym66
\mathchardef\eth="0\@ym67
\mathchardef\eqsim="3\@ym68
\mathchardef\beth="0\@ym69
\mathchardef\gimel="0\@ym6A
\mathchardef\daleth="0\@ym6B
\mathchardef\lessdot="3\@ym6C
\mathchardef\gtrdot="3\@ym6D
\mathchardef\ltimes="2\@ym6E
\mathchardef\rtimes="2\@ym6F
\mathchardef\shortmid="3\@ym70
\mathchardef\shortparallel="3\@ym71
\mathchardef\smallsetminus="2\@ym72
\mathchardef\thicksim="3\@ym73
\mathchardef\thickapprox="3\@ym74
\mathchardef\approxeq="3\@ym75
\mathchardef\succapprox="3\@ym76
\mathchardef\precapprox="3\@ym77
\mathchardef\curvearrowleft="3\@ym78
\mathchardef\curvearrowright="3\@ym79
\mathchardef\digamma="0\@ym7A
\mathchardef\varkappa="0\@ym7B
\mathchardef\hslash="0\@ym7D
\mathchardef\hbar="0\@ym7E
\mathchardef\backepsilon="3\@ym7F


\def\Bbb{\ifmmode\let\next\Bbb@\else
\def\next{\errmessage{Use \string\Bbb\space only in math mode}}\fi\next}
\def\Bbb@#1{{\Bbb@@{#1}}}
\def\Bbb@@#1{\fam\ymfam#1}
\fi


\def\Nulle{0} 
\def\Afe{1}   
\def\Hae{2}   
\def\Hbe{3}   
\def\Hce{4}   
\def\Hde{5}   


\newcount\LastMac       \LastMac=\Nulle

\newskip\half      \half=5.5pt plus 1.5pt minus 2.25pt
\newskip\one       \one=11pt plus 3pt minus 5.5pt
\newskip\onehalf   \onehalf=16.5pt plus 5.5pt minus 8.25pt
\newskip\two       \two=22pt plus 5.5pt minus 11pt

\def\Half{\addvspace{\half}}
\def\One{\addvspace{\one}}
\def\OneHalf{\addvspace{\onehalf}}
\def\Two{\addvspace{\two}}


\def\Raggedright{
  \rightskip=\z@ plus \hsize\relax
}

\def\Fullout{
  \rightskip=\z@\relax
}

\def\Hang#1#2{
  \hangindent=#1%
  \hangafter=#2\relax
}


\newif\ifsp@page
\def\pagestyle#1{\csname ps@#1\endcsname}
\def\thispagestyle#1{\global\sp@pagetrue\gdef\sp@type{#1}}

\def\ps@titlepage{%
  \def\@oddhead{\eightpoint\noindent \the\CatchLine
    \ifprod@font\else\qquad Printed\ \today\fi \hfil}%
  \let\@evenhead=\@oddhead
}

\def\ps@headings{%
  \def\@oddhead{\elevenpoint\it\noindent
    \hfill\the\RightHeader\hskip1.5em\rm\folio}%
  \def\@evenhead{\elevenpoint\noindent
    \folio\hskip1.5em\it\the\LeftHeader\hfill}%
}

\def\ps@plate{%
  \def\@oddhead{\eightpoint\noindent\plt@cap\hfil}%
  \def\@evenhead{\eightpoint\noindent\plt@cap\hfil}%
}



\def\title#1{
  \bgroup
    \vbox to 8pt{\vss}%
    \seventeenpoint
    \Raggedright
    \noindent \strut{\bf #1}\par
  \egroup
}

\def\author#1{
  \bgroup
    \ifnum\LastMac=\Afe \OneHalf\else \vskip 21pt\fi
    \fourteenpoint
    \Raggedright
    \noindent \strut #1\par
    \vskip 3pt%
  \egroup
}

\def\affiliation#1{
  \bgroup
    \vskip -4pt%
    \eightpoint
    \Raggedright
    \noindent \strut {\it #1}\par
  \egroup
  \LastMac=\Afe\relax
}

\def\acceptedline#1{
  \bgroup
    \Two
    \eightpoint
    \Raggedright
    \noindent \strut #1\par
  \egroup
}

\long\def\abstract#1{%
  \bgroup
    \vskip 20pt%
    \everypar{\Hang{11pc}{0}}%
    \noindent{\ninebf ABSTRACT}\par
    \tenpoint
    \Fullout
    \noindent #1\par
  \egroup
}

\long\def\keywords#1{
  \bgroup
    \Half
    \everypar{\Hang{11pc}{0}}%
    \tenpoint
    \Fullout
    \noindent\hbox{\bf Key words:}\ #1\par
  \egroup
}


\def\maketitle{%
  \EndOpening
  \ifsinglecol \else \MakePage\fi
}


\def\pageoffset#1#2{\hoffset=#1\relax\voffset=#2\relax}


\newif\ifAutoNumber \AutoNumberfalse
\newcount\Sec        
\newcount\SecSec
\newcount\SecSecSec

\Sec=\z@

\def\:{\let\@sptoken= } \:  
\def\:{\@xifnch} \expandafter\def\: {\futurelet\@tempc\@ifnch}

\def\@ifnextchar#1#2#3{%
  \let\@tempMACe #1%
  \def\@tempMACa{#2}%
  \def\@tempMACb{#3}%
  \futurelet \@tempMACc\@ifnch%
}

\def\@ifnch{%
\ifx \@tempMACc \@sptoken%
  \let\@tempMACd\@xifnch%
\else%
  \ifx \@tempMACc \@tempMACe%
    \let\@tempMACd\@tempMACa%
  \else%
    \let\@tempMACd\@tempMACb%
  \fi%
\fi%
\@tempMACd%
}

\def\@ifstar#1#2{\@ifnextchar *{\def\@tempMACa*{#1}\@tempMACa}{#2}}

\newskip\@tempskipb

\def\addvspace#1{%
  \ifvmode\else \endgraf\fi%
  \ifdim\lastskip=\z@%
    \vskip #1\relax%
  \else%
    \@tempskipb#1\relax\@xaddvskip%
  \fi%
}

\def\@xaddvskip{%
  \ifdim\lastskip<\@tempskipb%
    \vskip-\lastskip%
    \vskip\@tempskipb\relax%
  \else%
    \ifdim\@tempskipb<\z@%
      \ifdim\lastskip<\z@ \else%
        \advance\@tempskipb\lastskip%
        \vskip-\lastskip\vskip\@tempskipb%
      \fi%
    \fi%
  \fi%
}

\newskip\@tmpSKIP

\def\addpen#1{%
  \ifvmode
    \if@nobreak
    \else
      \ifdim\lastskip=\z@
        \penalty#1\relax
      \else
        \@tmpSKIP=\lastskip
        \vskip -\lastskip
        \penalty#1\vskip\@tmpSKIP
      \fi
    \fi
  \fi
}

\newcount\@clubpen   \@clubpen=\clubpenalty
\newif\if@nobreak    \@nobreakfalse

\def\@noafterindent{%
  \global\@nobreaktrue
  \everypar{\if@nobreak
              \global\@nobreakfalse
              \clubpenalty \@M
              {\setbox\z@\lastbox}%
              \LastMac=\Nulle\relax%
            \else
              \clubpenalty \@clubpen
              \everypar{}%
            \fi}
}

\newcount\gds@cbrk   \gds@cbrk=-300

\def\@nohdbrk{\interlinepenalty \@M\relax}

\let\@par=\par
\def\@restorepar{\def\par{\@par}}

\newif\if@endpe   \@endpefalse
 
\def\@doendpe{\@endpetrue \@nobreakfalse \LastMac=\Nulle\relax%
     \def\par{\@restorepar\everypar{}\par\@endpefalse}%
              \everypar{\setbox\z@\lastbox\everypar{}\@endpefalse}%
}

\def\section{\@ifstar{\@ssection}{\@section}}

\def\@section#1{
  \if@nobreak
    \everypar{}%
    \ifnum\LastMac=\Hae \addvspace{\half}\fi
  \else
    \addpen{\gds@cbrk}%
    \addvspace{\two}%
  \fi
  \bgroup
    \ninepoint\bf
    \Raggedright
    \ifAutoNumber
      \global\advance\Sec \@ne
      \noindent\@nohdbrk\number\Sec\hskip 1pc \uppercase{#1}\par
      \global\SecSec=\z@
    \else
      \noindent\@nohdbrk\uppercase{#1}\par
    \fi
  \egroup
  \nobreak
  \vskip\half
  \nobreak
  \@noafterindent
  \LastMac=\Hae\relax
}

\def\@ssection#1{
  \if@nobreak
    \everypar{}%
    \ifnum\LastMac=\Hae \addvspace{\half}\fi
  \else
    \addpen{\gds@cbrk}%
    \addvspace{\two}%
  \fi
  \bgroup
    \ninepoint\bf
    \Raggedright
    \noindent\@nohdbrk\uppercase{#1}\par
  \egroup
  \nobreak
  \vskip\half
  \nobreak
  \@noafterindent
  \LastMac=\Hae\relax
}

\def\subsection#1{
  \if@nobreak
    \everypar{}%
    \ifnum\LastMac=\Hae \addvspace{1pt plus 1pt minus .5pt}\fi
  \else
    \addpen{\gds@cbrk}%
    \addvspace{\onehalf}%
  \fi
  \bgroup
    \ninepoint\bf
    \Raggedright
    \ifAutoNumber
      \global\advance\SecSec \@ne
      \noindent\@nohdbrk\number\Sec.\number\SecSec \hskip 1pc\relax #1\par
      \global\SecSecSec=\z@
    \else
      \noindent\@nohdbrk #1\par
    \fi
  \egroup
  \nobreak
  \vskip\half
  \nobreak
  \@noafterindent
  \LastMac=\Hbe\relax
}

\def\subsubsection#1{
  \if@nobreak
    \everypar{}%
    \ifnum\LastMac=\Hbe \addvspace{1pt plus 1pt minus .5pt}\fi
  \else
    \addpen{\gds@cbrk}%
    \addvspace{\onehalf}%
  \fi
  \bgroup
    \ninepoint\it
    \Raggedright
    \ifAutoNumber
      \global\advance\SecSecSec \@ne
      \noindent\@nohdbrk\number\Sec.\number\SecSec.\number\SecSecSec
        \hskip 1pc\relax #1\par
    \else
      \noindent\@nohdbrk #1\par
    \fi
  \egroup
  \nobreak
  \vskip\half
  \nobreak
  \@noafterindent
  \LastMac=\Hce\relax
}

\def\paragraph#1{
  \if@nobreak
    \everypar{}%
  \else
    \addpen{\gds@cbrk}%
    \addvspace{\one}%
  \fi%
  \bgroup%
    \ninepoint\it
    \noindent #1\ \nobreak%
  \egroup
  \LastMac=\Hde\relax
  \ignorespaces
}


\let\tx=\relax 


\def\beginlist{%
  \par\if@nobreak \else\addvspace{\half}\fi%
  \bgroup%
    \ninepoint
    \let\item=\list@item%
}

\def\list@item{%
  \par\noindent\hskip 1em\relax%
  \ignorespaces%
}

\def\endlist{\par\egroup\addvspace{\half}\@doendpe}


\def\beginrefs{%
  \par
  \bgroup
    \eightpoint
    \Raggedright
    \let\bibitem=\bib@item
}

\def\bib@item{%
  \par\parindent=1.5em\Hang{1.5em}{1}%
  \everypar={\Hang{1.5em}{1}\ignorespaces}%
  \noindent\ignorespaces
}

\def\endrefs{\par\egroup\@doendpe}


\newtoks\CatchLine

\def\@journal{Mon.\ Not.\ R.\ Astron.\ Soc.\ }  
\def\@pubyear{1994}        
\def\@pagerange{000--000}  
\def\@volume{000}          
\def\@microfiche{}         %

\def\pubyear#1{\gdef\@pubyear{#1}\@makecatchline}
\def\pagerange#1{\gdef\@pagerange{#1}\@makecatchline}
\def\volume#1{\gdef\@volume{#1}\@makecatchline}
\def\microfiche#1{\gdef\@microfiche{and Microfiche\ #1}\@makecatchline}

\def\@makecatchline{%
  \global\CatchLine{%
    {\rm \@journal {\bf \@volume},\ \@pagerange\ (\@pubyear)\ \@microfiche}}%
}

\@makecatchline 

\newtoks\LeftHeader
\def\shortauthor#1{
  \global\LeftHeader{#1}%
}

\newtoks\RightHeader
\def\shorttitle#1{
  \global\RightHeader{#1}%
}

\def\PageHead{
  \begingroup
    \ifsp@page
      \csname ps@\sp@type\endcsname
      \global\sp@pagefalse
    \fi
    \ifodd\pageno
      \let\the@head=\@oddhead
    \else
      \let\the@head=\@evenhead
    \fi
    \vbox to \z@{\vskip-22.5\p@%
      \hbox to \PageWidth{\vbox to8.5\p@{}%
        \the@head
      }%
    \vss}%
  \endgroup
  \nointerlineskip
}

\def\today{%
  \number\day\space
  \ifcase\month\or January\or February\or March\or April\or May\or June\or
    July\or August\or September\or October\or November\or December\fi
  \space\number\year%
}

\def\PageFoot{} 

\def\authorcomment#1{%
  \gdef\PageFoot{%
    \nointerlineskip%
    \vbox to 22pt{\vfil%
      \hbox to \PageWidth{\elevenpoint\noindent \hfil #1 \hfil}}%
  }%
}


\newif\ifplate@page
\newbox\plt@box

\def\beginplatepage{%
  \let\plate=\plate@head
  \let\caption=\fig@caption
  \global\setbox\plt@box=\vbox\bgroup
  \TEMPDIMEN=\PageWidth 
  \hsize=\PageWidth\relax
}

\def\endplatepage{\par\egroup\global\plate@pagetrue}
\def\plate@head#1{\gdef\plt@cap{#1}}


\def\letters{%
  \gdef\folio{\ifnum\pageno<\z@ L\romannumeral-\pageno
    \else L\number\pageno \fi}%
}


\everydisplay{\displaysetup}

\newif\ifeqno
\newif\ifleqno

\def\displaysetup#1$${%
 \displaytest#1\eqno\eqno\displaytest
}

\def\displaytest#1\eqno#2\eqno#3\displaytest{%
 \if!#3!\ldisplaytest#1\leqno\leqno\ldisplaytest
 \else\eqnotrue\leqnofalse\def\eqn{#2}\def\eq{#1}\fi
 \generaldisplay$$}

\def\ldisplaytest#1\leqno#2\leqno#3\ldisplaytest{%
 \def\eq{#1}%
 \if!#3!\eqnofalse\else\eqnotrue\leqnotrue
  \def\eqn{#2}\fi}

\def\generaldisplay{%
\ifeqno \ifleqno 
   \hbox to \hsize{\noindent
     $\displaystyle\eq$\hfil$\displaystyle\eqn$}
  \else
    \hbox to \hsize{\noindent
     $\displaystyle\eq$\hfil$\displaystyle\eqn$}
  \fi
 \else
 \hbox to \hsize{\vbox{\noindent
  $\displaystyle\eq$\hfil}}
 \fi
}


\def\@notice{%
  \par\Two%
  \noindent{\b@ls{11pt}\ninerm This paper has been produced using the
    Blackwell Scientific Publications \TeX\ macros.\par}%
}

\outer\def\bye{\@notice\par\vfill\supereject\end}


\def\start@mess{%
  Monthly notices of the RAS journal style (\@typeface)\space
    v\@version,\space \@verdate.%
}

\everyjob{\Warn{\start@mess}}



\newif\if@debug \@debugfalse  

\def\Print#1{\if@debug\immediate\write16{#1}\else \fi}
\def\Warn#1{\immediate\write16{#1}}
\def\wlog#1{}

\newcount\Iteration 

\def\Single{0} \def\Double{1}                 
\def\Figure{0} \def\Table{1}                  

\def\InStack{0}  
\def\InZoneA{1}
\def\InZoneB{2}
\def\InZoneC{3}

\newcount\TEMPCOUNT 
\newdimen\TEMPDIMEN 
\newbox\TEMPBOX     
\newbox\VOIDBOX     

\newcount\LengthOfStack 
\newcount\MaxItems      
\newcount\StackPointer
\newcount\Point         
\newcount\NextFigure    
\newcount\NextTable     
\newcount\NextItem      

\newcount\StatusStack   
\newcount\NumStack      
\newcount\TypeStack     
\newcount\SpanStack     
\newcount\BoxStack      

\newcount\ItemSTATUS    
\newcount\ItemNUMBER    
\newcount\ItemTYPE      
\newcount\ItemSPAN      
\newbox\ItemBOX         
\newdimen\ItemSIZE      

\newdimen\PageHeight    
\newdimen\TextLeading   
\newdimen\Feathering    
\newcount\LinesPerPage  
\newdimen\ColumnWidth   
\newdimen\ColumnGap     
\newdimen\PageWidth     
\newdimen\BodgeHeight   
\newcount\Leading       

\newdimen\ZoneBSize  
\newdimen\TextSize   
\newbox\ZoneABOX     
\newbox\ZoneBBOX     
\newbox\ZoneCBOX     

\newif\ifFirstSingleItem
\newif\ifFirstZoneA
\newif\ifMakePageInComplete
\newif\ifMoreFigures \MoreFiguresfalse 
\newif\ifMoreTables  \MoreTablesfalse  

\newif\ifFigInZoneB 
\newif\ifFigInZoneC 
\newif\ifTabInZoneB 
\newif\ifTabInZoneC

\newif\ifZoneAFullPage

\newbox\MidBOX    
\newbox\LeftBOX
\newbox\RightBOX
\newbox\PageBOX   

\newif\ifLeftCOL  
\LeftCOLtrue

\newdimen\ZoneBAdjust

\newcount\ItemFits
\def\Yes{1}
\def\No{2}


\MaxItems=15
\NextFigure=\z@        
\NextTable=\@ne

\BodgeHeight=6pt
\TextLeading=11pt    
\Leading=11
\Feathering=\z@      
\LinesPerPage=61     
\topskip=\TextLeading
\ColumnWidth=20pc    
\ColumnGap=2pc       

\newskip\ItemSepamount  
\ItemSepamount=\TextLeading plus \TextLeading minus 4pt

\parskip=\z@ plus .1pt
\parindent=18pt
\widowpenalty=\z@
\clubpenalty=10000
\tolerance=1500
\hbadness=1500
\abovedisplayskip=6pt plus 2pt minus 2pt
\belowdisplayskip=6pt plus 2pt minus 2pt
\abovedisplayshortskip=6pt plus 2pt minus 2pt
\belowdisplayshortskip=6pt plus 2pt minus 2pt

\ninepoint 


\PageHeight=682pt

\PageWidth=2\ColumnWidth
\advance\PageWidth by \ColumnGap

\pagestyle{headings}




\newcount\DUMMY \StatusStack=\allocationnumber
\newcount\DUMMY \newcount\DUMMY \newcount\DUMMY 
\newcount\DUMMY \newcount\DUMMY \newcount\DUMMY 
\newcount\DUMMY \newcount\DUMMY \newcount\DUMMY
\newcount\DUMMY \newcount\DUMMY \newcount\DUMMY 
\newcount\DUMMY \newcount\DUMMY \newcount\DUMMY

\newcount\DUMMY \NumStack=\allocationnumber
\newcount\DUMMY \newcount\DUMMY \newcount\DUMMY 
\newcount\DUMMY \newcount\DUMMY \newcount\DUMMY 
\newcount\DUMMY \newcount\DUMMY \newcount\DUMMY 
\newcount\DUMMY \newcount\DUMMY \newcount\DUMMY 
\newcount\DUMMY \newcount\DUMMY \newcount\DUMMY

\newcount\DUMMY \TypeStack=\allocationnumber
\newcount\DUMMY \newcount\DUMMY \newcount\DUMMY 
\newcount\DUMMY \newcount\DUMMY \newcount\DUMMY 
\newcount\DUMMY \newcount\DUMMY \newcount\DUMMY 
\newcount\DUMMY \newcount\DUMMY \newcount\DUMMY 
\newcount\DUMMY \newcount\DUMMY \newcount\DUMMY

\newcount\DUMMY \SpanStack=\allocationnumber
\newcount\DUMMY \newcount\DUMMY \newcount\DUMMY 
\newcount\DUMMY \newcount\DUMMY \newcount\DUMMY 
\newcount\DUMMY \newcount\DUMMY \newcount\DUMMY 
\newcount\DUMMY \newcount\DUMMY \newcount\DUMMY 
\newcount\DUMMY \newcount\DUMMY \newcount\DUMMY

\newbox\DUMMY   \BoxStack=\allocationnumber
\newbox\DUMMY   \newbox\DUMMY \newbox\DUMMY 
\newbox\DUMMY   \newbox\DUMMY \newbox\DUMMY 
\newbox\DUMMY   \newbox\DUMMY \newbox\DUMMY 
\newbox\DUMMY   \newbox\DUMMY \newbox\DUMMY 
\newbox\DUMMY   \newbox\DUMMY \newbox\DUMMY

\def\wlog{\immediate\write\m@ne}


\def\GetItemAll#1{%
 \GetItemSTATUS{#1}
 \GetItemNUMBER{#1}
 \GetItemTYPE{#1}
 \GetItemSPAN{#1}
 \GetItemBOX{#1}
}

\def\GetItemSTATUS#1{%
 \Point=\StatusStack
 \advance\Point by #1
 \global\ItemSTATUS=\count\Point
}

\def\GetItemNUMBER#1{%
 \Point=\NumStack
 \advance\Point by #1
 \global\ItemNUMBER=\count\Point
}

\def\GetItemTYPE#1{%
 \Point=\TypeStack
 \advance\Point by #1
 \global\ItemTYPE=\count\Point
}

\def\GetItemSPAN#1{%
 \Point\SpanStack
 \advance\Point by #1
 \global\ItemSPAN=\count\Point
}

\def\GetItemBOX#1{%
 \Point=\BoxStack
 \advance\Point by #1
 \global\setbox\ItemBOX=\vbox{\copy\Point}
 \global\ItemSIZE=\ht\ItemBOX
 \global\advance\ItemSIZE by \dp\ItemBOX
 \TEMPCOUNT=\ItemSIZE
 \divide\TEMPCOUNT by \Leading
 \divide\TEMPCOUNT by 65536
 \advance\TEMPCOUNT \@ne
 \ItemSIZE=\TEMPCOUNT pt
 \global\multiply\ItemSIZE by \Leading
}


\def\JoinStack{%
 \ifnum\LengthOfStack=\MaxItems 
  \Warn{WARNING: Stack is full...some items will be lost!}
 \else
  \Point=\StatusStack
  \advance\Point by \LengthOfStack
  \global\count\Point=\ItemSTATUS
  \Point=\NumStack
  \advance\Point by \LengthOfStack
  \global\count\Point=\ItemNUMBER
  \Point=\TypeStack
  \advance\Point by \LengthOfStack
  \global\count\Point=\ItemTYPE
  \Point\SpanStack
  \advance\Point by \LengthOfStack
  \global\count\Point=\ItemSPAN
  \Point=\BoxStack
  \advance\Point by \LengthOfStack
  \global\setbox\Point=\vbox{\copy\ItemBOX}
  \global\advance\LengthOfStack \@ne
  \ifnum\ItemTYPE=\Figure 
   \global\MoreFigurestrue
  \else
   \global\MoreTablestrue
  \fi
 \fi
}


\def\LeaveStack#1{%
 {\Iteration=#1
 \loop
 \ifnum\Iteration<\LengthOfStack
  \advance\Iteration \@ne
  \GetItemSTATUS{\Iteration}
   \advance\Point by \m@ne
   \global\count\Point=\ItemSTATUS
  \GetItemNUMBER{\Iteration}
   \advance\Point by \m@ne
   \global\count\Point=\ItemNUMBER
  \GetItemTYPE{\Iteration}
   \advance\Point by \m@ne
   \global\count\Point=\ItemTYPE
  \GetItemSPAN{\Iteration}
   \advance\Point by \m@ne
   \global\count\Point=\ItemSPAN
  \GetItemBOX{\Iteration}
   \advance\Point by \m@ne
   \global\setbox\Point=\vbox{\copy\ItemBOX}
 \repeat}
 \global\advance\LengthOfStack by \m@ne
}


\newif\ifStackNotClean

\def\CleanStack{%
 \StackNotCleantrue
 {\Iteration=\z@
  \loop
   \ifStackNotClean
    \GetItemSTATUS{\Iteration}
    \ifnum\ItemSTATUS=\InStack
     \advance\Iteration \@ne
     \else
      \LeaveStack{\Iteration}
    \fi
   \ifnum\LengthOfStack<\Iteration
    \StackNotCleanfalse
   \fi
 \repeat}
}


\def\FindItem#1#2{%
 \global\StackPointer=\m@ne 
 {\Iteration=\z@
  \loop
  \ifnum\Iteration<\LengthOfStack
   \GetItemSTATUS{\Iteration}
   \ifnum\ItemSTATUS=\InStack
    \GetItemTYPE{\Iteration}
    \ifnum\ItemTYPE=#1
     \GetItemNUMBER{\Iteration}
     \ifnum\ItemNUMBER=#2
      \global\StackPointer=\Iteration
      \Iteration=\LengthOfStack 
     \fi
    \fi
   \fi
  \advance\Iteration \@ne
 \repeat}
}


\def\FindNext{%
 \global\StackPointer=\m@ne 
 {\Iteration=\z@
  \loop
  \ifnum\Iteration<\LengthOfStack
   \GetItemSTATUS{\Iteration}
   \ifnum\ItemSTATUS=\InStack
    \GetItemTYPE{\Iteration}
   \ifnum\ItemTYPE=\Figure
    \ifMoreFigures
      \global\NextItem=\Figure
      \global\StackPointer=\Iteration
      \Iteration=\LengthOfStack 
    \fi
   \fi
   \ifnum\ItemTYPE=\Table
    \ifMoreTables
      \global\NextItem=\Table
      \global\StackPointer=\Iteration
      \Iteration=\LengthOfStack 
    \fi
   \fi
  \fi
  \advance\Iteration \@ne
 \repeat}
}


\def\ChangeStatus#1#2{%
 \Point=\StatusStack
 \advance\Point by #1
 \global\count\Point=#2
}



\def\Zone{\InZoneA}

\ZoneBAdjust=\z@

\def\MakePage{
 \global\ZoneBSize=\PageHeight
 \global\TextSize=\ZoneBSize
 \global\ZoneAFullPagefalse
 \global\topskip=\TextLeading
 \MakePageInCompletetrue
 \MoreFigurestrue
 \MoreTablestrue
 \FigInZoneBfalse
 \FigInZoneCfalse
 \TabInZoneBfalse
 \TabInZoneCfalse
 \global\FirstSingleItemtrue
 \global\FirstZoneAtrue
 \global\setbox\ZoneABOX=\box\VOIDBOX
 \global\setbox\ZoneBBOX=\box\VOIDBOX
 \global\setbox\ZoneCBOX=\box\VOIDBOX
 \loop
  \ifMakePageInComplete
 \FindNext
 \ifnum\StackPointer=\m@ne
  \NextItem=\m@ne
  \MoreFiguresfalse
  \MoreTablesfalse
 \fi
 \ifnum\NextItem=\Figure
   \FindItem{\Figure}{\NextFigure}
   \ifnum\StackPointer=\m@ne \global\MoreFiguresfalse
   \else
    \GetItemSPAN{\StackPointer}
    \ifnum\ItemSPAN=\Single \def\Zone{\InZoneB}\relax
     \ifFigInZoneC \global\MoreFiguresfalse\fi
    \else
     \def\Zone{\InZoneA}
     \ifFigInZoneB \def\Zone{\InZoneC}\fi
    \fi
   \fi
   \ifMoreFigures\Print{}\FigureItems\fi
 \fi
\ifnum\NextItem=\Table
   \FindItem{\Table}{\NextTable}
   \ifnum\StackPointer=\m@ne \global\MoreTablesfalse
   \else
    \GetItemSPAN{\StackPointer}
    \ifnum\ItemSPAN=\Single\relax
     \ifTabInZoneC \global\MoreTablesfalse\fi
    \else
     \def\Zone{\InZoneA}
     \ifTabInZoneB \def\Zone{\InZoneC}\fi
    \fi
   \fi
   \ifMoreTables\Print{}\TableItems\fi
 \fi
   \MakePageInCompletefalse 
   \ifMoreFigures\MakePageInCompletetrue\fi
   \ifMoreTables\MakePageInCompletetrue\fi
 \repeat
 \ifZoneAFullPage
  \global\TextSize=\z@
  \global\ZoneBSize=\z@
  \global\vsize=\z@\relax
  \global\topskip=\z@\relax
  \vbox to \z@{\vss}
  \eject
 \else
 \global\advance\ZoneBSize by -\ZoneBAdjust
 \global\vsize=\ZoneBSize
 \global\hsize=\ColumnWidth
 \global\ZoneBAdjust=\z@
 \ifdim\TextSize<23pt
 \Warn{}
 \Warn{* Making column fall short: TextSize=\the\TextSize *}
 \vskip-\lastskip\eject\fi
 \fi
}

\def\MakeRightCol{
 \global\TextSize=\ZoneBSize
 \MakePageInCompletetrue
 \MoreFigurestrue
 \MoreTablestrue
 \global\FirstSingleItemtrue
 \global\setbox\ZoneBBOX=\box\VOIDBOX
 \def\Zone{\InZoneB}
 \loop
  \ifMakePageInComplete
 \FindNext
 \ifnum\StackPointer=\m@ne
  \NextItem=\m@ne
  \MoreFiguresfalse
  \MoreTablesfalse
 \fi
 \ifnum\NextItem=\Figure
   \FindItem{\Figure}{\NextFigure}
   \ifnum\StackPointer=\m@ne \MoreFiguresfalse
   \else
    \GetItemSPAN{\StackPointer}
    \ifnum\ItemSPAN=\Double\relax
     \MoreFiguresfalse\fi
   \fi
   \ifMoreFigures\Print{}\FigureItems\fi
 \fi
 \ifnum\NextItem=\Table
   \FindItem{\Table}{\NextTable}
   \ifnum\StackPointer=\m@ne \MoreTablesfalse
   \else
    \GetItemSPAN{\StackPointer}
    \ifnum\ItemSPAN=\Double\relax
     \MoreTablesfalse\fi
   \fi
   \ifMoreTables\Print{}\TableItems\fi
 \fi
   \MakePageInCompletefalse 
   \ifMoreFigures\MakePageInCompletetrue\fi
   \ifMoreTables\MakePageInCompletetrue\fi
 \repeat
 \ifZoneAFullPage
  \global\TextSize=\z@
  \global\ZoneBSize=\z@
  \global\vsize=\z@\relax
  \global\topskip=\z@\relax
  \vbox to \z@{\vss}
  \eject
 \else
 \global\vsize=\ZoneBSize
 \global\hsize=\ColumnWidth
 \ifdim\TextSize<23pt
 \Warn{}
 \Warn{* Making column fall short: TextSize=\the\TextSize *}
 \vskip-\lastskip\eject\fi
\fi
}

\def\FigureItems{
 \Print{Considering...}
 \ShowItem{\StackPointer}
 \GetItemBOX{\StackPointer} 
 \GetItemSPAN{\StackPointer}
  \CheckFitInZone 
  \ifnum\ItemFits=\Yes
   \ifnum\ItemSPAN=\Single
     \ChangeStatus{\StackPointer}{\InZoneB} 
     \global\FigInZoneBtrue
     \ifFirstSingleItem
      \hbox{}\vskip-\BodgeHeight
     \global\advance\ItemSIZE by \TextLeading
     \fi
     \unvbox\ItemBOX\ItemSep
     \global\FirstSingleItemfalse
     \global\advance\TextSize by -\ItemSIZE
     \global\advance\TextSize by -\TextLeading
   \else
    \ifFirstZoneA
     \global\advance\ItemSIZE by \TextLeading
     \global\FirstZoneAfalse\fi
    \global\advance\TextSize by -\ItemSIZE
    \global\advance\TextSize by -\TextLeading
    \global\advance\ZoneBSize by -\ItemSIZE
    \global\advance\ZoneBSize by -\TextLeading
    \ifFigInZoneB\relax
     \else
     \ifdim\TextSize<3\TextLeading
     \global\ZoneAFullPagetrue
     \fi
    \fi
    \ChangeStatus{\StackPointer}{\Zone}
    \ifnum\Zone=\InZoneC \global\FigInZoneCtrue\fi
  \fi
   \Print{TextSize=\the\TextSize}
   \Print{ZoneBSize=\the\ZoneBSize}
  \global\advance\NextFigure \@ne
   \Print{This figure has been placed.}
  \else
   \Print{No space available for this figure...holding over.}
   \Print{}
   \global\MoreFiguresfalse
  \fi
}

\def\TableItems{
 \Print{Considering...}
 \ShowItem{\StackPointer}
 \GetItemBOX{\StackPointer} 
 \GetItemSPAN{\StackPointer}
  \CheckFitInZone 
  \ifnum\ItemFits=\Yes
   \ifnum\ItemSPAN=\Single
    \ChangeStatus{\StackPointer}{\InZoneB}
     \global\TabInZoneBtrue
     \ifFirstSingleItem
      \hbox{}\vskip-\BodgeHeight
     \global\advance\ItemSIZE by \TextLeading
     \fi
     \unvbox\ItemBOX\ItemSep
     \global\FirstSingleItemfalse
     \global\advance\TextSize by -\ItemSIZE
     \global\advance\TextSize by -\TextLeading
   \else
    \ifFirstZoneA
    \global\advance\ItemSIZE by \TextLeading
    \global\FirstZoneAfalse\fi
    \global\advance\TextSize by -\ItemSIZE
    \global\advance\TextSize by -\TextLeading
    \global\advance\ZoneBSize by -\ItemSIZE
    \global\advance\ZoneBSize by -\TextLeading
    \ifFigInZoneB\relax
     \else
     \ifdim\TextSize<3\TextLeading
     \global\ZoneAFullPagetrue
     \fi
    \fi
    \ChangeStatus{\StackPointer}{\Zone}
    \ifnum\Zone=\InZoneC \global\TabInZoneCtrue\fi
   \fi
  \global\advance\NextTable \@ne
   \Print{This table has been placed.}
  \else
  \Print{No space available for this table...holding over.}
   \Print{}
   \global\MoreTablesfalse
  \fi
}


\def\CheckFitInZone{%
{\advance\TextSize by -\ItemSIZE
 \advance\TextSize by -\TextLeading
 \ifFirstSingleItem
  \advance\TextSize by \TextLeading
 \fi
 \ifnum\Zone=\InZoneA\relax
  \else \advance\TextSize by -\ZoneBAdjust
 \fi
 \ifdim\TextSize<3\TextLeading \global\ItemFits=\No
 \else \global\ItemFits=\Yes\fi}
}

\def\BeginOpening{%
  \thispagestyle{titlepage}%
  \global\setbox\ItemBOX=\vbox\bgroup%
    \hsize=\PageWidth%
    \hrule height \z@
    \ifsinglecol\vskip 6pt\fi 
}

\let\begintopmatter=\BeginOpening  

\def\EndOpening{%
  \One
  \egroup
  \ifsinglecol
    \box\ItemBOX%
    \vskip\TextLeading plus 2\TextLeading
    \@noafterindent
  \else
    \ItemNUMBER=\z@%
    \ItemTYPE=\Figure
    \ItemSPAN=\Double
    \ItemSTATUS=\InStack
    \JoinStack
  \fi
}


\newif\if@here  \@herefalse

\def\no@float{\global\@heretrue}
\let\nofloat=\relax 

\def\beginfigure{%
  \@ifstar{\global\@dfloattrue \@bfigure}{\global\@dfloatfalse \@bfigure}%
}

\def\@bfigure#1{%
  \par
  \if@dfloat
    \ItemSPAN=\Double
    \TEMPDIMEN=\PageWidth
  \else
    \ItemSPAN=\Single
    \TEMPDIMEN=\ColumnWidth
  \fi
  \ifsinglecol
    \TEMPDIMEN=\PageWidth
  \else
    \ItemSTATUS=\InStack
    \ItemNUMBER=#1%
    \ItemTYPE=\Figure
  \fi
  \bgroup
    \hsize=\TEMPDIMEN
    \global\setbox\ItemBOX=\vbox\bgroup
      \eightpoint\nostb@ls{10pt}%
      \let\caption=\fig@caption
      \ifsinglecol \let\nofloat=\no@float\fi
}

\def\fig@caption#1{%
  \vskip 5.5pt plus 6pt%
  \bgroup 
    \eightpoint\nostb@ls{10pt}%
    \setbox\TEMPBOX=\hbox{#1}%
    \ifdim\wd\TEMPBOX>\TEMPDIMEN
      \noindent \unhbox\TEMPBOX\par
    \else
      \hbox to \hsize{\hfil\unhbox\TEMPBOX\hfil}%
    \fi
  \egroup
}

\def\endfigure{%
  \par\egroup 
  \egroup
  \ifsinglecol
    \if@here \midinsert\global\@herefalse\else \topinsert\fi
      \unvbox\ItemBOX
    \endinsert
  \else
    \JoinStack
    \Print{Processing source for figure \the\ItemNUMBER}%
  \fi
}


\newbox\tab@cap@box
\def\tab@caption#1{\global\setbox\tab@cap@box=\hbox{#1\par}}

\newtoks\tab@txt@toks
\long\def\tab@txt#1{\global\tab@txt@toks={#1}\global\table@txttrue}

\newif\iftable@txt  \table@txtfalse
\newif\if@dfloat    \@dfloatfalse

\def\begintable{%
  \@ifstar{\global\@dfloattrue \@btable}{\global\@dfloatfalse \@btable}%
}

\def\@btable#1{%
  \par
  \if@dfloat
    \ItemSPAN=\Double
    \TEMPDIMEN=\PageWidth
  \else
    \ItemSPAN=\Single
    \TEMPDIMEN=\ColumnWidth
  \fi
  \ifsinglecol
    \TEMPDIMEN=\PageWidth
  \else
    \ItemSTATUS=\InStack
    \ItemNUMBER=#1%
    \ItemTYPE=\Table
  \fi
  \bgroup
    \eightpoint\nostb@ls{10pt}%
    \global\setbox\ItemBOX=\vbox\bgroup
      \let\caption=\tab@caption
      \let\tabletext=\tab@txt
      \ifsinglecol \let\nofloat=\no@float\fi
}

\def\endtable{%
  \par\egroup 
  \egroup
  \setbox\TEMPBOX=\hbox to \TEMPDIMEN{%
    \hss
    \vbox{%
      \hsize=\wd\ItemBOX
      \ifvoid\tab@cap@box
      \else
        \noindent\unhbox\tab@cap@box
        \vskip 5.5pt plus 6pt%
      \fi
      \box\ItemBOX
      \iftable@txt
        \vskip 10pt%
        \eightpoint\nostb@ls{10pt}%
        \noindent\the\tab@txt@toks
        \global\table@txtfalse
      \fi
    }%
    \hss
  }%
  \ifsinglecol
    \if@here \midinsert\global\@herefalse\else \topinsert\fi
      \box\TEMPBOX
    \endinsert
  \else
    \global\setbox\ItemBOX=\box\TEMPBOX
    \JoinStack
    \Print{Processing source for table \the\ItemNUMBER}%
  \fi
}

\def\UnloadZoneA{%
\FirstZoneAtrue
 \Iteration=\z@
  \loop
   \ifnum\Iteration<\LengthOfStack
    \GetItemSTATUS{\Iteration}
    \ifnum\ItemSTATUS=\InZoneA
     \GetItemBOX{\Iteration}
     \ifFirstZoneA \vbox to \BodgeHeight{\vfil}%
     \FirstZoneAfalse\fi
     \unvbox\ItemBOX\ItemSep
     \LeaveStack{\Iteration}
     \else
     \advance\Iteration \@ne
   \fi
 \repeat
}

\def\UnloadZoneC{%
\Iteration=\z@
  \loop
   \ifnum\Iteration<\LengthOfStack
    \GetItemSTATUS{\Iteration}
    \ifnum\ItemSTATUS=\InZoneC
     \GetItemBOX{\Iteration}
     \ItemSep\unvbox\ItemBOX
     \LeaveStack{\Iteration}
     \else
     \advance\Iteration \@ne
   \fi
 \repeat
}


\def\ShowItem#1{
  {\GetItemAll{#1}
  \Print{\the#1:
  {TYPE=\ifnum\ItemTYPE=\Figure Figure\else Table\fi}
  {NUMBER=\the\ItemNUMBER}
  {SPAN=\ifnum\ItemSPAN=\Single Single\else Double\fi}
  {SIZE=\the\ItemSIZE}}}
}

\def\ShowStack{%
 \Print{}
 \Print{LengthOfStack = \the\LengthOfStack}
 \ifnum\LengthOfStack=\z@ \Print{Stack is empty}\fi
 \Iteration=\z@
 \loop
 \ifnum\Iteration<\LengthOfStack
  \ShowItem{\Iteration}
  \advance\Iteration \@ne
 \repeat
}

\def\B#1#2{%
\hbox{\vrule\kern-0.4pt\vbox to #2{%
\hrule width #1\vfill\hrule}\kern-0.4pt\vrule}
}


\newif\ifsinglecol   \singlecolfalse

\def\onecolumn{%
  \global\output={\singlecoloutput}%
  \global\hsize=\PageWidth
  \global\vsize=\PageHeight
  \global\ColumnWidth=\hsize
  \global\TextLeading=12pt
  \global\Leading=12
  \global\singlecoltrue
  \global\let\onecolumn=\relax
  \global\let\footnote=\sing@footnote
  \global\let\vfootnote=\sing@vfootnote
  \ninepoint 
  \message{(Single column)}%
}

\def\singlecoloutput{%
  \shipout\vbox{\PageHead\pagebody\PageFoot}%
  \advancepageno
  \ifplate@page
    \shipout\vbox{%
      \sp@pagetrue
      \def\sp@type{plate}%
      \global\plate@pagefalse
      \PageHead\vbox to \PageHeight{\unvbox\plt@box\vfil}\PageFoot%
    }%
    \message{[plate]}%
    \advancepageno
  \fi
  \ifnum\outputpenalty>-\@MM \else\dosupereject\fi%
}

\def\ItemSep{\vskip\ItemSepamount\relax}

\def\ItemSepbreak{\par\ifdim\lastskip<\ItemSepamount
  \removelastskip\penalty-200\ItemSep\fi%
}


\let\@@endinsert=\endinsert 

\def\endinsert{\egroup 
  \if@mid \dimen@\ht\z@ \advance\dimen@\dp\z@ \advance\dimen@12\p@
    \advance\dimen@\pagetotal \advance\dimen@-\pageshrink
    \ifdim\dimen@>\pagegoal\@midfalse\p@gefalse\fi\fi
  \if@mid \ItemSep\box\z@\ItemSepbreak
  \else\insert\topins{\penalty100 
    \splittopskip\z@skip
    \splitmaxdepth\maxdimen \floatingpenalty\z@
    \ifp@ge \dimen@\dp\z@
    \vbox to\vsize{\unvbox\z@\kern-\dimen@}
    \else \box\z@\nobreak\ItemSep\fi}\fi\endgroup%
}


\def\gobbleone#1{}
\def\gobbletwo#1#2{}
\let\footnote=\gobbletwo 
\let\vfootnote=\gobbleone

\def\sing@footnote#1{\let\@sf\empty 
  \ifhmode\edef\@sf{\spacefactor\the\spacefactor}\/\fi
  \hbox{$^{\hbox{\eightpoint #1}}$}\@sf\sing@vfootnote{#1}%
}

\def\sing@vfootnote#1{\insert\footins\bgroup\eightpoint\b@ls{9pt}%
  \interlinepenalty\interfootnotelinepenalty
  \splittopskip\ht\strutbox 
  \splitmaxdepth\dp\strutbox \floatingpenalty\@MM
  \leftskip\z@skip \rightskip\z@skip \spaceskip\z@skip \xspaceskip\z@skip
  \noindent $^{\scriptstyle\hbox{#1}}$\hskip 4pt%
    \footstrut\futurelet\next\fo@t%
}

\def\footnoterule{\kern-3\p@ \hrule height \z@ \kern 3\p@}

\skip\footins=19.5pt plus 12pt minus 1pt
\count\footins=1000
\dimen\footins=\maxdimen


\def\landscape{%
  \global\TEMPDIMEN=\PageWidth
  \global\PageWidth=\PageHeight
  \global\PageHeight=\TEMPDIMEN
  \global\let\landscape=\relax
  \onecolumn
  \message{(landscape)}%
  \raggedbottom
}


\output{%
  \ifLeftCOL
    \global\setbox\LeftBOX=\vbox to \ZoneBSize{\box255\unvbox\ZoneBBOX}%
    \global\LeftCOLfalse
    \MakeRightCol
  \else
    \setbox\RightBOX=\vbox to \ZoneBSize{\box255\unvbox\ZoneBBOX}%
    \setbox\MidBOX=\hbox{\box\LeftBOX\hskip\ColumnGap\box\RightBOX}%
    \setbox\PageBOX=\vbox to \PageHeight{%
      \UnloadZoneA\box\MidBOX\UnloadZoneC}%
    \shipout\vbox{\PageHead\box\PageBOX\PageFoot}%
    \advancepageno
    \ifplate@page
      \shipout\vbox{%
        \sp@pagetrue
        \def\sp@type{plate}%
        \global\plate@pagefalse
        \PageHead\vbox to \PageHeight{\unvbox\plt@box\vfil}\PageFoot%
      }%
      \message{[plate]}%
      \advancepageno
    \fi
    \global\LeftCOLtrue
    \CleanStack
    \MakePage
  \fi
}


\Warn{\start@mess}


\catcode `\@=12 



%% file: ascaxrb_input.tex
\def\ergpercmpers{erg cm$^{-2}$ s$^{-1}$}

\def\uunit{keV cm$^{-2}$ s$^{-1}$ sr$^{-1}$ keV$^{-1}$}

\def\el{E_{\rm L}}
\def\eh{E_{\rm H}}

\def\logns{log$N$-log$S$}

\def\ausjpas{AJPAS}
\def\pasj{{PASJ}}
\def\mn{{MNRAS}}
\def\apj{{ApJ}}

\def\aa{{A \& A}}
\def\aaa{{ARA \& A}}
\def\phre{{Phys. Rep.}}

\def\etal{et al\ }
\def\eg{eg}
\def\ie{ie}

\def\qsf{QSF3 field}
\def\rs {ROSAT}
\def\ginga {{\it Ginga}}
\def\asca{ASCA}
\def\heao1a2{HEAO-1 A2}
\def\abc{ABC Guide}

\def\ts{$kT_{\rm s}$}
\def\th{$kT_{\rm h}$}
\def\gs{$\Gamma_{\rm s}$}
\def\gh{$\Gamma_{\rm h}$}
\def\ns{$A_{\rm s}$}
\def\nh{$A_{\rm h}$}
\def\rchi{$\chi^2_{\nu}$}

\font\rmeight=cmr8 
\def\moda{{\rmeight MODEL~A}}
\def\modb{{\rmeight MODEL~B}}
\def\modc{{\rmeight MODEL~C}}

\def\srca{{\rmeight S1}}
\def\srcb{{\rmeight Q1}}
\def\srcc{{\rmeight Q2}}

\def\nssa{non-stellar spectrum 1}
\def\nssb{non-stellar spectrum 2}
\def\nssa{{\rmeight RS\_QA}}
\def\nssb{{\rmeight RS\_QB}}


\def\axrb{A_{\rm XRB}}
\def\gxrb{\Gamma_{\rm XRB}}
\def\asrc{A_{\rm src}}
\def\gsrc{\Gamma_{\rm src}}
\def\ath{A_{T_{\rm h}}}

\def\cagn{C_{\rm src}}

\def\sgxrb{\alpha_{\rm XRB}}
\def\sgsrc{\alpha_{\rm src}}

\def\ixrb{I_{\rm XRB}}
\def\isrc{I_{\rm src}}
\def\itot{I_{\rm tot}}

\def\asquare{\ht0=.4pc \dp0=.14pc \wd0=.6pc\makeblankbox{.2pt}{.2pt}\ }

\def\safxrb{steep-AGN-free XRB}


\def\img#1{\hbox{\psfig{figure=#1,width=9cm,%
bbllx=78bp,bblly=117bp,bburx=489bp,bbury=527bp,clip=t
}}}

\def\figga{
\beginfigure*{1}
\centerline{\img{}\img{b.ps} }
\medskip
\centerline{\img{b.ps}\img{b.ps} }
\caption{{\bf Figure.~1 (a)} The \asca\  SIS0 full field of view of the \qsf.  
The SIS is composed of four CCD chips (their outlines are shown by the dark solid lines).  For SIS0 in (a)--(b), they are chip~3, 
2, 1, 0 from top clockwise and chip~1, 0, 3, 2 for SIS1 in (c)--(d); each 
chip has an $11\times 11$~arcmin$^2$ field of view.  The pointing center is 
at RA (J2000) $55^{\circ}.44$, Dec (J2000) $-44^{\circ}.12$.    All 
the bad data have been screened out except  the particle background and  stray 
light.  
Each image has been binned into $\sim 30\times 30$~arcsec$^2$ pixels 
and smoothed with a Gaussian filter of dispersion $\sim 50$~arcsec.  
The grey-scale bar is in units of counts in the corresponding energy band per 
unit pixel per exposure time.  The levels of the contours are the values at the lower-right corner of each image in the same unit of the grey-scale bar.  
(a) The \qsf\  in 0.5--2~keV, three bright point sources can be seen in the 
upper left quarter of this image (sources 11, 37, and 41, white triangles); other sources detected by the PSPC within the SIS field of view are indicated by black triangles (the sources are numbered according to the PSPC source list of Table~2a  and Fig~2a).   (b) The \qsf\  in 2--8~keV.  The strong
 soft sources become much fainter.   (c)--(d) Same as (a) and (b) but observed 
with SIS1.  
 10 sources are detected in the merged (SIS0+SIS1) soft SIS image and 6 in the hard image.  See Table~2b and Sec~3.1.1 for details. }   
\endfigure}

\def\figgaa{
\beginfigure*{2}
\centerline{\img{b.ps}\img{b.ps} }
\caption{{\bf Figure.~1 (e)--(f)} The  merged \asca\  SIS images in the soft (Fig.~1e) and hard band (Fig.~1f).  The wider gap between 2 pairs of CCD chips disappears because of the offset along the SIS $y$-axis of the detector coordinate between the two SIS during the observation.
}   
\endfigure}


\def\figgab{
\beginfigure*{3}
\centerline{\hbox{\psfig{figure=b.ps,width=9cm,%
bbllx=51bp,bblly=134bp,bburx=453bp,bbury=497bp,clip=t}}
\hbox{\psfig{figure=b.ps,width=9cm,%
bbllx=51bp,bblly=134bp,bburx=453bp,bbury=497bp,clip=t}}
}\caption{{\bf Figure.~1 (g)--(h)} The pixel-to-pixel comparison of SIS0 and SIS1 image count rates in the hard (the diamonds $\diamond$) and soft (the crosses $\times$) bands.  The unit pixel size for the images is about $4'15''\times 4'15''$.   Each data set is fit by a linear equation $y=Ax+B$ ($x$ for the SIS0 count rate and $y$ for the SIS1).  The dotted and dashed straight lines are the best-fits in the hard and soft bands, respectively; the curved lines represent the $1\sigma$ intervals. 
  Fig~1h shows the soft SIS count rate against the hard SIS count rate (diamond/dotted lines for SIS0 and crosses/dashes lines for  SIS1).  Note that the bright pixels have a soft excess.}
\endfigure}

\def\figgac{
\beginfigure{4}
\centerline{\hbox{\psfig{figure=b.ps,width=9cm,%
bbllx=51bp,bblly=134bp,bburx=453bp,bbury=497bp,clip=t}}
}\caption{{\bf Figure.~1 (i)} The soft/hard ratio as a function of the photon index of the power-law model plus the Galactic absorption toward the QSF3 field.
}\endfigure}

\def\figgba{
\beginfigure*{5}
\centerline{\hbox{\psfig{figure=b.ps,width=15cm,%
bbllx=79bp,bblly=124bp,bburx=495bp,bbury=526bp,clip=t}}}
\caption{{\bf Figure.~2 (a)} The merged \rs\ PSPC-B+PSPC-C central field of view of the \qsf\ 
with pixel size $\sim 15\times 15$~arcsec$^2$ (exposure time $\sim$ 57~ks).  Only the central 16~arcmin 
region is used for our spectral analysis (the inner circular line).   The outer circular line indicates the region that was used in Shanks \etal\ (1991), and the solid lines outline the field of view of the 4 SIS1 CCD chips.   92 sources are detected in this image and are numbered from north to south  (also see Table~2a).   Sources reported and identified as QSO/non-QSO by Shanks \etal\  are represented by squares (\asquare)/triangles ($\triangle$),  and the new detected sources  by the crosses ($+$).  The three brightest sources are sources 11, 41, and 37.  
The thick crosses in the center of the image are the pointing positions of the PSPC (the eastern one) and SIS (the western one) observations.  
The same image processing described in Fig.~1a has been applied on the image.  
}\endfigure}

\def\figgbb{
\beginfigure{6}
\centerline{\hbox{\psfig{figure=b.ps,width=8cm,%
bbllx=95bp,bblly=87bp,bburx=458bp,bbury=587bp,clip=t}}}
\caption{{\bf Figure.~2 (b)} The hardness ratio of 92 sources detected by the PSPC in the 1--2~keV band (Table~2a).  The symbol of each data point follows the same definition used in Fig~2a.  The dotted line is the mean hardness ratio when the points are grouped, and the solid lines represent the 1~$\sigma$ interval.   The hardness ratio is defined as $(H-S)/(H+S)$}
\endfigure}

\def\figgbc{
\beginfigure{7}
\centerline{\hbox{\psfig{figure=b.ps,width=8cm,%
bbllx=67bp,bblly=410bp,bburx=544bp,bbury=776bp,clip=t}}}
\caption{{\bf Figure.~2 (c)} The cross SIS and PSPC hardness ratio of sources detected by SIS (see Table~2b).    Error bars are reduced by a factor of 5 for display purposes.}
\endfigure}

\def\figgc{
\beginfigure*{8}
\centerline{\hbox{\psfig{figure=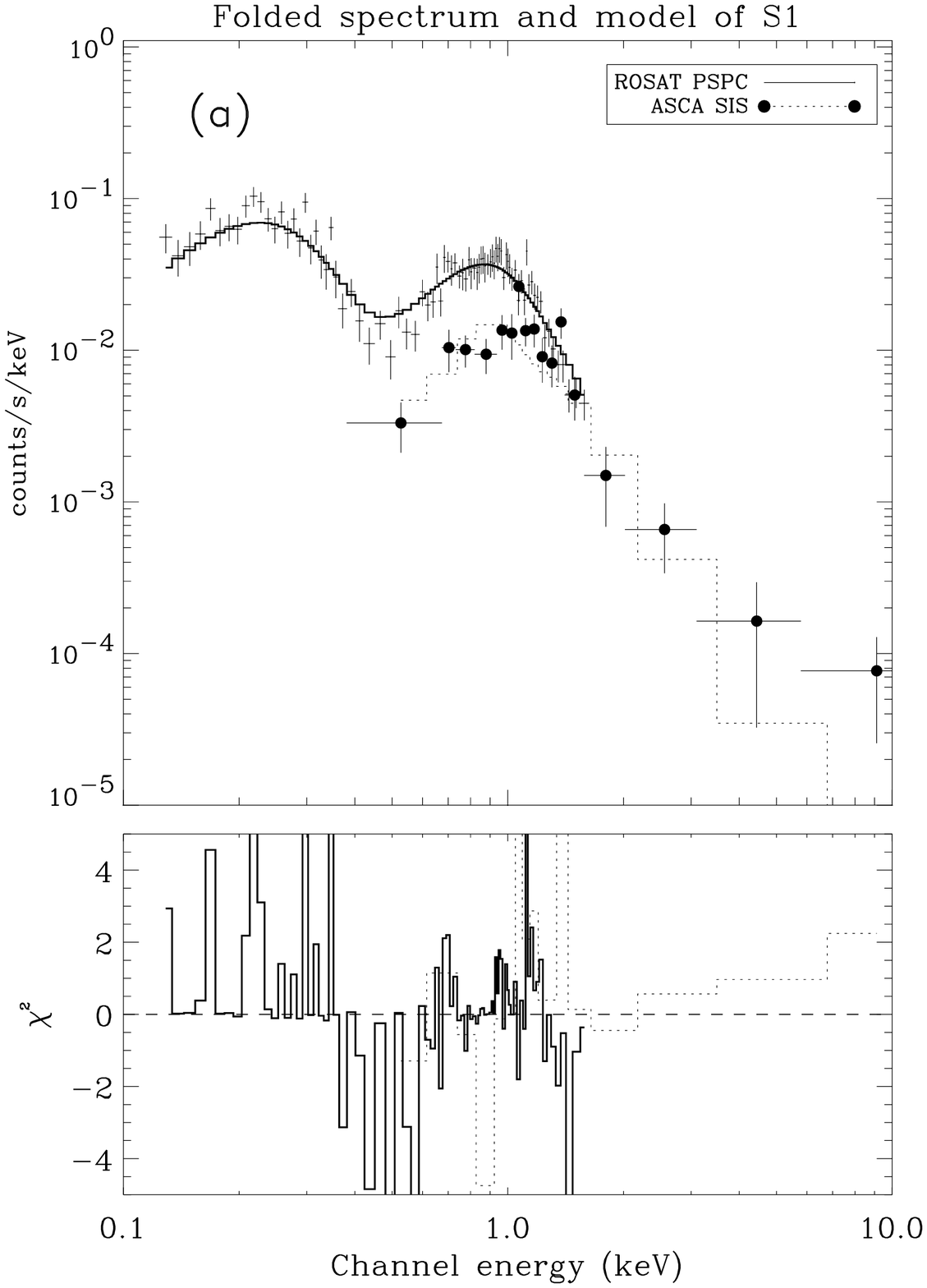,height=8.8cm}
\psfig{figure=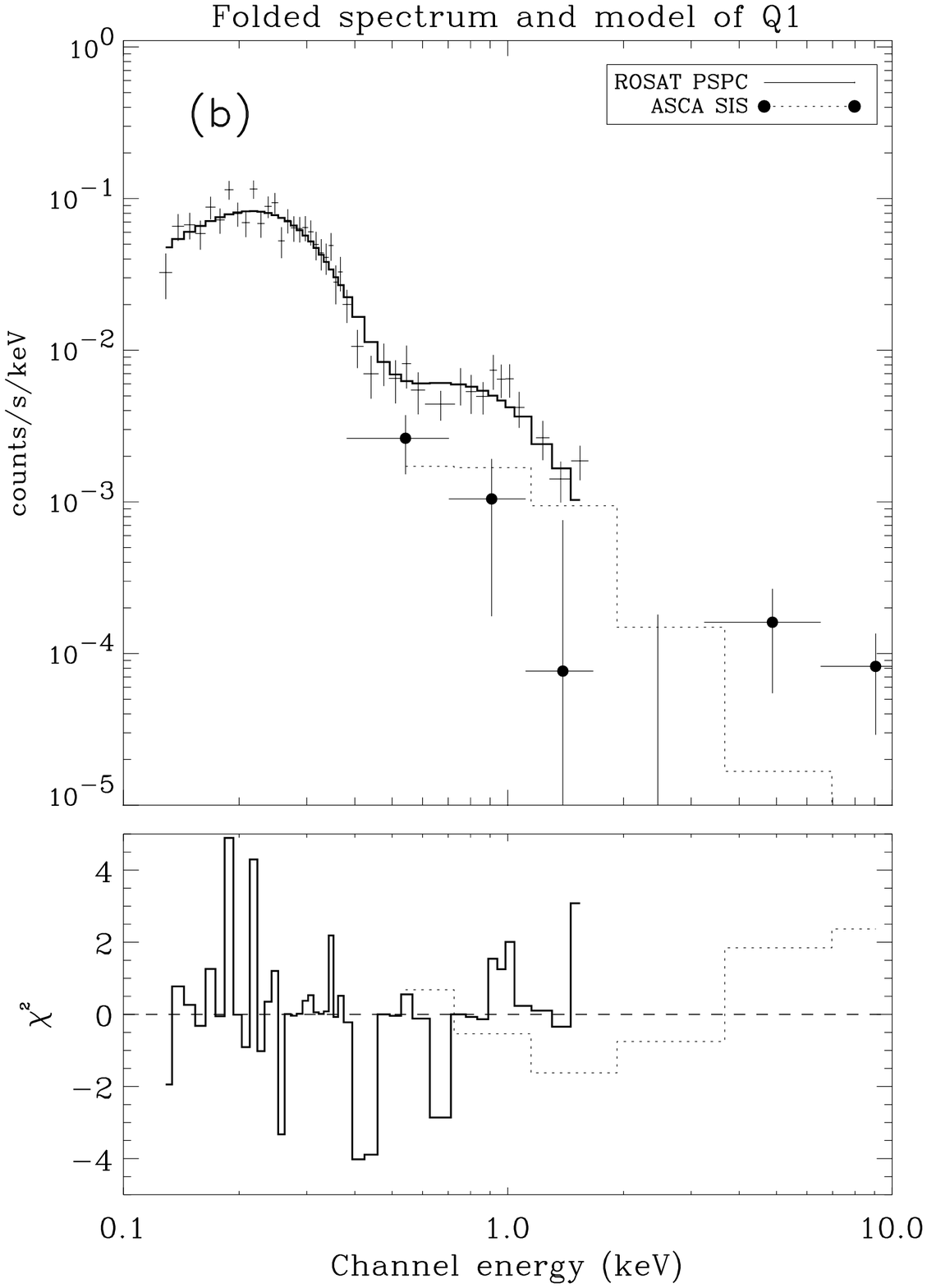,height=8.8cm}
\psfig{figure=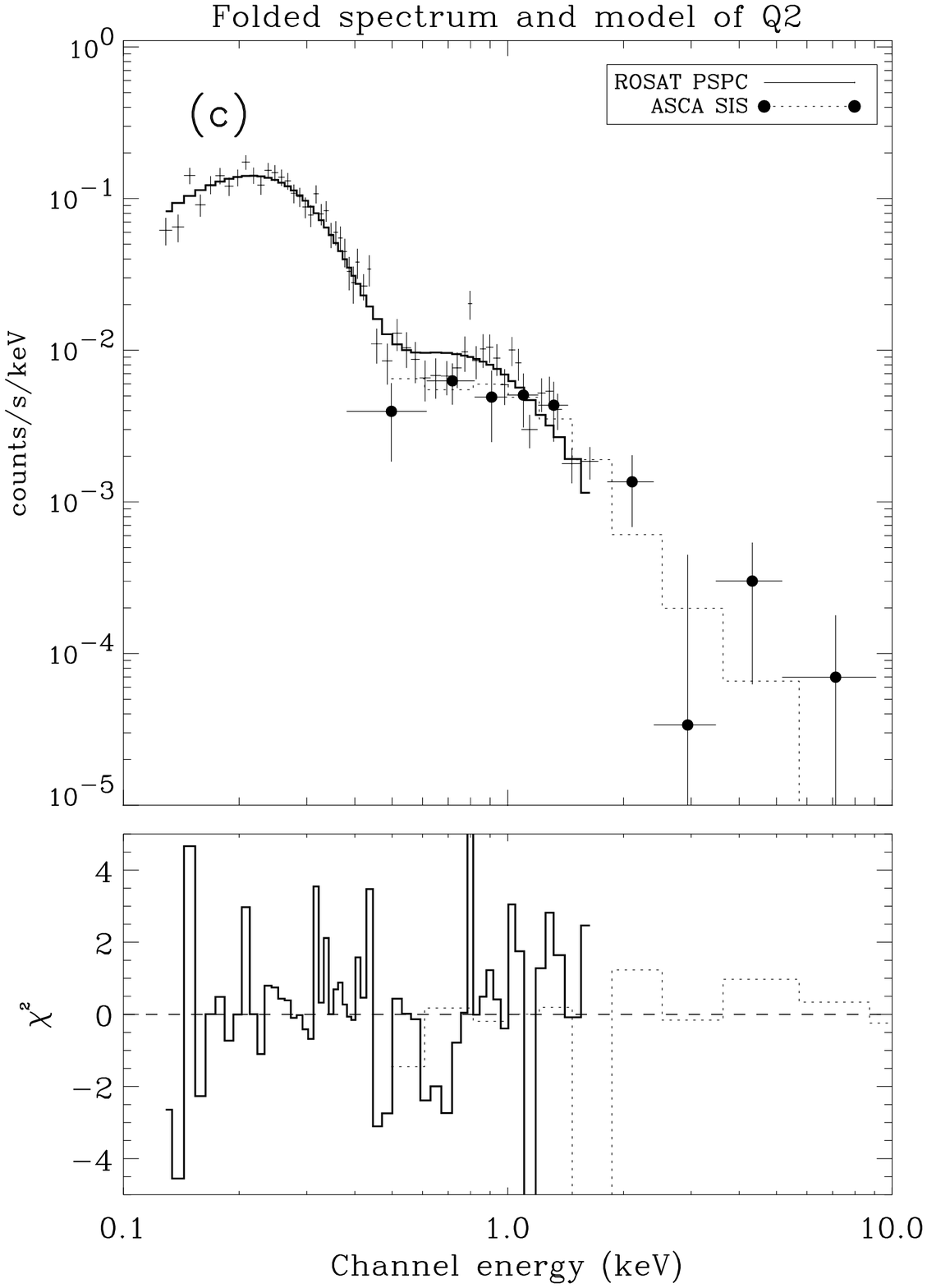,height=8.8cm}}}
\caption{{\bf Figure.~3} Spectra of three sources resolved by the SIS.  
(a) Source \srca: a star (b) Source \srcb: a $z$=0.6 quasar 
(c) Source \srcc: a $z$=0.3 quasar.  In each figure, the count spectrum and 
folded model are shown in  the upper panel and the $\chi^2$ contributions to
each bin in the 
bottom panel.  The filled circles and dots represent the folded 
spectrum measured by the \asca\ SIS and the \rs\ PSPC, 
respectively, and
the dotted and solid lines are the respective best-fitting 
 model.  See Table~3a for the best-fitting results. 
}
\endfigure}

\def\figgd{
\beginfigure{9}
\centerline{\hbox{\psfig{figure=b.ps,width=9.cm,%
bbllx=76bp,bblly=97bp,bburx=534bp,bbury=568bp,clip=t}}}
\caption{{\bf Figure.~4} Folded XRB model (\moda) and spectrum  
measured by the SIS (0.4--7~keV) and the PSPC (0.1--2~keV).  Data 
divided by the best-fitting model are shown in the 2 panels at the bottom.  
Note that the amplitudes of the \rs\ spectrum and model have been reduced 
by a factor of 20 for display purposes.  
} 
\endfigure}

\def\figgda{
\beginfigure{10}
\centerline{\hbox{\psfig{figure=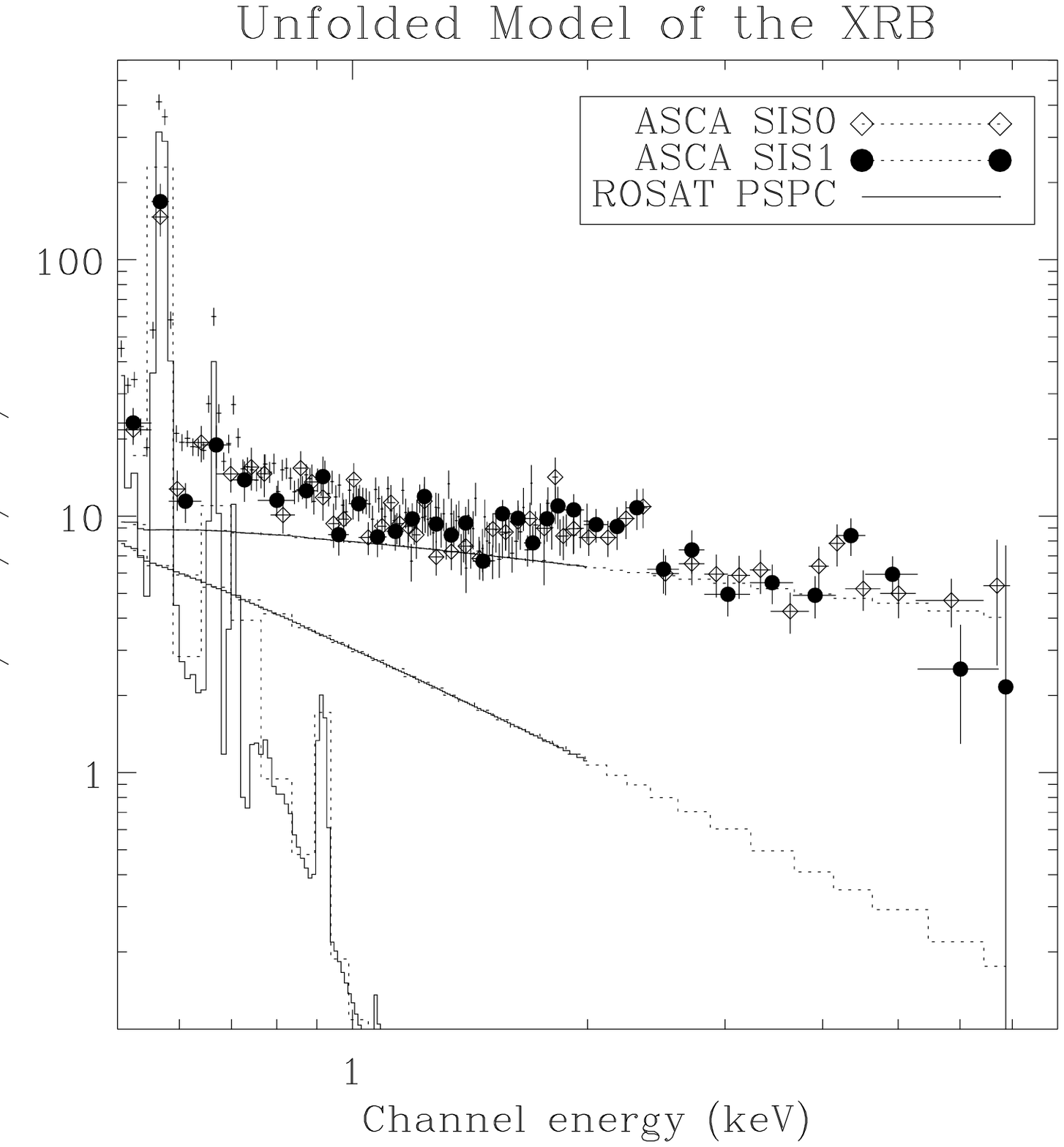,width=9.cm}}}
\caption{{\bf Figure.~5} Unfolded XRB spectrum above 0.5~keV measured 
by the SIS  and the PSPC.  The dotted and solid lines represent the 
best-fitting model components (\modb) of the \asca\ and \rs\ data 
respectively.  The model consists of three components to take  into account contributions 
from Galactic emission (the lines with line features), AGN-like sources 
(the steep lines with photon index $\gsrc$ fixed at 2.5) and the yet unknown 
XRB sources (the flat lines).}
\endfigure}

\def\figsixa{
\beginfigure{11}
\centerline{\hbox{\psfig{figure=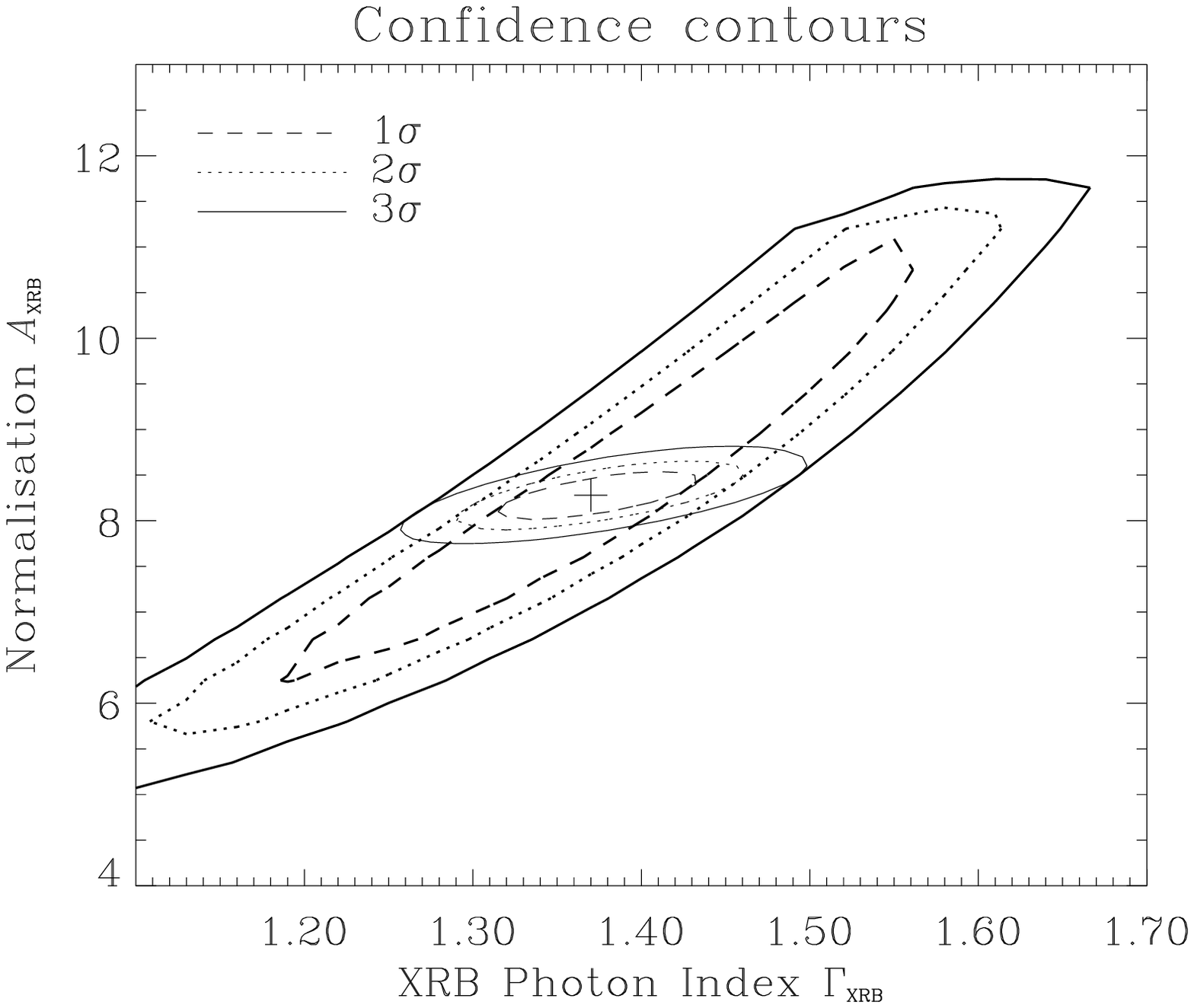,width=9.cm}}}
\caption{{\bf Figure.~6a} Confidence contours of the XRB photon index and 
normalisation for the spectral fitting in Fig.~5.  The cross indicates the best-fit and the small contours are calculated with the steep power-law 
component fixed at $\gsrc=2.5$ and $\asrc=3.16$.} 
\endfigure}

\def\figsixb{
\beginfigure{12}
\centerline{\hbox{\psfig{figure=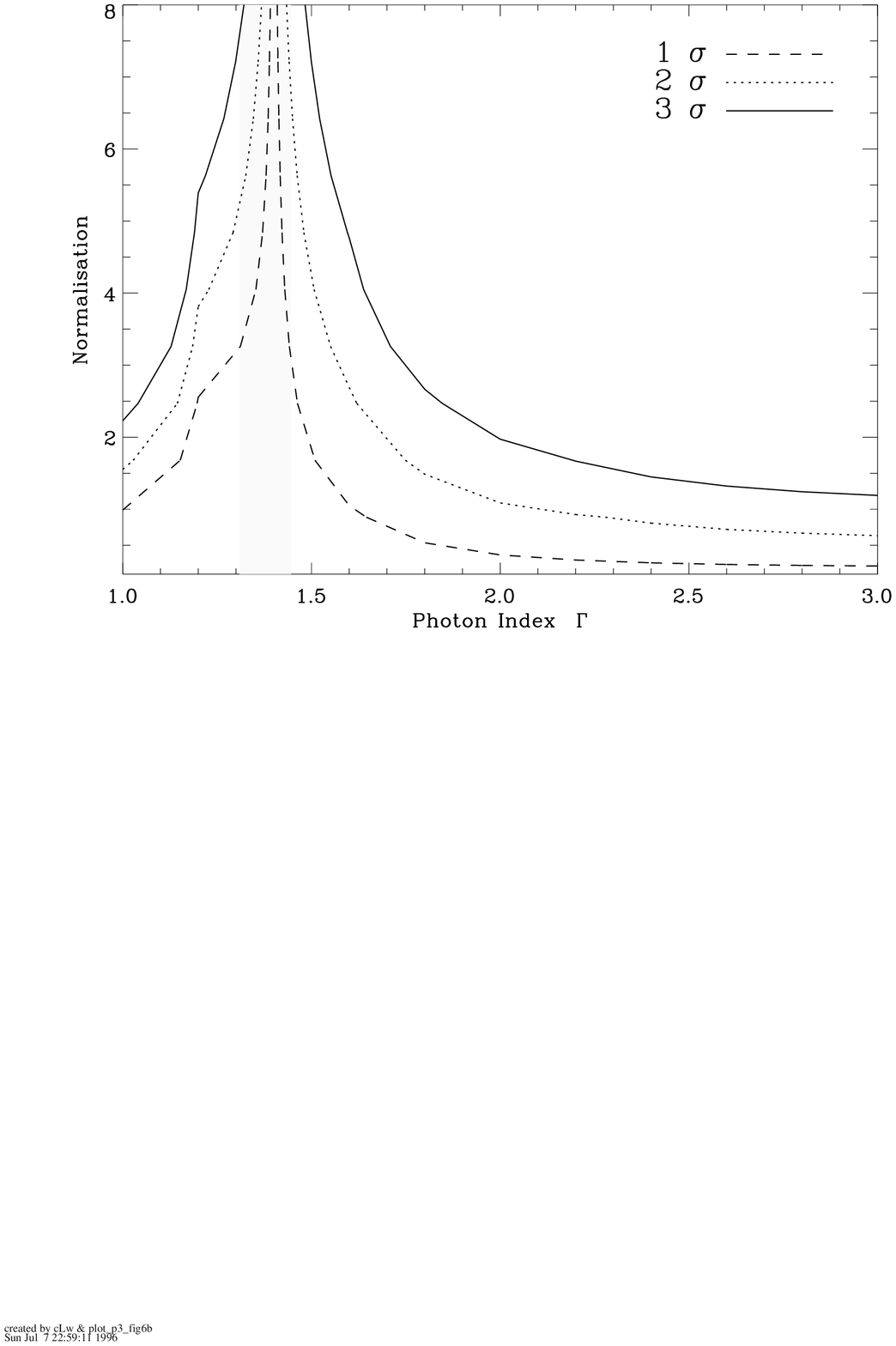,width=9.cm,%
bbllx=80bp,bblly=419bp,bburx=537bp,bbury=765bp,clip=t}}}
\caption{{\bf Figure.~6b} Confidence contours of the photon index and 
normalisation of the third power-law component added in \modb\ (see Sec.~3.4.3).  The shadowed area represents the 90~per cent confidence interval of the XRB photon index  when the XRB spectrum is fit by \modc.} 
\endfigure}

\def\figgdc{
\beginfigure*{13}
\centerline{\hbox{\psfig{figure=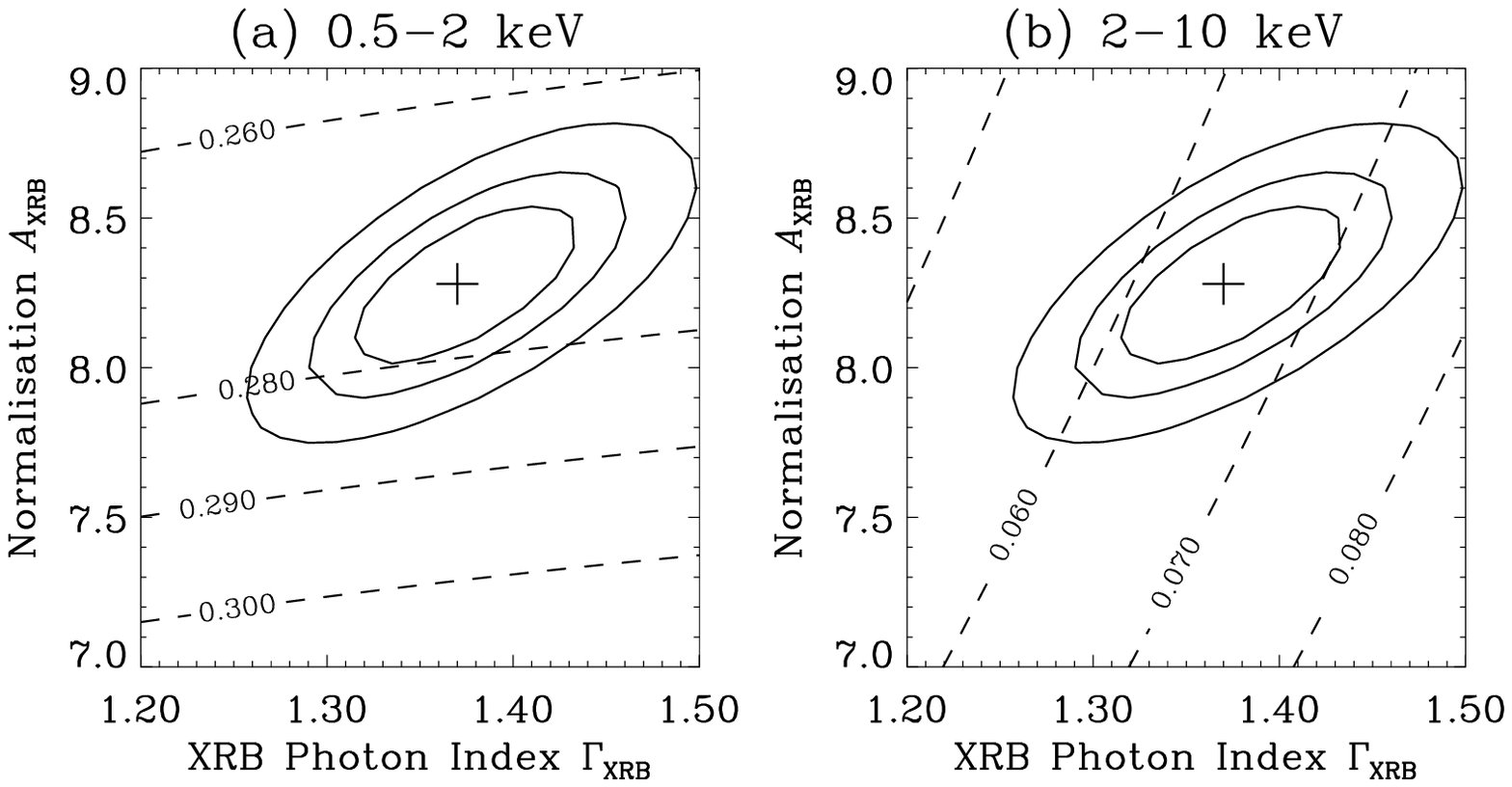,width=18cm,%
bbllx=35bp,bblly=486bp,bburx=515bp,bbury=733bp,clip=t
}}}
\caption{{\bf Figure.~7} The ratio of the AGN contribution to 
the XRB.  The solid contour lines are the same small 
contours in Fig.~6a.  With $\gsrc$ and $\asrc$ fixed, the ratio of 
$\isrc/(\isrc+\ixrb)$ is calculated over the $\gxrb$, $\axrb$ space, 
which is represented by the dashed contours.  
(a) Ratio calculated in the 0.5--2~keV band (b) in the 2--10~keV band.} 
\endfigure}

\def\figge{
\beginfigure{14}
\vskip -.5cm
\centerline{\hbox{\psfig{figure=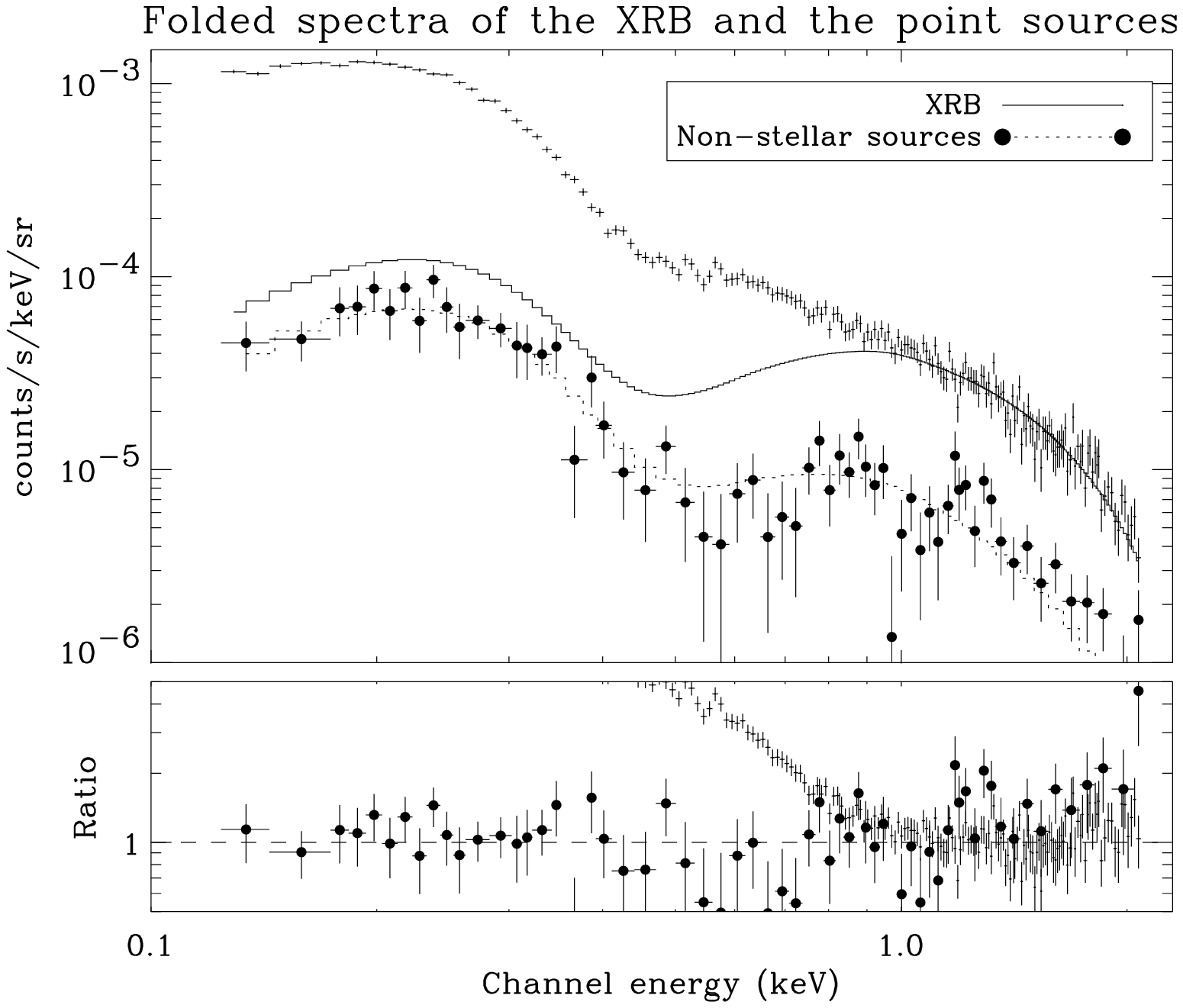,width=9.cm,%
bbllx=48bp,bblly=87bp,bburx=497bp,bbury=473bp,clip=t}}}
\caption{{\bf Figure.~8} The folded spectrum of the XRB and the accumulated 
non-stellar spectrum \nssb\ (the filled circles) from the same \rs\ data.  
The dotted line is the best-fitting power-law model of the non-stellar 
spectrum and the solid line is the power-law component of the best-fitting 
XRB model (the best-fit of $\cagn=1$ in Table~4).  The deviation of the folded XRB spectrum to the solid line shows a large soft excess of the XRB below 1~keV.  The ratio of the data to the power-law component is shown in the 
bottom panel.}
\endfigure}

\def\figgf{
\beginfigure{15}
\centerline{\hbox{\psfig{figure=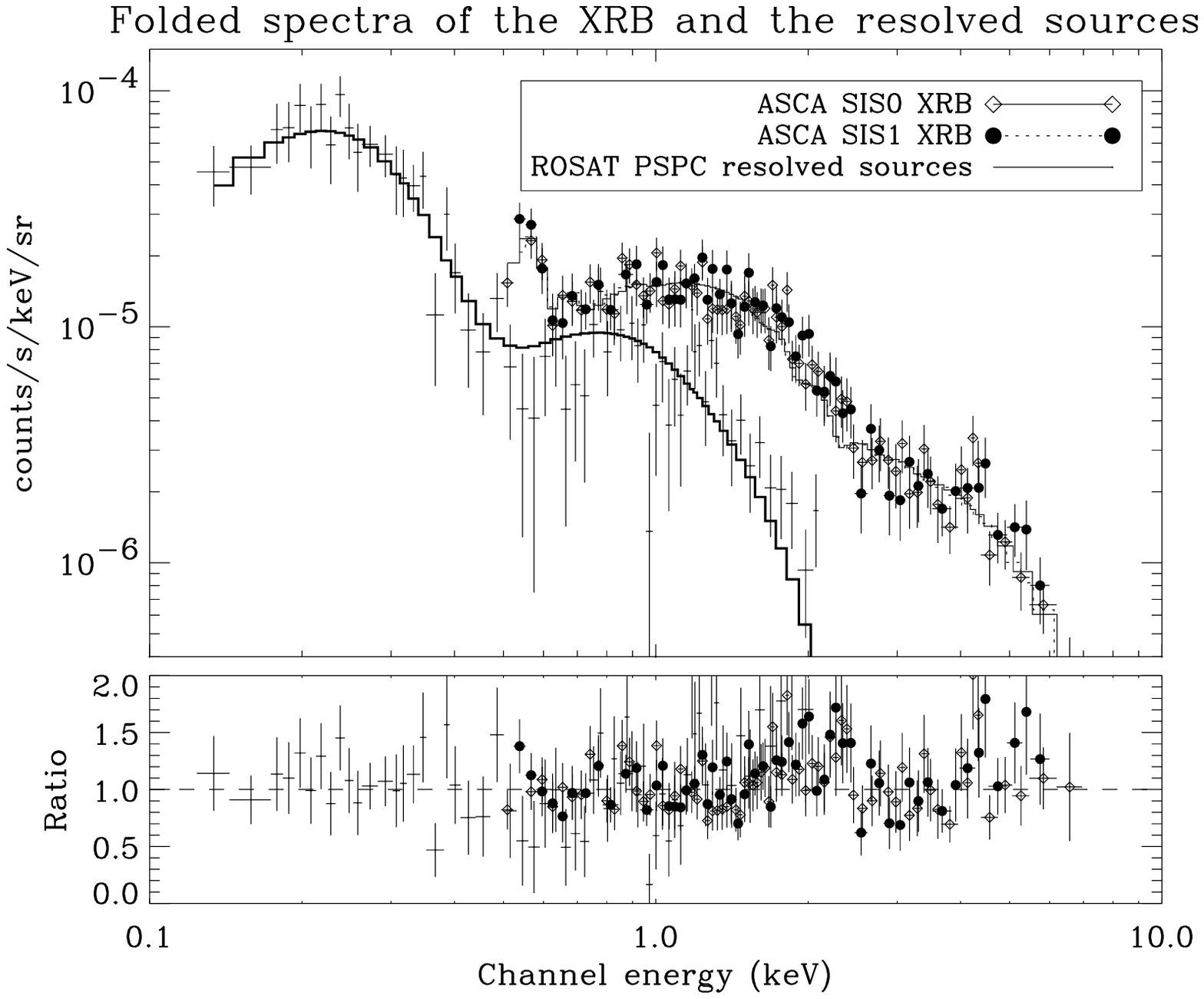,width=9.cm,%
bbllx=33bp,bblly=84bp,bburx=521bp,bbury=478bp,clip=t}}}
\caption{{\bf Figure.~9} Folded spectra of the SIS XRB (0.4--7~keV) and the 
PSPC non-stellar point sources \nssb\ (0.1--2~keV).}
\endfigure}

\def\figgg{
\beginfigure{16}
\centerline{\hbox{\psfig{figure=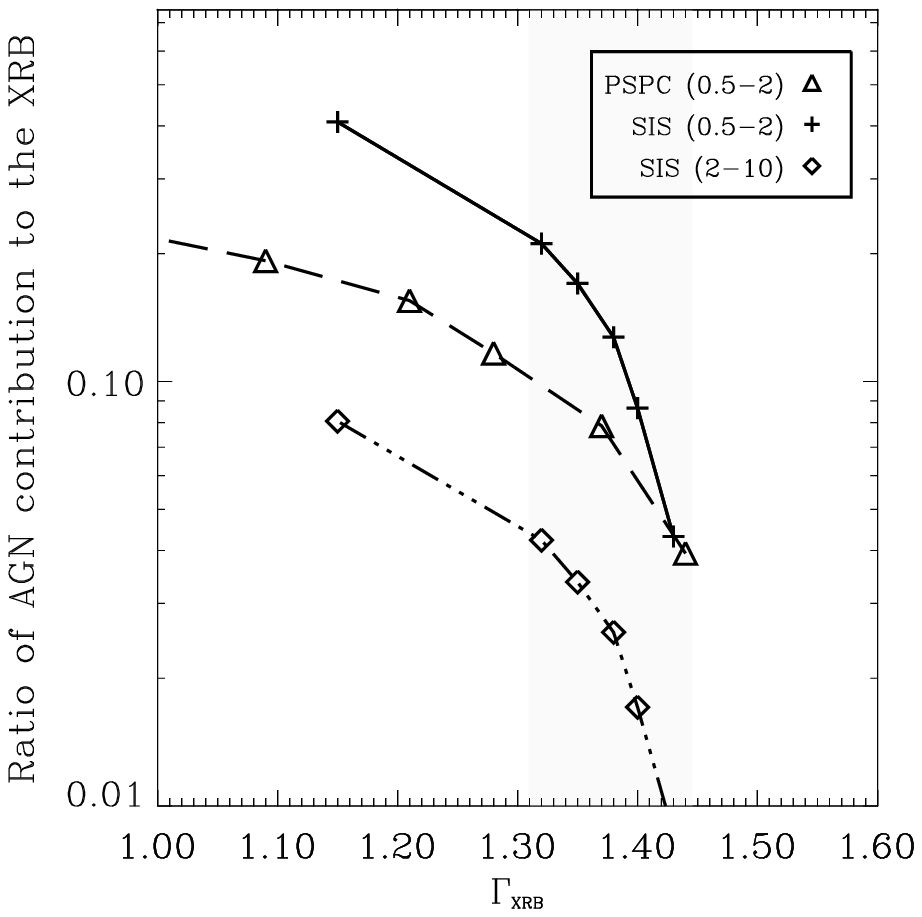,width=8.cm,%
bbllx=156bp,bblly=205bp,bburx=434bp,bbury=479bp,clip=t}}}
\caption{{\bf Figure.~10} The fractional contribution of the steep AGN to the XRB as a function of the best-fitting $\gxrb$ based on 
Tables~4--5.  PSPC refers to the fitting results of the
 PSPC \safxrb\ and PSPC AGN \nssb\ spectra (Table~4); 
SIS (0.5--2~keV) and SIS (2--10~keV) are both obtained from fitting the PSPC XRB and PSPC AGN spectra (Table~5) and the power-law model of the XRB.  The shadowed area indicates the 90 per cent confidence interval of $\gxrb$ as obtained from \modc.}
\endfigure}

 \def\figgi{
\beginfigure{17}
\centerline{\hbox{\psfig{figure=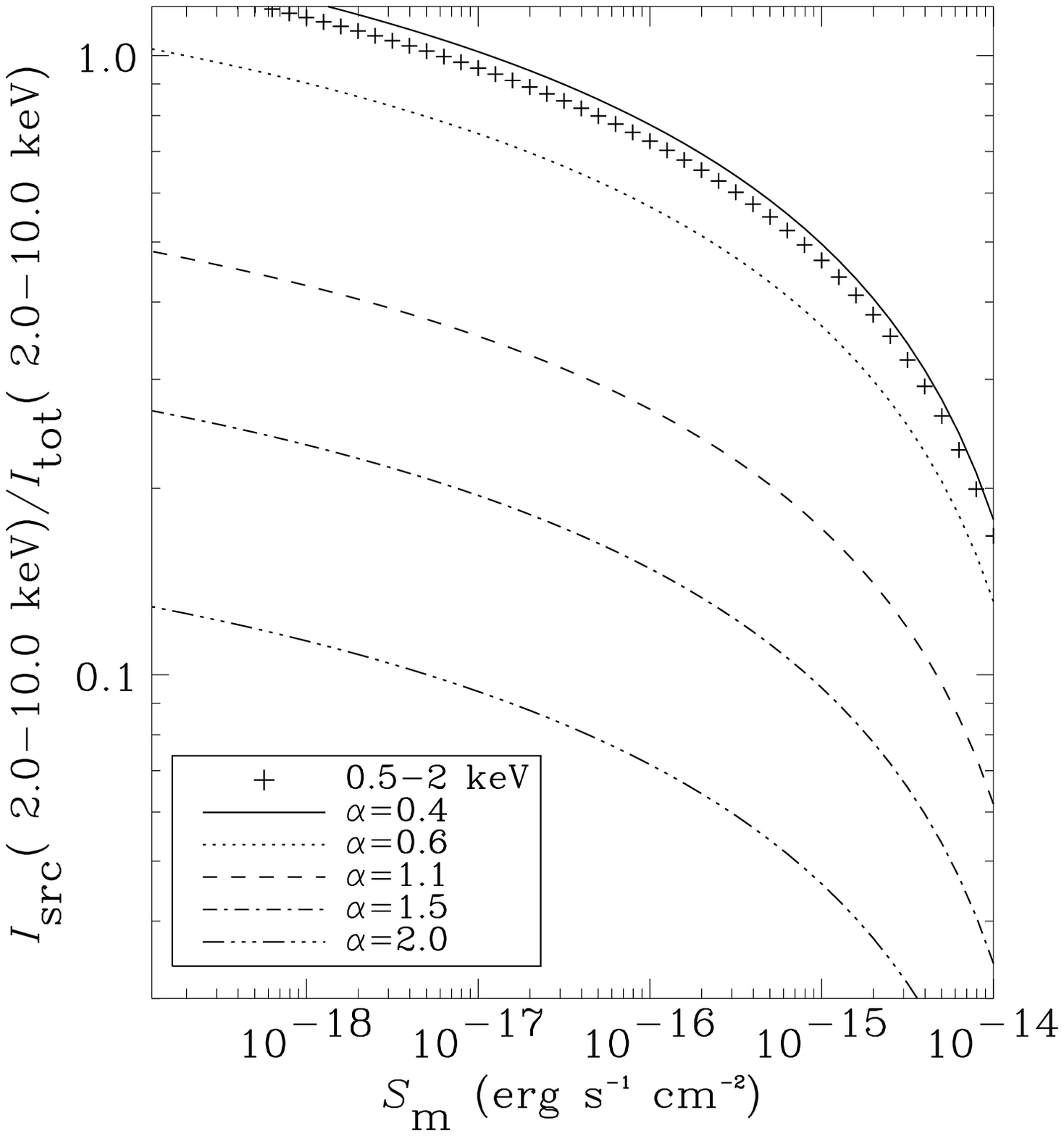,width=8cm,%
bbllx=54bp,bblly=65bp,bburx=501bp,bbury=540bp,clip=t}}}
\caption{{\bf Figure.~11} The contribution  of  discrete sources to the XRB.  
The ratio is $\isrc(\el-\eh)/\itot(\el-\eh)$ according to the observed  
\logns\ (Hasinger \etal\ 1993) and the XRB spectrum (this work).  The 
crosses represent the ratio for $(\el,\eh)=(0.5,2)$~keV which is 
$\sgsrc$-independent, the other lines are ratios for different $\sgsrc$ 
 in the 2--10~keV band as indicated in the legend.  
Note $S_{\rm m}$ is always in the 0.5--2~keV band.}
\endfigure}


\def\abcg{the ABC Guide}
\def\tbf{\rm}

\def\tableaa{
\begintable*{1}
\caption{{\bf Table 1 (a) }\asca/\rs\ observations of the \qsf}
\vbox{
\offinterlineskip\halign{
 \strut
##\quad     & ##\quad    & ##\hfil\quad  & ##\quad   & ##\quad\cr
\noalign{\vskip 4pt\hrule\vskip 4pt}
\omit Instrument &  Dates     & Pointing Centre & Detector&Time$^*$ \cr
\omit      &                & RA, Dec (J2000) &         &               \cr
\asca & 93-Jul-11/12 &$55^{\circ}.44,-44^{\circ}.12$& SIS0 & 21303         \cr
      &                &                 & SIS1    & 17317         \cr
\rs   & 90-Jul-23/24 &$55^{\circ}.56,-44^{\circ}.13$ & PSPC-C  & 21935     \cr
      & 93-Aug-04/05 &$55^{\circ}.55,-44^{\circ}.13$ & PSPC-B  & 34735     \cr
\noalign{\vskip 2pt\hrule}
}}
\tabletext{\item{*}Exposure time is in seconds and calculated after screening 
described in Sec.~2.}
\endtable
}


\def\tableab{
\begintable{2}
\caption{{\bf Table 1 (b) }Modes and screening criteria of \asca\ observations (see Sec~2.2.1 for description)}
\vbox{
\offinterlineskip\halign{
 \strut
##\hfil \quad& \hfil ## \quad& \hfil##\cr
\noalign{\hrule\vskip 4pt}
\omit\tbf  Mode/HK Parameter& \tbf Setting & {\tbf Reference}$^*$\cr
\noalign{\vskip 4pt\hrule\vskip 4pt}
Clock Mode  & 4CCD            & 5.4.1\cr
Data Mode   & faint, bright   & 5.4.2\cr
Grade       &0, 2, 3, 4       & 5.4.4\cr
Bit-rate    & high, medium    & 2.2\cr
            &                 &  \cr
\tt SAA          &  0         & 5.2.2 \cr
\tt T\_SAA      &  $>$60 s      & 5.2.2 \cr
\tt COR\_MIN    &  $>$8 GeV/$c$ & 5.2.4 \cr
\tt T\_DY\_NT  &  $>$200 s     & 5.2.4 \cr
\tt ELV\_MIN    &  $>$5 deg     & 5.2.5 \cr
\tt BR\_EARTH   &  $>$20 deg    & 5.2.5 \cr
\noalign{\vskip 2pt\hrule}
}}
\tabletext{$^*$Section of \abcg.}
\endtable
}


\def\tableb{
\begintable*{6}
\caption{{\bf Table 3 (a).} Best-fit results for the resolved sources}
\vbox{
\offinterlineskip\halign{
 \strut
##\hfil\quad& ##\hfil\quad&  ##\quad& \hfil ##\hfil& ##\quad& 
                                            ##\hfil\quad& ##\hfil\quad\cr
\noalign{\hrule\vskip 4pt}
Source &  &  & Flux     &         & Best-fit & \rchi \cr
   &    &{\it a} & {\it b} & {\it c}  &              &\cr
\noalign{\hrule\vskip 4pt}
\srca\ (star) & SIS+PSPC & $28.0\pm 1.8$ & $35.3^{+5.8}_{-5.1} $ &$ 3.5^{+0.6}_{-0.5}$ & $kT=0.81^{+0.05}_{-0.06}$~keV, $0.05\pm 0.01 Z_{\odot}$ &1.44\cr
&  & &   & & $kT_{\rm bremss}=1.01$~keV & 2.07 \cr
 & SIS   & &         &    & $kT=1.12^{+0.24}_{-0.15}$~keV, $0.13_{-0.09}^{+0.17} Z_{\odot}$   & 1.15  \cr
&  & &   & & $kT_{\rm bremss}=1.02^{+0.34}_{-0.22}$~keV & 1.35 \cr
  &PSPC   & &         &    & $kT=0.74^{+0.07}_{-0.05}$~keV, $0.05\pm 0.01 Z_{\odot}$   & 0.87  \cr
&  & &   & & $kT_{\rm bremss}=1.01$~keV & 2.35 \cr
\srcb\ ($z=0.64$ QSO)  & SIS+PSPC 
    & $4.8\pm 0.7$ & $4.5^{+2.3}_{-3.1}$   &$1.1^{+0.6}_{-0.8}$ & $\Gamma=3.08\pm 0.10 $ & 1.18\cr
 &SIS         &  &    &   & $\Gamma=3.71^{+1.93}_{-1.73} $ & 1.39\cr
 &PSPC         &  &    &   & $\Gamma=2.98^{+0.11}_{-0.09}  $ & 0.70\cr
\srcc\ ($z=0.38$ QSO) & SIS+PSPC 
& $7.6^{+0.7}_{-0.8}$ & $8.0\pm 2.3$ & $1.9\pm 0.5$ & $\Gamma=3.15\pm 0.07 $ &1.36\cr
 &SIS         &  &    &   & $\Gamma=2.75^{+0.68}_{-0.66} $ & 1.16\cr
 &PSPC         &  &    &   & $\Gamma=3.2^{+0.08}_{-0.06}   $ & 1.46\cr
\nssa &PSPC   &&&& $\Gamma=2.76_{-0.06}^{+0.07}$, $A=3.09\pm 0.23$ & 1.1\cr
\nssb &PSPC   &&&& $\Gamma= 2.53_{-0.11}^{+0.10}$, $A=2.24\pm 0.23$ &0.96\cr
\noalign{\vskip 2pt\hrule}
}}
\tabletext{Flux is in unit of $10^{-14}$~\ergpercmpers.  {\it a} in the PSPC 0.5--2~keV 
band, {\it b} in the SIS 0.5--2~keV band, {\it c} in the SIS 2--10~keV band.  
\srca\ is fit by  a Raymond-Smith gas ($T$) and thermal bremsstrahlung ($T_{\rm bremss}$) model; \srcb\ and \srcc\ are fit by a power-law.  
\nssa\ and \nssb\ are the accumulated \rs\ non-stellar spectra as defined at
the end of Sec.~3.1.   $A$ is the normalisation in units of \uunit.  
}\endtable
}


\def\xx{$\bullet$}
\def\xx{$-$}

\def\tablec{
\begintable*{7}
\hrule\nofloat
 \caption{{\bf Table 3 (b).} Best-fitting parameters of various XRB spectrum models}
\vbox{
\offinterlineskip\halign{
 \strut
       ##\hfil&\quad 
        ##\hfil&\quad ##\hfil&\quad ##\hfil&\quad ##\hfil&\quad ##\hfil&\quad           ##\hfil &\quad ##\hfil\cr
\noalign{\hrule\vskip 4pt}
MODEL & \gs & \ns & \ts & \gh & \nh & \th & \rchi  \cr
\noalign{\hrule\vskip 4pt}
{\bf \asca\ SIS}  &  &     &  & & & &     \cr
P (1--7~keV)  &\xx&\xx&\xx  &$1.43\pm 0.08$ 
             &$10.0\pm 0.6       $ &\xx & 1.12 \cr
P (1--3~keV) &\xx&\xx&\xx  &$1.30\pm 0.13$ 
             &$9.6\pm 0.7$ &\xx & 0.96 \cr
P (3--7~keV) &\xx&\xx&\xx  &$1.40\pm 0.45$ 
             &$9.0_{-4.4}^{+8.7}$ &\xx & 1.40\cr
W (P + R)    &\xx&\xx  &\xx & $1.44_{ -0.06}^{+  0.07} $
             &$10.2\pm 0.5$  & 0.12& 1.24 \cr
W (P$_{\rm h}$ + R) + P$_{\rm s}$&
    0.36 & $0 $  &\xx & $1.44\pm 0.07$  & $10.2\pm 0.5$ & 0.12 &1.26      \cr
W (P + R$_{\rm h}$) + B    &\xx
    &\xx  & 0.21 & $1.44\pm 0.07$  & $10.2\pm 0.5$ & 0.12    & 1.26 \cr
W (P$_{\rm h}$) + P$_{\rm s}$&
  $4.33_{ -0.94}^{+  1.29}$ & $1.36_{ -0.86}^{+  1.64}$ &\xx&
  $1.41_{ -0.14}^{+  0.11}$ & $9.6_{ -1.6}^{+  1.1}$ & \xx&1.70     \cr
&  &     &  && &     &  \cr    
\moda&  &     &  & & &     &  \cr     
W (P + R$_{\rm h}$) + R$_{\rm s}$&\xx
&\xx& 0.02 & $1.44\pm 0.07$ & $10.16_{-0.62}^{+0.44}$ & 0.13 &1.23     \cr
\noalign{\vskip 2pt\hrule\vskip 4pt}
{\bf \asca$+$\rs}  &  &     &  &&  &     &  \cr
P (1--7~keV) &\xx&\xx&\xx &$1.48\pm 0.07$ &$10.5\pm 0.4$ &\xx & 1.14\cr
P (1--3~keV)&\xx&\xx&\xx  & $1.39\pm 0.11$ &$10.3\pm 0.4$ &\xx & 1.07\cr
W (P$_{\rm h}$ + R) + P$_{\rm s}$&
     $4.77_{ -0.70}^{+  0.90}$ & $0.08_{ -0.00}^{+  0.23} $  &\xx
     &$1.45\pm 0.06$ & $10.4_{ -0.3}^{+  0.4}$ & 0.12 &1.13     \cr
W (P + R$_{\rm h}$) + B &\xx
 &\xx & 0.1 &$1.42_{ -0.05}^{+  0.06}$& $10.2_{ -0.3}^{+  0.4}$ & 0.14 &1.13 \cr
&  &     &  && & &       \cr    
\moda&  &     &  && &      &  \cr       
W (P + R$_{\rm h}$) + R$_{\rm s}$&\xx
&\xx & 0.05 &$1.46\pm 0.06$& $10.5\pm 0.3$ & 0.12 &1.12\cr
&  &     &  && & &      \cr      
\modb& $\gsrc$ (fixed) & $\asrc$ &  & $\gxrb$ & $\axrb$ & \th & \rchi \cr       
(above 0.5~keV)&  &     &  && &      &  \cr   
W (P$_{\rm XRB}$ + P$_{\rm src}$ + R$_{\rm h}$)  
&  2.5 & $3.2_{-2.0}^{+1.5}$ &  & $1.37_{ -0.21}^{+0.22}$ & $8.3_{ -3.2}^{+2.2}$ & 0.09 & 1.46 \cr     
&  3.0 & $1.8_{ -1.1}^{+1.0}$ &  & $1.44\pm 0.15$ & $9.6_{ -1.6}^{+0.0}$ & 0.09 & 1.46 \cr     
&  &     &  && & &      \cr      
\modc& $\gsrc$ (fixed) & $\asrc$ (fixed)&  & $\gxrb$ & $\axrb$ &  & \rchi \cr       (above 1~keV)&  &     &  && &      &  \cr   
W (P$_{\rm XRB}$ + P$_{\rm src}$)  
&  2.53 & 2.24 &  & $1.38\pm 0.07 $ &  $8.7_{-0.4}^{+0.3} $ & \xx  & 1.15 \cr     
\noalign{\vskip 2pt\hrule\vskip 4pt}
}}

 \tabletext{
\item{1} Codes of the model: 
\itemitem{} P--power-law, 
\itemitem{} R--Raymond-Smith hot plasma, with redshift=0 and abundance=1  fixed
\itemitem{} W--Galactic HI absorption, fixed at 1.66$\times 10^{20}$~cm$^{-2}$
\itemitem{} B--thermal bremsstrahlung model.

\item{2} Symbols of the model component (subscript s/h indicates the soft/hard  component): 
\itemitem{} $\Gamma$--photon index of the power-law component, 
\itemitem{} $A$--normalization of the power-law component in \uunit,
\itemitem{} $kT$--temperature of the thermal model (either Raymond-Smith or  
bremsstrahlung) in keV.
\itemitem{}``\xx"-- component not involved.
\item{3}The boundary of the soft/hard band in this model fitting is  
around 0.5~keV where the significance of the Galactic  absorption changes.
}
\endtable
}


\def\tablefour{
\begintable{8}
\nofloat
 \caption{{\bf Table 4.} Best fit results of the PSPC \safxrb\ and PSPC 
steep-AGN \nssb\ spectra}
\vbox{
 \halign{&##\hfil&\quad
 \strut
        ##\hfil&\quad ##\hfil&\quad ##\hfil&\quad ##\hfil&\quad ##\hfil
         &\quad ##\hfil\cr
\noalign{\vskip 4pt\hrule\vskip 4pt}
$\cagn$& $\gsrc$ & $\asrc$ & $\gxrb$ &  $\axrb$   &   $\ath$& \rchi\cr
0.0   & 2.53 & 2.24 & 1.53 & 11.3  & 38.6 &  0.96 \cr
0.2 & 2.53 & 2.24 & 1.44 & 10.9  & 38.7 &  0.96 \cr
0.4 & 2.53 & 2.24 & 1.37 & 10.4  & 38.5 &  0.96 \cr
0.6 & 2.53 & 2.24 & 1.28 & 10.0  & 38.2 &  0.95 \cr
0.8 & 2.53 & 2.24 & 1.21 &  9.5  & 37.8 &  0.95 \cr
1.0   & 2.53 & 2.24 & 1.09 &  9.0  & 37.6 &  0.95 \cr
2.0   & 2.53 & 2.24 & 0.20 &  2.1  &  35.1&  0.96 \cr
\noalign{\vskip 2pt\hrule}
}}
\tabletext{
*The degree of freedom is 282 for all the fits.}
\endtable
}

\def\tablefive{
\begintable{9}
\nofloat
 \caption{{\bf Table 5.} Best fit results of the PSPC steep-AGN \nssb\ spectrum and the SIS XRB spectrum}
\vbox{
 \halign{&##\hfil&\quad
 \strut
          ##\hfil&\quad ##\hfil&\quad ##\hfil&\quad ##\hfil&\quad ##\hfil
          &\quad ##\hfil\cr
\noalign{\vskip 4pt\hrule\vskip 4pt}
$\cagn$& $\gsrc$ & $\asrc$ & $\gxrb$ &  $\axrb$   &   $\ath$& \rchi\cr
0.0   & 2.53 & 2.24 & 1.45 & 10.3  & 25.6 &  1.01 \cr
0.2   & 2.53 & 2.24 & 1.43 &  9.9  & 25.0 &  1.01 \cr
0.4   & 2.53 & 2.24 & 1.40 &  9.4  &  24.7&  1.01 \cr
0.6   & 2.52 & 2.23 & 1.38 &  9.1  &  24.1&   1.02\cr
0.8   & 2.53 & 2.23 & 1.35 &  8.6  &  23.8&  1.02 \cr
1.0   & 2.52 & 2.23 & 1.32 &  8.2  &  23.4&  1.02 \cr
2.0  & 2.52 & 2.22 & 1.15 &  6.2  &  21.0 &  1.05 \cr
\noalign{\vskip 2pt\hrule}
}}
 \tabletext{
*The degree of freedom is 293 for all the fits.}
\endtable
}

\def\tablesix{
\begintable{10}
\nofloat
 \caption{{\bf Table 6.} Best fit results of the PSPC XRB and PSPC steep-AGN \nssb\ spectra}
\vbox{
 \halign{&##\hfil&\quad
 \strut
         ##\hfil&\quad ##\hfil&\quad ##\hfil&\quad ##\hfil&\quad ##\hfil
         &\quad ##\hfil\cr
\noalign{\vskip 4pt\hrule\vskip 4pt}
$\cagn$& $\gsrc$ & $\asrc$ & $\gxrb$ &  $\axrb$   &   $\ath$& \rchi\cr
0.0   & 2.53 & 2.24 & 1.60 & 11.3  & 84.4 &  0.96 \cr
0.2 & 2.53 & 2.24 & 1.57 & 10.9  & 84.1 &  0.96 \cr
0.4 & 2.53 & 2.24 & 1.54 & 10.4  & 83.8 &  0.96 \cr
0.6 & 2.53 & 2.24 & 1.51 & 10.0  & 83.4 &  0.96 \cr
0.8 & 2.53 & 2.24 & 1.47 &  9.5  & 83.1 &  0.96 \cr
1.0   & 2.53 & 2.24 & 1.40 &  9.0  & 83.2 &  0.96 \cr
2.0   & 2.53 & 2.24 & 1.14 &  6.8  & 80.7 &  0.96 \cr
\noalign{\vskip 2pt\hrule}
}}
\tabletext{
*The degree of freedom is 282 for all the fits.}
\endtable
}







\def\tablei{
\begintable*{3}
\caption{{\bf Table 2 (a) }Sources detected by PSPC in 1--2~keV}
\vbox{
\offinterlineskip\halign{
 \strut
##\quad& \hfil ## \quad& \hfil## \quad & \hfil ## \quad& \hfil ## \quad& \hfil ## \quad& \hfil ## \quad & ##\hfil \quad & \hfil ## \hfil\quad \cr
\noalign{\hrule\vskip 4pt}
\omit\tbf No. \hfil& \tbf RA\hfil &\tbf Dec\hfil &\tbf Source \hfil &\tbf Bkgd \hfil &\tbf  SNR \hfil& \tbf Name \hfil&\hfil \tbf Note \hfil & \tbf Hardness \cr
\omit\tbf \hfil& \tbf \hfil &\tbf \hfil &\tbf Count\hfil &\tbf Count\hfil &\tbf   \hfil& \hfil&\hfil \tbf \hfil & \tbf Ratio\cr
\noalign{\vskip 4pt}
\omit \hfil (1) &  \hfil & (2) \hfil & (3)   \hfil& (4) \hfil& (5) \hfil & 
(6) \hfil & (7) \hfil &  (8) \cr
\noalign{\vskip 4pt\hrule\vskip 4pt}
  1 & $  55.55$ & $ -43.81$ &  41.7$\pm$  8.0 &   3.6$\pm$  0.7 &   19.9 &       & +   & $   0.22\pm   0.14$\cr
   2 & $  55.56$ & $ -43.82$ &  14.9$\pm$  4.5 &   6.0$\pm$  0.9 &    3.6       &          & +   & $  -0.15\pm   0.19$\cr
   3 & $  55.37$ & $ -43.84$ &  26.4$\pm$  6.2 &   5.2$\pm$  0.8 &    9.2 &  XSF3:19 & +   & $  -0.05\pm   0.17$\cr
   4 & $  55.64$ & $ -43.84$ &  18.2$\pm$  5.0 &   5.1$\pm$  0.9 &    5.7 &          & +   & $  -0.19\pm   0.21$\cr
   5 & $  55.55$ & $ -43.85$ &  35.4$\pm$  6.8 &   5.7$\pm$  0.8 &   12.4 &  XSF3:17 & +   & $  -0.04\pm   0.13$\cr
   6 & $  55.46$ & $ -43.85$ &  13.5$\pm$  4.1 &   5.4$\pm$  0.9 &    3.5 &          & + $\odot$& $   0.98\pm   1.07$\cr
   7 & $  55.37$ & $ -43.86$ &  58.2$\pm$  8.9 &   5.4$\pm$  0.8 &   22.8 &          & +   & $   0.04\pm   0.11$\cr
   8 & $  55.58$ & $ -43.86$ &  17.5$\pm$  4.8 &   4.7$\pm$  0.8 &    5.9 &          &     & $   0.41\pm   0.53$\cr
   9 & $  55.25$ & $ -43.87$ &  20.9$\pm$  6.0 &   4.1$\pm$  0.7 &    8.2 &          & +   & $  -0.37\pm   0.23$\cr
  10 & $  55.65$ & $ -43.88$ &  33.9$\pm$  6.8 &   7.2$\pm$  0.8 &   10.0 &  XSF3:20 &     & $  -0.02\pm   0.16$\cr
   &  &  &     &      &       &   &     &\cr
  11 (S1)& $  55.46$ & $ -43.88$ & 488.8$\pm$ 24.3 &   4.9$\pm$  0.6 &  218.3 &  XSF3:21 &   $\odot$& $  -0.20\pm   0.03$\cr
  12 & $  55.63$ & $ -43.88$ &  43.0$\pm$  7.6 &   6.7$\pm$  0.8 &   14.0 &          &     & $  -0.01\pm   0.14$\cr
  13 & $  55.35$ & $ -43.89$ &  15.8$\pm$  4.6 &   4.0$\pm$  0.7 &    5.9 &          & +   & $  -0.37\pm   0.23$\cr
  14 & $  55.70$ & $ -43.90$ &  12.4$\pm$  4.1 &   5.3$\pm$  0.8 &    3.1 &          & +   & $  -0.09\pm   0.61$\cr
  15 & $  55.49$ & $ -43.91$ &   9.7$\pm$  3.4 &   4.1$\pm$  0.6 &    2.7 &          &   $\odot$& $   0.60\pm   0.91$\cr
  16 & $  55.68$ & $ -43.92$ &  13.0$\pm$  4.1 &   5.6$\pm$  0.8 &    3.1 &          &     & $   0.25\pm   0.50$\cr
  17 & $  55.37$ & $ -43.92$ &  11.4$\pm$  3.8 &   3.6$\pm$  0.6 &    4.1 &          & +   & $  -0.07\pm   0.34$\cr
  18 & $  55.76$ & $ -43.92$ &  67.1$\pm$  9.7 &   6.5$\pm$  0.8 &   23.7 &  XSF3:25 & +   & $  -0.12\pm   0.09$\cr
  19 & $  55.58$ & $ -43.92$ &  11.0$\pm$  3.7 &   3.9$\pm$  0.6 &    3.6 &  XSF3:23 &     & $  -0.48\pm   0.22$\cr
  20 & $  55.83$ & $ -43.92$ &  30.0$\pm$  6.7 &   6.1$\pm$  0.9 &    9.6 &  XSF3:24 & +   & $  -0.21\pm   0.15$\cr
   &  &  &     &      &       &   &     &\cr
  21 & $  55.24$ & $ -43.95$ &  24.6$\pm$  5.6 &   4.3$\pm$  0.6 &    9.8 &  XSF3:28 & +   & $   0.21\pm   0.20$\cr
  22 & $  55.87$ & $ -43.95$ &  13.5$\pm$  4.8 &   5.4$\pm$  0.9 &    3.5 &          & +   & $  -0.61\pm   0.33$\cr
  23 & $  55.31$ & $ -43.96$ &  20.8$\pm$  5.0 &   3.7$\pm$  0.6 &    8.9 &  XSF3:29 & +   & $   0.11\pm   0.24$\cr
  24 & $  55.48$ & $ -43.96$ &  11.7$\pm$  3.7 &   4.0$\pm$  0.6 &    3.8 &          &     & $   1.11\pm   1.56$\cr
  25 & $  55.89$ & $ -43.96$ &  14.5$\pm$  5.1 &   4.7$\pm$  0.8 &    4.6 &          & +   & $  -0.25\pm   0.45$\cr
  26 & $  55.67$ & $ -43.97$ &  17.6$\pm$  4.7 &   4.6$\pm$  0.6 &    6.0 &          &     & $   0.37\pm   0.40$\cr
  27 & $  55.61$ & $ -44.00$ &  18.9$\pm$  4.7 &   5.2$\pm$  0.7 &    6.0 &          &   $\odot$& $   0.15\pm   0.38$\cr
  28 & $  55.13$ & $ -44.00$ &  16.6$\pm$  5.3 &   5.9$\pm$  0.8 &    4.4 &          & +   & $  -0.15\pm   0.46$\cr
  29 & $  55.58$ & $ -44.01$ &  31.7$\pm$  6.1 &   5.0$\pm$  0.6 &   11.9 &          &   $\odot$& $   0.13\pm   0.18$\cr
  30 & $  55.24$ & $ -44.03$ &  24.4$\pm$  5.3 &   3.6$\pm$  0.6 &   11.0 &  XSF3:32 & +Q  & $  -0.22\pm   0.18$\cr
   &  &  &     &      &       &   &     &\cr
  31 & $  55.33$ & $ -44.03$ &  20.9$\pm$  4.9 &   5.0$\pm$  0.6 &    7.1 &          &     & $  -0.07\pm   0.22$\cr
  32 & $  55.65$ & $ -44.03$ &  16.3$\pm$  4.3 &   4.3$\pm$  0.7 &    5.7 &          &   $\odot$& $   0.54\pm   0.69$\cr
  33 & $  55.28$ & $ -44.03$ &  21.0$\pm$  4.9 &   3.8$\pm$  0.6 &    8.9 &  XSF3:33 & +   & $   0.31\pm   0.25$\cr
  34 & $  55.42$ & $ -44.04$ &  15.8$\pm$  4.2 &   4.4$\pm$  0.6 &    5.4 &          &     & $   0.54\pm   0.41$\cr
  35 & $  55.16$ & $ -44.05$ &  13.2$\pm$  4.2 &   4.5$\pm$  0.6 &    4.1 &  XSF3:35 & +   & $   0.04\pm   0.41$\cr
  36 & $  55.94$ & $ -44.05$ &  21.1$\pm$  5.9 &   5.0$\pm$  0.7 &    7.3 &          & +   & $   0.48\pm   0.31$\cr
  37 (Q1)& $  55.50$ & $ -44.05$ & 152.2$\pm$ 13.1 &   4.3$\pm$  0.6 &   71.6 &  XSF3:36 &  Q$\odot$& $  -0.15\pm   0.06$\cr
  38 & $  55.96$ & $ -44.05$ &  35.2$\pm$  7.7 &   4.9$\pm$  0.8 &   13.7 &  XSF3:37 & +   & $   0.21\pm   0.15$\cr
  39 & $  55.45$ & $ -44.05$ &  11.2$\pm$  3.5 &   3.1$\pm$  0.5 &    4.6 &          &     & $   0.28\pm   0.45$\cr
  40 & $  55.61$ & $ -44.07$ &  19.8$\pm$  4.5 &   3.8$\pm$  0.6 &    8.2 &          &   $\odot$& $   0.49\pm   0.36$\cr
   &  &  &     &      &       &   &     &\cr
  41 (Q2)& $  55.65$ & $ -44.07$ & 180.7$\pm$ 13.9 &   5.0$\pm$  0.6 &   78.7 &  XSF3:38 &  Q$\odot$& $  -0.20\pm   0.05$\cr
  42 & $  55.72$ & $ -44.08$ &  11.9$\pm$  3.6 &   3.4$\pm$  0.5 &    4.6 &          &     & $   0.39\pm   0.53$\cr
  43 & $  55.29$ & $ -44.09$ &  14.6$\pm$  4.1 &   4.2$\pm$  0.6 &    5.0 &  XSF3:42 &  Q  & $  -0.33\pm   0.27$\cr
  44 & $  55.96$ & $ -44.09$ &  16.6$\pm$  5.2 &   6.5$\pm$  0.9 &    4.0 &          & +   & $   0.25\pm   0.45$\cr
  45 & $  55.85$ & $ -44.10$ &  45.3$\pm$  7.5 &   4.5$\pm$  0.6 &   19.3 &  XSF3:44 & +Q  & $   0.20\pm   0.15$\cr
  46 & $  55.25$ & $ -44.11$ &  22.0$\pm$  5.0 &   4.1$\pm$  0.6 &    8.9 &  XSF3:45 & +Q  & $  -0.37\pm   0.20$\cr
  47 & $  55.61$ & $ -44.12$ &  22.9$\pm$  5.0 &   4.6$\pm$  0.6 &    8.6 &  XSF3:46 &   $\odot$& $  -0.13\pm   0.22$\cr
  48 & $  55.95$ & $ -44.13$ &  21.4$\pm$  5.9 &   5.0$\pm$  0.8 &    7.3 &          & +   & $   0.39\pm   0.34$\cr
  49 & $  55.95$ & $ -44.15$ &  12.4$\pm$  4.7 &   4.3$\pm$  0.7 &    3.9 &          & +   & $   1.11\pm   1.20$\cr
  50 & $  55.51$ & $ -44.16$ &  22.3$\pm$  5.0 &   3.5$\pm$  0.5 &   10.0 &  XSF3:48 &  Q  & $   0.04\pm   0.20$\cr
\noalign{\vskip 2pt\hrule}
}}
\endtable
}

\def\tablej{
\begintable*{4}
\caption{{\bf Table 2 (a) } -- {\it continued} }
\vbox{
\offinterlineskip\halign{
 \strut
##\quad& \hfil ## \quad& \hfil## \quad & \hfil ## \quad& \hfil ## \quad& \hfil ## \quad& \hfil ## \quad & ##\hfil \quad & \hfil ## \hfil\quad \cr
\noalign{\hrule\vskip 4pt}
\omit\tbf No. \hfil& \tbf RA\hfil &\tbf Dec\hfil &\tbf Source \hfil &\tbf Bkgd \hfil &\tbf  SNR \hfil& \tbf Name \hfil&\hfil \tbf Note \hfil & \tbf Hardness \cr
\omit\tbf \hfil& \tbf \hfil &\tbf \hfil &\tbf Count\hfil &\tbf Count\hfil &\tbf   \hfil& \hfil&\hfil \tbf \hfil & \tbf Ratio\cr
\noalign{\vskip 4pt}
\omit \hfil (1) &  \hfil & (2) \hfil & (3)   \hfil& (4) \hfil& (5) \hfil & 
(6) \hfil & (7) \hfil & (8) \cr
\noalign{\vskip 4pt\hrule\vskip 4pt}
  51 & $  55.38$ & $ -44.16$ &  11.7$\pm$  3.7 &   5.0$\pm$  0.7 &    3.0 &          &     & $  -0.25\pm   0.37$\cr
  52 & $  55.81$ & $ -44.16$ &  17.5$\pm$  4.7 &   6.2$\pm$  0.7 &    4.5 &          &     & $  -0.23\pm   0.39$\cr
  53 & $  55.48$ & $ -44.17$ &  12.5$\pm$  3.8 &   5.2$\pm$  0.6 &    3.2 &          &   $\odot$& $   0.48\pm   0.39$\cr
  54 & $  55.32$ & $ -44.17$ &  70.9$\pm$  9.0 &   4.5$\pm$  0.6 &   31.3 &  XSF3:51 &   $\odot$& $   0.23\pm   0.11$\cr
  55 & $  55.78$ & $ -44.17$ &  14.2$\pm$  4.1 &   4.4$\pm$  0.6 &    4.7 &          &     & $  -0.05\pm   0.31$\cr
  56 & $  55.39$ & $ -44.17$ &  16.3$\pm$  4.4 &   4.3$\pm$  0.6 &    5.8 &  XSF3:53 &  Q  & $  -0.25\pm   0.21$\cr
  57 & $  55.70$ & $ -44.18$ &  13.1$\pm$  3.8 &   4.5$\pm$  0.6 &    4.0 &          &     & $   0.37\pm   0.62$\cr
  58 & $  55.56$ & $ -44.19$ &  26.5$\pm$  5.5 &   4.8$\pm$  0.6 &    9.8 &          &     & $  -0.17\pm   0.16$\cr
  59 & $  55.59$ & $ -44.19$ &  18.1$\pm$  4.5 &   5.4$\pm$  0.6 &    5.5 &          &     & $   0.71\pm   0.66$\cr
  60 & $  55.80$ & $ -44.19$ &  29.6$\pm$  6.1 &   5.8$\pm$  0.7 &    9.8 &          &     & $   0.45\pm   0.30$\cr
   &  &  &     &      &       &   &     &\cr
  61 & $  55.98$ & $ -44.19$ &  20.1$\pm$  6.1 &   6.0$\pm$  0.9 &    5.7 &  XSF3:56 & +Q  & $  -0.04\pm   0.28$\cr
  62 & $  55.26$ & $ -44.19$ &  67.1$\pm$  8.8 &   4.0$\pm$  0.6 &   31.8 &  XSF3:54 & +Q  & $  -0.08\pm   0.09$\cr
  63 & $  55.56$ & $ -44.20$ &  43.7$\pm$  7.1 &   4.6$\pm$  0.6 &   18.2 &  XSF3:57 &  Q  & $  -0.20\pm   0.12$\cr
  64 & $  55.75$ & $ -44.22$ &  19.5$\pm$  4.9 &   5.3$\pm$  0.7 &    6.2 &          &     & $   0.29\pm   0.53$\cr
  65 & $  55.47$ & $ -44.23$ &  28.9$\pm$  5.8 &   5.7$\pm$  0.7 &    9.8 &  XSF3:58 &  Q  & $  -0.14\pm   0.19$\cr
  66 & $  55.37$ & $ -44.24$ &  19.6$\pm$  4.8 &   4.6$\pm$  0.6 &    7.0 &  XSF3:59 &  Q  & $   0.36\pm   0.34$\cr
  67 & $  55.23$ & $ -44.25$ &  16.4$\pm$  4.4 &   8.0$\pm$  0.9 &    3.0 &  XSF3:63 & +   & $   0.45\pm   0.77$\cr
  68 & $  55.79$ & $ -44.25$ &  17.4$\pm$  4.6 &   4.0$\pm$  0.6 &    6.7 &          &     & $   0.44\pm   0.34$\cr
  69 & $  55.26$ & $ -44.26$ &  11.7$\pm$  3.7 &   4.9$\pm$  0.7 &    3.1 &  XSF3:61 & +Q  & $   0.33\pm   0.43$\cr
  70 & $  55.76$ & $ -44.26$ &  32.3$\pm$  6.2 &   4.9$\pm$  0.6 &   12.3 &  XSF3:64 &     & $  -0.63\pm   0.11$\cr
  71 & $  55.54$ & $ -44.27$ &  41.0$\pm$  6.9 &   3.7$\pm$  0.6 &   19.3 &  XSF3:65 &  Q  & $   0.07\pm   0.17$\cr
  72 & $  55.47$ & $ -44.27$ &  53.4$\pm$  8.0 &   5.0$\pm$  0.7 &   21.6 &  XSF3:66 &  Q$\odot$& $   0.12\pm   0.13$\cr
  73 & $  55.59$ & $ -44.28$ &  30.6$\pm$  6.1 &   3.7$\pm$  0.6 &   14.0 &  XSF3:68 &     & $   0.17\pm   0.19$\cr
  74 & $  55.53$ & $ -44.28$ &  25.1$\pm$  5.5 &   3.9$\pm$  0.7 &   10.7 &          &     & $   0.31\pm   0.24$\cr
  75 & $  55.74$ & $ -44.29$ &  14.9$\pm$  4.3 &   4.3$\pm$  0.6 &    5.1 &          &     & $  -0.22\pm   0.35$\cr
  76 & $  55.28$ & $ -44.29$ &  34.5$\pm$  6.4 &   5.2$\pm$  0.7 &   12.8 &  XSF3:67 & +Q  & $  -0.10\pm   0.14$\cr
  77 & $  55.89$ & $ -44.29$ &  15.1$\pm$  5.0 &   6.4$\pm$  0.9 &    3.5 &          & +   & $   0.18\pm   0.48$\cr
  78 & $  55.43$ & $ -44.30$ &  12.3$\pm$  3.9 &   4.9$\pm$  0.7 &    3.3 &          &     & $  -0.00\pm   0.42$\cr
  79 & $  55.55$ & $ -44.30$ &   9.9$\pm$  3.5 &   2.6$\pm$  0.5 &    4.5 &          &     & $   0.30\pm   0.36$\cr
  80 & $  55.60$ & $ -44.30$ &  22.7$\pm$  5.4 &   4.7$\pm$  0.7 &    8.3 &          &     & $  -0.04\pm   0.17$\cr
   &  &  &     &      &       &   &     &\cr
  81 & $  55.22$ & $ -44.31$ &  17.6$\pm$  4.7 &   6.0$\pm$  0.9 &    4.7 &          & +   & $   0.25\pm   0.23$\cr
  82 & $  55.91$ & $ -44.31$ &  13.6$\pm$  4.8 &   5.6$\pm$  0.8 &    3.4 &          & +   & $   0.05\pm   0.38$\cr
  83 & $  55.21$ & $ -44.32$ &  11.9$\pm$  4.0 &   4.8$\pm$  0.8 &    3.3 &          & +   & $   0.94\pm   0.64$\cr
  84 & $  55.80$ & $ -44.33$ &  25.9$\pm$  6.1 &   4.8$\pm$  0.7 &    9.6 &  XSF3:70 & +Q  & $  -0.00\pm   0.17$\cr
  85 & $  55.49$ & $ -44.34$ &  21.9$\pm$  5.2 &   5.4$\pm$  0.7 &    7.1 &          &     & $  -0.06\pm   0.22$\cr
  86 & $  55.36$ & $ -44.35$ &  16.3$\pm$  4.5 &   4.6$\pm$  0.6 &    5.4 &          & +   & $  -0.26\pm   0.22$\cr
  87 & $  55.44$ & $ -44.36$ &  20.0$\pm$  5.0 &   6.4$\pm$  0.8 &    5.3 &          &     & $   0.33\pm   0.36$\cr
  88 & $  55.48$ & $ -44.36$ &  31.1$\pm$  6.2 &   5.0$\pm$  0.7 &   11.6 &  XSF3:71 &  Q  & $  -0.01\pm   0.17$\cr
  89 & $  55.72$ & $ -44.37$ &  56.6$\pm$  9.1 &   4.1$\pm$  0.6 &   26.0 &  XSF3:72 & +Q  & $  -0.30\pm   0.10$\cr
  90 & $  55.52$ & $ -44.41$ &  13.1$\pm$  4.1 &   4.9$\pm$  0.7 &    3.7 &  XSF3:78 & +   & $  -0.47\pm   0.37$\cr
   &  &  &     &      &       &   &     &\cr
  91 & $  55.60$ & $ -44.41$ &  17.7$\pm$  4.9 &   4.0$\pm$  0.6 &    6.9 &          & +   & $   0.78\pm   0.65$\cr
  92 & $  55.37$ & $ -44.42$ &  20.4$\pm$  5.5 &   5.8$\pm$  1.1 &    6.1 &  XSF3:77 & +Q  & $  -0.30\pm   0.21$\cr
\noalign{\vskip 2pt\hrule}
}}
\tabletext{(1) Sources numbered as shown in Figs~1--2.  (2) Coordinates in J2000.  (3) PSPCB+PSPCC source counts collected from a $0.75$~arcmin radius region in 1--2~keV.  (4)  Background counts collected from a ring centred on the source with the inner radius 2~arcmin and the outer radius 3~arcmin, and normalised by  the same area size as that of the source count.  (5) Signal-to-noise ratio in 1--2~keV, the detection cell size is 0.75~arcmin radius.  (6) Survey name used in Table~1 of Shanks \etal (1991), sources newly found in this work are blank in this column.   (7) Q: sources identified as QSO by Shanks \etal (1991); $\odot$: sources also detected by the ASCA SIS (Table~2b).  `+' Sources located between the the inner 16~arcmin and the 20~arcmin radius region of the PSPC.  Sources 11, 37, 41 are named as \srca, \srcb\ and \srcc, respectively in this work.  }
\endtable
}


\def\tablek{
\begintable{5}
\caption{{\bf Table 2 (b) }Sources detected by ASCA SIS}
\vbox{
\offinterlineskip\halign{
 \strut
 ##\quad& \hfil ## \quad& \hfil## \quad & \hfil ## \hfil \quad& \hfil ## \hfil\cr
\noalign{\hrule\vskip 4pt}
\omit\tbf No. \hfil &\tbf Count  \hfil &\tbf Count \hfil &\tbf SNR  & \tbf SNR   \cr
\omit \hfil &\tbf (soft) \hfil &\tbf (hard) \hfil &\tbf (soft) & \tbf (hard) \cr
\noalign{\vskip 4pt\hrule\vskip 4pt}
  6 &   44.01$\pm$   7.64 &    5.66$\pm$   5.27 &   7.0 &      \cr
 11 (S1) &  246.86$\pm$  11.99 &   33.59$\pm$   5.41 &  39.2 &   6.8\cr
 15 &   55.15$\pm$   6.93 &    1.63$\pm$   3.66 &   8.8 &      \cr
 27 &   22.14$\pm$   5.83 &    7.89$\pm$   5.49 &   3.5 &      \cr
 29 &   19.25$\pm$   5.71 &   10.45$\pm$   5.64 &   3.1 &   2.1\cr
 32 &   13.61$\pm$   5.25 &    3.78$\pm$   3.81 &   2.2 &      \cr
 37 (Q1) &   30.06$\pm$   6.24 &   10.03$\pm$   4.34 &   4.8 &   2.0\cr
 40 &   73.12$\pm$   7.55 &    6.63$\pm$   3.98 &  11.6 &      \cr
 41 (Q2) &  100.08$\pm$   8.40 &    6.63$\pm$   3.98 &  15.9 &      \cr
 47 &   27.19$\pm$   5.84 &    9.62$\pm$   4.17 &   4.3 &      \cr
 53 &   10.82$\pm$   5.49 &   19.44$\pm$   5.19 &       &   3.9\cr
 54 &   11.86$\pm$   5.39 &   11.68$\pm$   4.48 &       &   2.4\cr
 72 &    9.21$\pm$   5.01 &   12.62$\pm$   4.34 &       &   2.6\cr
\noalign{\vskip 2pt\hrule}
}}
\tabletext{Sources  detected above 2 $\sigma$ on the merged SIS0+SIS1 images (exposure time $\sim$ 40~ks)  in either 0.5--2~keV or 2--8~keV band.  Source counts are accumulated from a 1.5~arcmin radius region (background counts are subtracted in this table); the background counts are estimated from the bright-source-free region of the whole image, which are $39.71\pm 0.84$ and $24.33\pm 0.64$ for the soft and hard band, respectively.  We use a 1.5~arcmin radius detection cell to derive the  significance of source detection (SNR).}  
\endtable
}